\pdfoutput=1
\documentclass[11pt]{article}
\usepackage[utf8]{inputenc}
\usepackage[T1]{fontenc}
\usepackage{amsmath,amssymb}
\usepackage{textcomp}
\usepackage[margin=1in]{geometry}
\usepackage{graphicx}
\usepackage{booktabs}
\usepackage{longtable}
\usepackage{array}
\usepackage{ragged2e}
\usepackage[font=small,labelfont=bf]{caption}
\usepackage{setspace}
\usepackage[hidelinks]{hyperref}
\usepackage{xcolor}
\providecommand{\tightlist}{\setlength{\itemsep}{0pt}\setlength{\parskip}{0pt}}

\title{\textbf{Simulating a Post-Automation Economy}\\[4pt]
\large An agent-based, stock-flow-consistent model of automation, microfounded wealth concentration, behavioural taxation, and capital mobility}
\author{Leonard Wossnig \\ {\normalsize Department of Computer Science, University College London, London, UK} \\ {\normalsize \texttt{leonard.wossnig.17@ucl.ac.uk}}}
\date{This version: June 2026}
\begin{document}
\maketitle
\begin{abstract}
\noindent We develop an agent-based, stock-flow-consistent model of an economy undergoing automation, built to ask which fiscal instrument reaches the durable surplus that artificial intelligence creates. The model separates two channels: a competitive return on reproducible robotic capital, and a mobile, foreign-held intellectual-property rent earned by AI. Production is an endogenous nested-CES technology; wealth concentration is microfounded through heterogeneous, persistent returns to wealth; and taxation and capital mobility are modelled as behavioural responses. The central result is that the durable surplus is the foreign-held AI rent, a cross-border licence fee that corporate, robot, and compute or token taxes largely miss and that only a source-based levy (a digital-services-style tax or a withholding) reaches. The appropriate policy depends decisively on whether a country owns the automation or imports it: for a host that owns the rent the problem is domestic inequality, reached by progressive and wealth taxes; for a rent-importing host the problem is base erosion and a gradual transfer of capital ownership abroad, which a residence-based wealth tax cannot reach. We report conditional orderings, stress-tested with global (Sobol) sensitivity and a formal stability analysis, rather than point forecasts.
\end{abstract}

\vspace{0.3em}

\vspace{0.2em}
\vspace{0.6em}

\hypertarget{overview}{%
\section{Overview}\label{overview}}

This report documents a refined model of an automating economy, in which
artificial intelligence and robotics progressively shift income from
labour to capital, the ownership of that capital may sit with
households, the state, or abroad, and the state may fund a universal
basic income (UBI) through taxation. Because the right policy turns on
whether a country owns the automation or imports it, the analysis is run
for both: a country that hosts the AI owners (the United States) and one
that pays the rent abroad (the United Kingdom, or the European Union),
which, as section 8 shows, need almost opposite responses.

The refined version (v3) was built specifically to answer the most
damaging criticisms of the first-generation model: that its wealth tax
was frictionless, that its concentration engine was an imposed kinetic
kernel rather than an economic mechanism, and that its calibration was
not disciplined. It replaces the kinetic kernel with microfounded
heterogeneous returns, makes the wealth tax induce avoidance and capital
flight, replaces the ad-hoc investment rule with a stable
differential-saving closure, and adds a formal local-stability analysis.
Two further hardening steps, a per-agent offshore-wealth metric and a
global sensitivity analysis, sharpen the central finding and are
reported in sections 6 and 9.

The headline conclusion is threefold, and is throughout a positive,
closure-conditional comparison rather than an optimal-tax or welfare
claim. First, stock taxes (a wealth tax, or a progressive wealth tax)
compress the wealth Gini sharply, from about 0.63 under laissez-faire to
about 0.15, while flow taxes (income tax plus UBI) leave it near the
laissez-faire level. Second, solvency is a distinct axis that turns on
the adequacy of taxation, not on ownership form: adequately-taxed
regimes stay solvent and even accumulate a sovereign equity fund out of
surpluses, whereas a welfare state promising a universal income on too
thin a tax base is pushed onto an explosive r-greater-than-g debt path.
Third, a behavioural wealth tax introduces only a modest wedge between
measured and true inequality once fled capital is attributed back, and
the distributional results are governed mainly by the primitive
dispersion of returns to wealth rather than by the behavioural
elasticities that are hardest to estimate (the global sensitivity
analysis quantifies this, with the caveat that those
elasticities\textquotesingle{} indices are themselves imprecisely
estimated at the sample size used). All are stated carefully below,
including a correction to a more dramatic but mis-calibrated earlier
version of the composition result. They hold under this
model\textquotesingle s closure (output at potential, investment as the
goods-market residual, equity valued at book) and should be read as
numerically supported within it, not as general theorems. Throughout,
these are conditional results that depend on the model\textquotesingle s
assumptions; those assumptions and limitations, and which of the
conclusions are robust to them, are set out in full in Appendix A.

\textbf{Reproducibility.} All results come from Monte Carlo runs over 10
to 12 seeds. Two accounting invariants (the four
sectors\textquotesingle{} deposits sum to zero; the sum of sector net
worths equals the capital stock) hold to machine precision across the
reported runs and tested configurations, and are asserted in a suite of
142 passing tests. Level quantities are reported as a ratio to
per-capita output, which is scale-free.

\hypertarget{the-live-policy-debate-this-speaks-to}{%
\subsection{The live policy debate this speaks
to}\label{the-live-policy-debate-this-speaks-to}}

This is not an abstract exercise. Through 2025 and 2026 a concrete
version of the question moved into mainstream politics on two fronts.
The first is displacement: Anthropic\textquotesingle s Dario Amodei
warned that AI could remove a large share of entry-level white-collar
roles and push unemployment into double digits within a few years
(Axios, 2025), and a run of layoffs at firms including Amazon, Meta and
UPS, announced alongside record AI spending, revived a long-standing
worry that fewer taxed workers mean a thinner fiscal base (EL País,
2025). The second is what, if anything, to tax in response, a question
that goes back to the Gates and Phelps robot-tax proposals.

The 2026 proposals cluster around a levy on AI usage. Mark Cuban floated
a federal per-token charge on providers; the Michigan Senate candidate
Mallory McMorrow built a "Token Tax" into a worker-protection plan
funding an AI Workforce Reinvestment Fund; Elizabeth Warren argued in
Time for an excise on data-centre energy with the door left open to
bolder measures; DuckDuckGo\textquotesingle s Gabriel Weinberg proposed
a ten per cent surcharge on token charges (roughly the employer
payroll-tax rate) held in a lockbox for displaced workers; and Amodei
himself mooted a three per cent revenue levy on model usage. The
first-principles case behind them is that across the OECD a worker costs
about a third of labour cost in tax, so when an agent does the same job
the value re-appears as margin or capital gain that is taxed far less,
eroding the base (a risk the IMF flagged in 2024).

The pushback is equally sharp. Dave Friedman\textquotesingle s response
to Cuban makes four arguments against a token base: a token is an
internal accounting unit, not a unit of value, work or energy; the base
is endogenous to the taxpayer, since providers write their own
tokenizers and can shrink the count without changing the activity;
per-token prices have fallen roughly two-hundred-fold a year, so a fixed
rate is either confiscatory or quickly indexed to nothing; and a
provider-level tax on domestic firms is in effect a subsidy for foreign
inference. Palmer Luckey makes the territorial point bluntly, that such
a tax mostly makes foreign models more attractive while building the
apparatus to monitor AI usage, and others (Sinofsky, Petersen) reject
the unit outright. The academic reviews are cautious rather than
hostile: the IMF (2024) and Brookings both decline to single out AI,
preferring broader capital or consumption bases, with Brookings leaning
toward a consumption-side levy folded into VAT with business-to-business
use exempted.

This paper takes no side on whether to tax AI. It asks the question
underneath the slogans: in an economy where AI shifts income from labour
to a capital-and-IP layer that can sit abroad, which tax base actually
reaches the durable surplus, and who ends up owning the automated
capital. The two-channel results (section 6) bear directly on the choice
of base, on the territoriality objection, and on the displaced-worker
funding question, and section 8 reads the model against these specific
proposals. One note on perspective: most of the proposals, and most of
the pushback, are framed for the United States, the home of the
providers, whereas the model is framed for a host that imports the AI
rent (read the United Kingdom). That difference turns out to matter for
whether the territoriality objection bites.

Because that difference is the organising idea of the paper, we make it
explicit by separating two regimes. In the first the AI is owned at
home, the case of the United States, where providers such as OpenAI and
Anthropic are resident: the rent stays in the country, and the policy
question is how its concentration among domestic owners is shared. In
the second the AI is owned abroad, the case of the United Kingdom with a
United States or Chinese owner: the rent leaves as a licence fee, and
the policy question is how, if at all, the host can reach it and slow
the transfer of ownership. We take the domestic-owner case first, as the
simpler benchmark with a familiar policy menu, then turn to the
foreign-owner case, which is this paper\textquotesingle s main
contribution and occupies most of section 6. A single parameter, the
foreign-owned share of the AI IP, moves the model between the two.

A second dimension, beyond where the rent goes, is how large it becomes,
and it is central to how the results should be read. The experiments
hold the rent at a fixed share of cognitive value added, a roughly
constant eighth of output, which is deliberately the conservative case.
Two forces are more likely to push it up than to hold it flat. As the
technology becomes essential and the market concentrates, the
owner\textquotesingle s pricing power grows, so the markup need not stay
fixed. And as AI-native production does whole jobs rather than assisting
inside an otherwise-human firm, the AI cluster captures a rising share
of total output rather than a fixed slice of one cluster, which is
arguably already visible in software. When the rent rises with
automation in this way the surplus roughly doubles to a quarter of
output, or approaches half in the extreme, and the foreign-ownership
drift accelerates from about seventy per cent of the capital stock
toward almost the whole of it (section 6, Figure 26). The flat-rent
results reported everywhere else are therefore a floor, not a central
estimate. The policy stakes scale with the rent: the same instruments
still reach it, but they capture far more, the ownership transfer they
curb is larger and faster, and the cost of waiting is correspondingly
higher (section 8).

\hypertarget{prior}{}
\hypertarget{prior-and-related-work}{%
\section{Prior and related work}\label{prior-and-related-work}}

The model sits at the intersection of five literatures.

\hypertarget{automation-and-the-factor-distribution}{%
\subsection{Automation and the factor
distribution}\label{automation-and-the-factor-distribution}}

Acemoglu and Restrepo (2022) model tasks allocated between labour and
capital, with automation expanding the set of tasks performed by
capital. The CES task block here is the reduced macro form of that
mechanism. The closest analytic antecedent for the wealth dimension is
Moll, Rachel and Restrepo (2022), who link automation to the personal
income and wealth distributions through a return gap that generalises
Piketty\textquotesingle s r $-$ g, with top-tail inequality governed by a
random growth process. The refined model\textquotesingle s concentration
engine is built directly on that random-growth logic. We note the
empirical debate the model must respect: Rognlie (2015) shows much of
the measured rise in the capital share is housing and markup, not
productive capital, and the recent AI-labour evidence is mixed on
whether displacement is yet visible in aggregate; section 4 frames the
automation scenario accordingly.

\hypertarget{heterogeneous-returns-to-wealth}{%
\subsection{Heterogeneous returns to
wealth}\label{heterogeneous-returns-to-wealth}}

The refined concentration engine rests on the empirical regularity that
returns to wealth are heterogeneous and persistent across households
(Fagereng, Guiso, Malacrino and Pistaferri 2020). The resulting Pareto
tail is the random-growth result of Benhabib, Bisin and Zhu, in which
idiosyncratic persistent returns plus a demographic reset produce a
stationary heavy-tailed wealth distribution. This replaces the
econophysics kernel used in the first generation.

\hypertarget{behavioural-responses-to-wealth-taxation}{%
\subsection{Behavioural responses to wealth
taxation}\label{behavioural-responses-to-wealth-taxation}}

The behavioural wealth tax is calibrated to the empirical literature on
avoidance and mobility, and the two channels are kept distinct because
they have different welfare and revenue implications. The first is base
erosion: Brülhart and co-authors estimate a bunching elasticity of
declared taxable wealth with respect to the net-of-tax rate of around
0.7 to 0.8. The model maps this to the avoidance\_elasticity parameter
through a constant-elasticity base factor, taxable\_base = (1 $-$
$\tau$\textsubscript{w})\textsuperscript{$\varepsilon$\textsubscript{a}} with
$\varepsilon$\textsubscript{a} = 0.75, so a higher rate mechanically shrinks the
base it applies to (under-declaration, reclassification, legal
sheltering) without any wealth physically leaving. The second is
relocation: Jakobsen, Kleven, Kolsrud, Landais and Muñoz find
international migration responses to wealth taxes, on the order of a two
per cent reduction in the stock of top wealth-holders per percentage
point of rate. The model maps this to migration\_semi\_elast = 0.02 as a
target OFFSHORE share, target\_offshore = 0.02 $\times$ ($\tau$\textsubscript{w} in
pp) of household equity, towards which the offshore stock adjusts
partially each period, drawn disproportionately from the top of the
distribution. The distinction matters: avoidance reduces measured
revenue but the wealth stays in the domestic distribution, whereas
migration removes wealth from the domestic base but the per-agent
offshore account still attributes it to its owner, so the ``true'' Gini
(section 6) counts it while the ``measured'' Gini does not. Treating the
two separately is what lets experiment B decompose the behavioural
response into an avoidance leg and a mobility leg.

\hypertarget{stock-flow-consistent-and-agent-based-macro}{%
\subsection{Stock-flow-consistent and agent-based
macro}\label{stock-flow-consistent-and-agent-based-macro}}

The accounting backbone follows Godley and Lavoie (2007); the
agent-based stock-flow-consistent benchmark is Caiani and co-authors
(2016). The closest agent-based prior art is Carvalho and Di Guilmi
(2020), which studies technological unemployment and inequality but
locates inequality on the income and credit side rather than in the
ownership of automated capital. The standard methodological critique of
this class (over-parameterisation, stylised-fact-only validation) is
addressed in section 5 through a Sobol sensitivity analysis and the
stationary-baseline and stability checks.

\hypertarget{optimal-taxation-and-open-economy}{%
\subsection{Optimal taxation and open
economy}\label{optimal-taxation-and-open-economy}}

Guerreiro, Rebelo and Teles (2022) study the optimal taxation of robots;
the present model is positive rather than optimal-tax. The
foreign-ownership dimension uses a rest-of-world balance-sheet sector,
motivated by the scale of profit repatriation documented by Parnreiter,
Steinwärder and Kolhoff (2024).

\hypertarget{model}{}
\hypertarget{the-model}{%
\section{The model}\label{the-model}}

\hypertarget{the-model-at-a-glance}{%
\subsection{The model at a glance}\label{the-model-at-a-glance}}

The diagram shows the five parties and the flows between them. The firm
produces in two channels: a routine cluster (routine workers plus
robotic capital) and a cognitive cluster (cognitive workers plus AI
compute). Wages and dividends flow to domestic households; taxes flow to
the state, which pays a UBI back; and a foreign owner abroad (read the
United States or China) supplies the robots and AI compute, owns equity,
and collects the AI IP rent as a licence fee that leaves the country.
Each tax lever (in sienna) acts at a different point, which is the whole
reason the instruments differ.

\begin{figure}[htbp]
\centering
\includegraphics[width=0.9\linewidth]{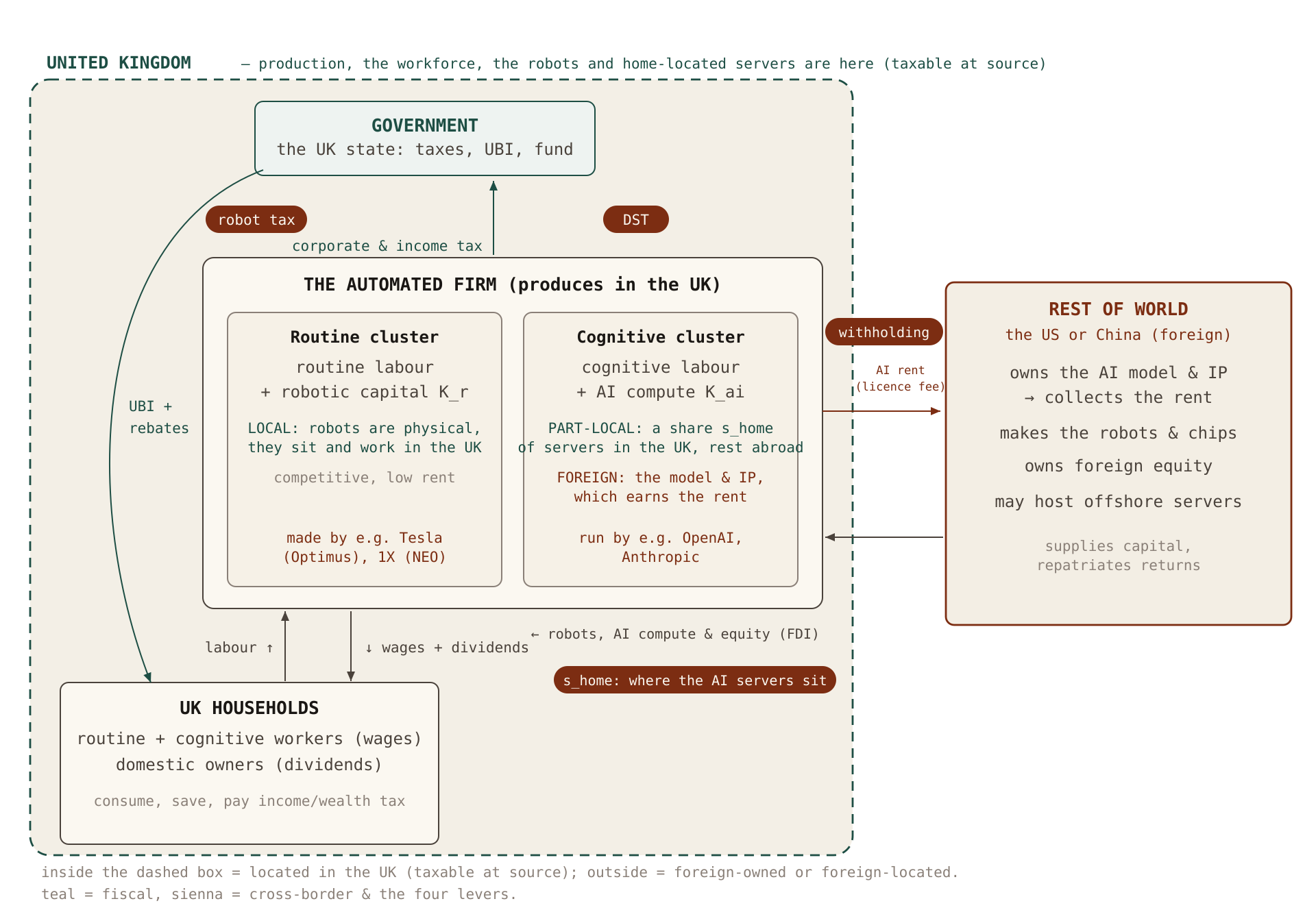}
\caption*{\textbf{Figure 1a} The parties and their relations, and what is
local versus foreign. Everything inside the dashed boundary is located
in the United Kingdom and is therefore reachable by tax at source: the
workforce, the firm\textquotesingle s operations, the physical robots,
and the home-located share of the AI servers. Outside the boundary sits
the foreign owner (read the United States or China), which owns the AI
model and its intellectual property (the source of the rent), makes the
robots and chips, owns foreign equity, and may host the offshore share
of the servers. The routine cluster pairs routine workers with robotic
capital that is fully physical and domestic; the cognitive cluster pairs
cognitive workers with AI compute whose servers are split between home
and abroad by the share s\_home, while the model and IP behind it stay
foreign. Grey flows are real and income flows (labour, wages and
dividends between households and the firm); teal flows are fiscal (tax
up to the state, UBI and rebates back to households); sienna flows cross
the border (the AI rent leaving as a licence fee, and the inward supply
of robots, compute and equity as foreign direct investment). The four
sienna levers act at different points, which is the whole reason the
instruments differ: the robot tax on the physical robotic income inside
the UK, the DST on AI revenue earned in the UK, the withholding on the
rent as it crosses the border, and s\_home on how much of the AI compute
is located, and so taxable, at home.}
\end{figure}

\hypertarget{sectors-and-invariants}{%
\subsection{Sectors and invariants}\label{sectors-and-invariants}}

Four sectors carry balance sheets: households (N agents), the firm or
capital-owning sector, the government, and the rest of the world (RoW).
Deposits are a liquid claim that sums to zero across sectors; equity
represents claims on the capital stock K, split between households, the
state, and RoW. Two identities hold every period by construction and are
tested to machine precision: the four sectors\textquotesingle{} deposit
balances sum to zero, and the sum of sector net worths equals real
assets (the capital stock K, with equity claims summing exactly to K and
the firm holding no residual cash).

The two matrices below make the stock and flow structure inspectable in
the stock-flow-consistent tradition (Godley and Lavoie 2007; Caiani et
al. 2016). In the balance-sheet matrix every row sums to its
economy-wide total (deposits to zero, equity to the capital stock once
the firm\textquotesingle s issued-equity liability is netted) and the
net-worth row sums to the capital stock. In the transaction-flow matrix
every row sums to zero (every payment is someone\textquotesingle s
receipt) and each column\textquotesingle s current and capital entries
balance, which is what the deposit and net-worth invariants verify
numerically.

\begin{longtable}[]{@{}llllll@{}}
\caption*{Balance-sheet matrix (end-of-period stocks; a plus sign is an
asset, a minus sign a liability).}\tabularnewline
\toprule\noalign{}
Stock & Households & Firm & Government & RoW & $\Sigma$ \\
\midrule\noalign{}
\endfirsthead
\toprule\noalign{}
Stock & Households & Firm & Government & RoW & $\Sigma$ \\
\midrule\noalign{}
\endhead
\bottomrule\noalign{}
\endlastfoot
Capital & & +K & & & +K \\
Deposits / bonds & +D\textsubscript{h} & +D\textsubscript{f}$\approx$0 &
+D\textsubscript{g} & +D\textsubscript{r} & 0 \\
Equity (claims on K) & +E\textsubscript{h} & $-$K & +E\textsubscript{s} &
+E\textsubscript{r} & 0 \\
\textbf{Net worth} & E\textsubscript{h}+D\textsubscript{h} & $\approx$0 &
E\textsubscript{s}+D\textsubscript{g} &
E\textsubscript{r}+D\textsubscript{r} & +K \\
\end{longtable}

Deposits sum to zero (a government bond is a household/RoW asset and a
government liability); equity claims sum to K, so once the
firm\textquotesingle s issued equity ($-$K) is netted against the capital
it holds (+K) the firm\textquotesingle s net worth is just its residual
cash, which the closure keeps at zero. Household deposits and equity are
both non-negative by construction (the no-overdraft rule and the equity
floor), so household net worth is non-negative and the Gini needs no
negative-value convention.

\begin{longtable}[]{@{}lllllll@{}}
\caption*{Transaction-flow matrix (within-period flows; a plus sign is a
receipt/source, a minus sign a payment/use; the firm is split into a
current and a capital account).}\tabularnewline
\toprule\noalign{}
Flow & Households & Firm~(cur.) & Firm~(cap.) & Gov\textquotesingle t &
RoW & $\Sigma$ \\
\midrule\noalign{}
\endfirsthead
\toprule\noalign{}
Flow & Households & Firm~(cur.) & Firm~(cap.) & Gov\textquotesingle t &
RoW & $\Sigma$ \\
\midrule\noalign{}
\endhead
\bottomrule\noalign{}
\endlastfoot
Consumption & $-$C & +C & & & & 0 \\
Government spending & & +G & & $-$G & & 0 \\
Wages & +W & $-$W & & & & 0 \\
Profit (after tax) & & $-$$\Pi$ & +$\Pi$ & & & 0 \\
Dividends & +Div\textsubscript{h} & & $-$Div & +Div\textsubscript{s} &
+Div\textsubscript{r} & 0 \\
Corporate tax & & $-$T\textsubscript{c} & & +T\textsubscript{c} & & 0 \\
Income + wealth tax & $-$T\textsubscript{y,w} & & & +T\textsubscript{y,w}
& & 0 \\
UBI & +U & & & $-$U & & 0 \\
Interest on deposits & +rD\textsubscript{h} & & & +rD\textsubscript{g} &
+rD\textsubscript{r} & 0 \\
Investment & & & $-$I & & & $-$I \\
$\Delta$ Equity (issuance) & $-$$\Delta$E\textsubscript{h} & & +$\Delta$E &
$-$$\Delta$E\textsubscript{s} & $-$$\Delta$E\textsubscript{r} & 0 \\
$\Delta$ Deposits & $-$$\Delta$D\textsubscript{h} & & $-$$\Delta$D\textsubscript{f} &
$-$$\Delta$D\textsubscript{g} & $-$$\Delta$D\textsubscript{r} & 0 \\
\end{longtable}

The investment row sits on the firm\textquotesingle s capital account:
gross saving (retained profit plus the equity savers subscribe) finances
investment, so the capital-account column balances and the $\Delta$-deposit row
closes the period. Retained profit grows existing
owners\textquotesingle{} claims pro-rata; new issuance is subscribed by
whichever sectors are saving (households, the surplus government, RoW),
which is how a persistent surplus is converted into an equity stake
rather than an ever-growing deposit. The no-overdraft rule is the one
place a household that cannot fund its allocation sells equity at book
to a surplus sector, a matched swap that leaves both the $\Delta$-equity and
$\Delta$-deposit rows summing to zero.

\hypertarget{production-and-factor-prices}{%
\subsection{Production and factor
prices}\label{production-and-factor-prices}}

Output is a CES aggregate of a labour-task bundle and a capital-task
bundle:

Y = A \textperiodcentered{} {[} (1 $-$ I)\^{}(1/e) \textperiodcentered{} (A\_L L)\^{}((e$-$1)/e) ~+~ I\^{}(1/e) \textperiodcentered{}
(A\_K K)\^{}((e$-$1)/e) {]}\^{}(e/(e$-$1))

The automation index I is the share of tasks performed by capital and e
the elasticity of substitution. Factor prices are marginal products, so
the capital share rises endogenously with automation, and the net return
r and growth rate g are endogenous. With gross complements (e below
one), potential output is bounded in capital. Output is
supply-determined at potential, investment is the residual that clears
the goods market (S = I), and consumption follows the
differential-saving rule (high propensity out of labour income, low out
of capital income, small wealth effect on equity). That rule pins the
capital-output ratio at a stable interior value and removes the
first-generation output blow-up, because scarcity rents on capital are
saved and reinvested rather than consumed.

\hypertarget{concentration-heterogeneous-persistent-returns}{%
\subsection{Concentration: heterogeneous persistent
returns}\label{concentration-heterogeneous-persistent-returns}}

Wealth concentrates because returns to equity are heterogeneous and
persistent across households, an AR(1) return type per agent calibrated
to the returns-to-wealth literature. A small demographic turnover (death
and replacement by low-wealth entrants) pins a stationary Pareto tail.
The additive inflow that prevents runaway condensation is the saving of
wage earners who purchase equity. There is no imposed kinetic kernel and
no artificial conservation: the tail is produced by the process, and the
only rescaling enforces the balance-sheet identity that household claims
sum to the household share of K.

The engine is disciplined against the standard cross-sectional
calibration targets rather than only producing ``a'' heavy tail. The
return primitives are set to the Fagereng et al. estimates: a
cross-sectional standard deviation of household wealth returns of about
five percentage points (ret\_sigma = 0.05) and a high annual persistence
of the return type (ret\_persist = 0.92, in the range those panel
studies report). With those primitives the laissez-faire benchmark
settles at a wealth Gini of about 0.63, a top-1\% share of about 24\%
and a top-10\% share of about 53\%, which bracket the empirical wealth
concentration of advanced economies, and the upper tail follows an
approximate power law whose slope corresponds to a Pareto exponent of
roughly 1.5, consistent with the 1.4 to 1.7 range estimated for wealth.
The return-dispersion comparative static (experiment C) is the explicit
sensitivity to this calibration: SD of 0.03 / 0.05 / 0.07 maps to
top-1\% shares of about 12\% / 24\% / 39\%, so the targets above are
matched at the central value rather than by construction.

\hypertarget{behavioural-policy-and-mobility}{%
\subsection{Behavioural policy and
mobility}\label{behavioural-policy-and-mobility}}

The wealth tax is levied on the stock and split into an equity leg
(ownership transferred to the state) and a cash leg. It is no longer
frictionless: the taxable base erodes with the rate (avoidance,
calibrated to a bunching elasticity of 0.75), and domestically-held
equity relocates abroad in response to the tax (capital flight,
calibrated to a two per cent per percentage point semi-elasticity). The
foreign ownership share is therefore an endogenous state variable that
the tax itself moves. Other instruments (corporate and income tax,
progressive wealth brackets, UBI, a citizens\textquotesingle{} wealth
fund that rebates state dividends per capita) carry over from the first
generation.

\hypertarget{baseline-calibration}{%
\subsection{Baseline calibration}\label{baseline-calibration}}

{\renewcommand{\_}{\textunderscore\allowbreak}%
\begin{longtable}[]{@{}>{\raggedright\arraybackslash}p{0.20\linewidth}>{\raggedright\arraybackslash}p{0.18\linewidth}>{\raggedright\arraybackslash}p{0.52\linewidth}@{}}
\toprule\noalign{}
Parameter & Value & Meaning \\
\midrule\noalign{}
\endhead
\bottomrule\noalign{}
\endlastfoot
N / periods & 2000 / 600 & agents and horizon \\
eps (e) & 0.6 & elasticity of substitution (gross complements,
empirically defensible) \\
I\_base $\rightarrow$ max & 0.50 $\rightarrow$ \textasciitilde0.95 & capital-task share before
and after the AI ramp \\
K0 per capita & 5.0 & initial capital stock (no-automation steady state
K/Y \textasciitilde{} 2.9) \\
c\_income & 0.85 & propensity to consume out of labour income \\
c\_profit & 0.35 & propensity to consume out of capital income
(differential saving) \\
c\_wealth & 0.03 & propensity to consume out of equity wealth \\
ret\_sigma & 0.05 & cross-sectional SD of idiosyncratic wealth returns
(Fagereng et al) \\
ret\_persist & 0.92 & persistence of the return type \\
demographic\_reset & 0.02 & turnover that pins the stationary tail \\
avoidance\_elasticity & 0.75 & base erosion w.r.t. the net-of-tax rate
(Brülhart et al) \\
migration\_semi\_elast & 0.02 & capital flight per pp of wealth tax
(Jakobsen et al. 2024) \\
init\_wealth\_sigma & 1.15 & seed log-dispersion (initial Gini
\textasciitilde{} 0.6) \\
gov\_cost / r\_debt & 0.10 / 0.02 & government running cost (0 in the
no-government laissez-faire benchmark); interest on balances \\
\end{longtable}}

\hypertarget{two-channel-parameters-and-why-they-take-these-values}{%
\subsection{Two-channel parameters and why they take these
values}\label{two-channel-parameters-and-why-they-take-these-values}}

The two-channel extension replaces the single capital-task CES with a
nested three-cluster CES: a routine cluster pairing routine labour with
robotic capital, a cognitive cluster pairing cognitive labour with AI
compute, and a top CES between the clusters. Every new parameter is set
either to reproduce the single-channel baseline (so nothing in the
existing results moves when the channels are switched off), to match an
empirical anchor, or as a clearly-flagged illustrative policy level. The
reasoning is recorded in full below so the choices are auditable rather
than buried.

{\renewcommand{\_}{\textunderscore\allowbreak}%
\begin{longtable}[]{@{}>{\raggedright\arraybackslash}p{0.20\linewidth}>{\raggedright\arraybackslash}p{0.18\linewidth}>{\raggedright\arraybackslash}p{0.52\linewidth}@{}}
\toprule\noalign{}
Parameter & Value & Why this value \\
\midrule\noalign{}
\endhead
\bottomrule\noalign{}
\endlastfoot
e\_top & 1.20 & Elasticity between the routine and cognitive clusters.
Set above the within-cluster elasticities so robots and AI substitute
more readily with each other (production can shift between routine- and
cognitive-intensive output) than with their own labour, which is the
nesting choice. Mild gross substitutes at the broad-task level. \\
e\_routine, e\_cog & 0.60 & Within-cluster capital-labour elasticities,
gross complements, matching the single-channel value and the empirically
defensible sub-unity range (Acemoglu-Restrepo 2022; Oberfield-Raval).
Automation raises the wage on the remaining tasks in a cluster rather
than collapsing it. \\
theta\_cog & 0.50 & Top-level weight on the cognitive cluster. Symmetric
at baseline; no prior reason to tilt, and symmetry lets the baseline
match the single-channel calibration exactly. \\
cognitive\_share & 0.50 & Fraction of the workforce doing cognitive
work, taken as the top half by skill. Roughly the knowledge-vs-routine
split of employment in an advanced economy; putting AI against the
higher-skill half encodes that AI displaces white-collar work, the
reverse of routine-biased robotics. \\
a\_r\_base, a\_ai\_base & 0.50 & Capital-task share in each cluster
before the ramps. Set to the single-channel I\_base so the no-automation
economy reproduces the v3 baseline (capital share 0.255, K/Y near 3).
This is the calibration anchor that makes the two-channel model collapse
to the validated single-channel macro baseline; lower values starve
capital of income and the economy fails to sustain itself. \\
a\_r ramp & start 60, speed 0.05, +0.45 & Robots ramp earlier and more
slowly, to about 0.95. Industrial robotics has diffused for decades, so
the robotic channel is the more mature, gradual one. \\
a\_ai ramp & start 110, speed 0.10, +0.49 & AI ramps later, faster, and
reaches further (to about 0.99). Encodes the recent, rapid LLM wave
reaching deep into cognitive tasks; the later-faster-higher profile is
the empirical sequencing relative to robotics. \\
robot\_capital\_share0 & 0.50 & Initial split of the capital stock
between robots and AI compute. Symmetric start; the allocation rule then
moves it endogenously. \\
depreciation\_r & 0.05 & Robotic / physical-equipment depreciation, the
standard macro equipment rate, matching the single-channel model. \\
depreciation\_ai & 0.18 & AI compute depreciates far faster. Frontier
accelerators obsolesce in roughly two to three years and models turn
over in about eighteen months, while the data-centre shell lasts fifteen
to thirty years; 0.18 (about a five- to six-year effective life) is a
deliberate blend of fast-obsolescing accelerators and longer-lived
facilities, set well above the robotic rate. The qualitative results are
robust to the exact figure. \\
invest\_damping & 0.25 & Speed of the return-equalising reallocation
between the two stocks: the investment share moves up to a quarter of
the way toward the return-equalising target each period. Chosen for
stability (no all-or-nothing flips) while still equalising net returns
over several periods. \\
mu\_frac & 0.25 (illustrative) & The AI IP rent as a share of
cognitive-cluster value added, the markup the model owner charges over
the competitive cost of compute and labour. Anchored on the markup
evidence: US markups rose from about 21\% over marginal cost in 1980 to
roughly 61\% by 2016, concentrated in the upper-tail, IP-heavy firms (De
Loecker, Eeckhout and Unger 2020). A quarter of AI value added as pure
rent is a moderate level for a dominant IP owner; the markup literature
is contested on magnitude (Foster-Haltiwanger-Tuttle), so this is
presented as illustrative, not estimated. \\
s\_home & 1.0 vs 0.2 & Fraction of AI compute located on home (or
European) servers, which fixes how much of the physical AI-capital
income is source-taxable. Bracketed between fully onshored and largely
offshore to represent the sovereign-compute choice; the IP rent stays
mobile regardless of where the servers sit. \\
robot\_tax & 0.15 (illustrative) & Source levy on robotic capital
income, the Gates-style tax on the profits of labour-saving automation
(Gates 2017; South Korea\textquotesingle s 2017 incentive change; the
rejected 2017 EU Parliament motion). Robots are physical and located
where they operate, so the base is fully collectable. \\
dst\_ai & 0.10 (illustrative) & A digital-services-style levy on AI
revenue (value added plus rent), collected at source. In the spirit of
the UK DST, a 2\% levy on gross digital revenue introduced in 2020. A
gross-revenue base of this kind is harder to shift offshore than profit,
which is part of the rationale for such levies (HM Treasury, 2020),
though it is also criticised for falling on firms regardless of
profitability (ITIF, 2025). The base here is AI value added rather than
gross revenue, so the rate is not directly comparable and is
illustrative; the shared property that matters is reaching local value
rather than shiftable profit. \\
tax\_repat & 0.30 (illustrative) & Withholding on the rent as it is
repatriated as a licence fee. Double-taxation treaties commonly cap
cross-border royalty and dividend withholding at low single-digit to
mid-teens rates, so 0.30 sits deliberately above typical treaty ceilings
to show the instrument\textquotesingle s reach; it is illustrative, not
a currently-available rate. \\
\end{longtable}}

Three choices carry most of the weight and deserve a sentence each. The
elasticity ordering (e\_top above the within-cluster elasticities) is
what makes the two capitals substitutes for each other but complements
with labour, so automating one channel pulls investment toward it
without simply collapsing the wage. The fast AI-compute depreciation is
what keeps the AI capital stock from accumulating without limit and is
grounded in the observed obsolescence of accelerators. And the AI rent
(mu\_frac) is the single addition that gives the mature economy
something durable to tax: without it, competitive factor pricing
competes the net return to zero and there is no permanent surplus, which
is exactly why a robot tax or a corporate tax reaches so little of the
AI value in steady state.

\hypertarget{why-ai-value-is-modelled-as-a-mobile-rent}{%
\subsection{Why AI value is modelled as a mobile
rent}\label{why-ai-value-is-modelled-as-a-mobile-rent}}

The central modelling choice, that the durable AI surplus is a mobile IP
rent rather than a return on physical capital, reflects how the
providers actually operate. The scarce, hard-to-reproduce asset is the
trained model and the brand, not the compute: anyone can rent broadly
similar accelerators, so the return on the physical layer is competed
down, while the model and its reputation command a markup. The
commercial model is built around exactly this. Closed providers keep the
weights behind an API and charge per unit of use, a markup over the
marginal cost of inference, which is the rent in the
model\textquotesingle s sense. And it is recognised abroad: a UK or
European customer of a provider such as OpenAI contracts with its Irish
entity or its US parent rather than with a UK taxable presence, so the
value leaves as a cross-border charge that ordinary domestic corporate
tax does not reach, which is precisely the deductible licence fee the
model represents. The robotic channel is treated differently, as
physical capital earning a competed-down return, because robots are
located where they operate and their hardware is a reproducible good, so
the lasting margin there accrues to the maker abroad rather than
persisting as a domestic rent.

One qualification belongs here. The rent is, at the time of writing,
prospective rather than realised: the frontier labs run at a loss, with
OpenAI reportedly spending well over a dollar for every dollar of
revenue, and do not expect to be consistently profitable until roughly
2029 or 2030 (Anthropic targets a little earlier). The durable markup
the model assumes is the mature-market structure the
sector\textquotesingle s valuations are betting on, not a description of
current profit and loss, and its eventual size depends on the market
staying concentrated rather than being commoditised (subsection O). The
model should therefore be read as analysing the steady state the
industry is investing toward, with the rent\textquotesingle s existence
and magnitude treated as a calibrated assumption about that future
rather than an estimate from today.

\hypertarget{upgrades}{}
\hypertarget{key-mechanisms-and-why-they-are-modelled-this-way}{%
\section{Key mechanisms and why they are modelled this
way}\label{key-mechanisms-and-why-they-are-modelled-this-way}}

Three mechanisms do most of the work in the model, and each is a
deliberate design choice with an empirical anchor rather than a
convenience. Setting them out explicitly is part of justifying the
approach: the wealth tax is behavioural rather than frictionless,
concentration is generated rather than imposed, and the economy is
closed in a way that produces a bounded, convergent transition.

\hypertarget{behavioural-wealth-tax-and-endogenous-capital-mobility}{%
\subsection{1. Behavioural wealth tax and endogenous capital
mobility}\label{behavioural-wealth-tax-and-endogenous-capital-mobility}}

The wealth tax is behavioural: it erodes its own base and triggers
capital flight, both calibrated to the empirical elasticities (Brülhart
et al. 2022; Jakobsen et al. 2024), and the foreign ownership share
responds endogenously to it. A per-agent offshore-wealth account then
lets the model measure true inequality alongside the measured domestic
figure. This matters because a wealth tax that assumed away avoidance
and relocation would overstate its own effectiveness; building the
responses in turns the central result into a tested, calibrated finding
(section 6) rather than an artefact of frictionless redistribution.

\hypertarget{concentration-generated-from-heterogeneous-returns}{%
\subsection{2. Concentration generated from heterogeneous
returns}\label{concentration-generated-from-heterogeneous-returns}}

The wealth distribution\textquotesingle s tail is produced from
heterogeneous, persistent returns to wealth, a documented empirical fact
(Fagereng et al. 2020), in line with the Moll--Rachel--Restrepo (2022)
return-gap mechanism. The distribution is therefore an output of the
model, not an assumption baked into a kinetic kernel. This matters
because a concentration result is only informative if the concentration
was not hard-wired: here the same return heterogeneity that the data
show is what generates the tail, so the inequality dynamics can be read
as a consequence of the mechanism rather than of the modelling choice.

\hypertarget{a-stable-stock-flow-consistent-closure}{%
\subsection{3. A stable, stock-flow-consistent
closure}\label{a-stable-stock-flow-consistent-closure}}

The economy is closed in the standard neoclassical way so that the
automation transition is bounded and convergent. Output is
supply-determined (gross complements bound potential output in capital),
investment is the residual that clears the goods market (the
saving-equals-investment identity made operational), and consumption
follows a differential-saving rule in the Kaldor (1957) / Pasinetti
(1962) tradition: a high propensity to consume out of labour income and
a low one out of capital income, plus a small wealth effect on equity.
Differential saving is what makes the steady state stable rather than
merely calibrated: when capital becomes scarce its return spikes, and
because that windfall is mostly saved it is reinvested and rebuilds the
stock, so the capital-output ratio converges from any starting point.
Output then grows by a large multiple over the long horizon (K/Y rising
from about 2.9 to roughly 8 to 10 as the economy automates), but the
path is bounded and convergent rather than explosive, and the
no-automation baseline is a verified stationary equilibrium. A Sobol
sensitivity analysis (section 6E) quantifies which parameters drive the
results.

Scope of the claims

Consistent with the empirical debate, the model treats near-complete
automation as a long-run scenario rather than a documented trend, and
presents the falling labour share as one contested channel (alongside
housing and market power) rather than a settled mechanism. The policy
results are positive, not welfare-optimal: statements about one regime
being preferable are framed in terms of measured inequality and
solvency, not social welfare, since the model does not yet contain an
explicit welfare criterion.

\hypertarget{method}{%
\section{Method}\label{method}}

Each experiment fixes the baseline calibration and varies one lever.
Outcomes are read at the end of the 600-period horizon, averaged over
the final 30 to 40 periods for distributional metrics and taken as the
terminal value for stocks, then averaged across seeds with reported
bands. Robustness is assessed three ways: Monte Carlo seed replication
with confidence bands; a global (Sobol) sensitivity analysis decomposing
the variance of the headline outcomes across the free parameters; and a
formal stability analysis of the deterministic skeleton (section 6,
experiment D). The no-automation baseline is a verified stationary
equilibrium.

\hypertarget{experiments}{}
\hypertarget{experiments-and-results}{%
\section{Experiments and results}\label{experiments-and-results}}

The experiments build in a deliberate order, each one answering a
question the previous one raises. The first group (A to E)
re-establishes and stress-tests the single-channel distributional
engine: that stock taxes compress inequality and flow taxes do not (A),
that the compression is largely genuine rather than an artefact of
capital fleeing offshore (B), that the concentration is microfounded
rather than imposed (C), that the closure is formally stable (D), and
which parameters actually drive the results (E). The second group (F to
H) introduces the distinction the AI-tax debate turns on, between the
foreign-owned, mobile AI IP rent and the competitive return on physical
robots, and asks which tax base reaches the durable surplus and who ends
up owning the capital (F, G), then makes those answers concrete for
Britain (H). The third group tests the robustness and the costs of those
answers: to the rent\textquotesingle s size and to opening the economy
to goods trade (I), and to the labour-market response, first as a
changing labour share (J), then as measured unemployment and the funding
of a safety net (L), with the AI-tax debate\textquotesingle s two
strongest objections put directly to the model in between (K) and the
whole set of added parameters calibrated and sensitivity-tested (M). The
final group (N to P) hardens the central recommendation against a
strategic adversary and two alternatives: a foreign owner that actively
shifts where the rent is booked (N), competition that erodes the rent
without anyone taxing it (O), and the value of the rent as a capitalised
stock rather than a flow (P). The thread throughout is a single
question, which base reaches the durable surplus and who owns the
automated capital, asked of progressively more demanding versions of the
world.

\hypertarget{a-the-behavioural-policy-frontier}{%
\subsection{A \textperiodcentered{} The behavioural policy
frontier}\label{a-the-behavioural-policy-frontier}}

\textbf{Question.} Once the wealth tax leaks abroad and erodes its own
base, on a microfounded concentration engine, under disciplined
calibration, is the first-generation finding that stock taxes compress
inequality more effectively than flow taxes still numerically supported
under this closure?

\begin{figure}[htbp]
\centering
\includegraphics[width=0.9\linewidth]{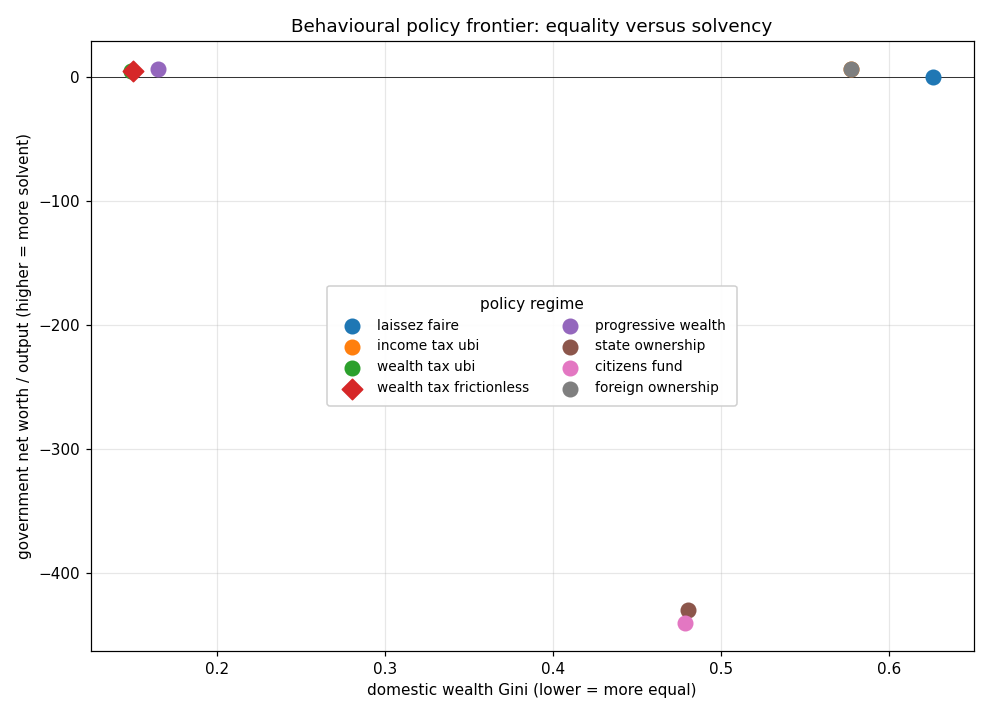}
\caption*{\textbf{Figure 1} Behavioural policy frontier: domestic wealth
Gini against government net worth relative to output. Ten-seed means.
The frictionless wealth tax (diamond) sits close to the behavioural one
at the baseline rate; they diverge at higher rates (Figure 2).}
\end{figure}

\textbf{Result.} The stock-tax regimes occupy the favourable region (low
inequality, solvent): the wealth tax brings the domestic Gini to about
0.15 and the progressive variant to about 0.17, both while accumulating
positive government net worth (a sovereign equity fund built from fiscal
surpluses, worth roughly five to seven times annual per-capita output by
the end). The flow instruments do not compress: income-tax-plus-UBI
leaves the Gini at about 0.58, essentially the laissez-faire level of
about 0.63. The insolvent regimes are the under-taxed welfare states:
state ownership and the citizens fund pair a 15\% income tax with a
universal income and a 10\%-of-output running cost, run persistent
primary deficits, and end with deeply negative government net worth
(about $-$420 times per-capita output) as the r-greater-than-g debt path
compounds. Laissez-faire, by contrast, runs no government at all and so
sits at zero net worth: it is the most unequal regime but not a fiscal
one.

\textbf{Interpretation.} The qualitative ranking is robust within the
closure: stock instruments compress the distribution, flow instruments
do not. Solvency is a separate axis and turns on whether taxation is
adequate to the spending commitment, not on ownership form. A
redistributive state that taxes enough (the 30\% income tax, or either
wealth tax) not only stays solvent but gradually accumulates a majority
equity stake out of its surpluses, an emergent socialisation of capital;
a state that promises a universal income on a 15\% income base is
insolvent regardless of how much equity it nominally owns. Two caveats
bound the strength of the compression result. First, part of it is
mechanical: the wealth tax transfers equity in kind to the state at BOOK
value, with no asset-price response, no liquidity or fire-sale discount,
and no enforcement gap beyond the calibrated avoidance and migration
channels, so the instrument moves ownership almost frictionlessly in a
way a real wealth tax would not. Second, whether the measured equality
is genuine or partly compositional (capital fled offshore but still
owned domestically) depends on the rate, which experiment B quantifies.
Neither overturns the ranking, but both mean the magnitudes are an upper
bound on what an equivalent real-world instrument would achieve.

\hypertarget{b-the-composition-effect-measured-vs-true-inequality}{%
\subsection{B \textperiodcentered{} The composition effect --- measured vs true
inequality}\label{b-the-composition-effect-measured-vs-true-inequality}}

\textbf{Question.} The wealth tax lowers the measured domestic Gini, but
some of that capital has left the country rather than been
redistributed. How much of the apparent equality is real, once we
attribute fled capital back to its original owners?

\begin{figure}[htbp]
\centering
\includegraphics[width=0.9\linewidth]{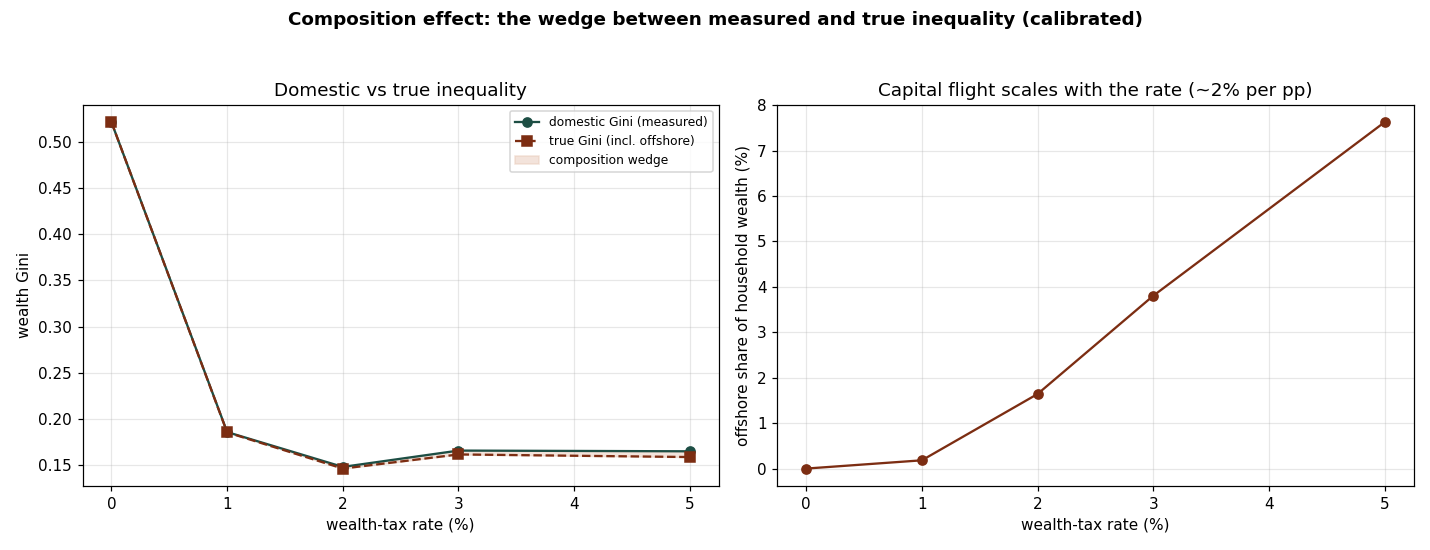}
\caption*{\textbf{Figure 2} Left: the domestic (measured) wealth Gini
against the true Gini that includes each household\textquotesingle s
offshore holdings attributed to it, across wealth-tax rates. Right: the
offshore share of household wealth, which scales at about two per cent
per percentage point of tax, matching the migration semi-elasticity it
is calibrated to.}
\end{figure}

\textbf{Result.} The model tracks each household\textquotesingle s
relocated capital in a per-agent offshore account, so the true
distribution can be reconstructed. The wealth tax compresses the
domestic Gini sharply at the first percentage point (from about 0.63
untaxed to about 0.19 at 1\%, then about 0.15 at 2\%), after which
further rate increases barely compress at all (about 0.17 at both 3\%
and 5\%). The wedge between measured and true inequality stays small
throughout: at a 2\% tax both are about 0.15, and even at 5\% the
measured Gini is about 0.17 against a true Gini of about 0.16, with
roughly 7.6\% of household wealth offshore. The offshore share scales
close to linearly with the rate (about two points per percentage point),
as the calibration to the migration evidence requires.

\textbf{Interpretation.} The composition effect is real but modest at
empirically-anchored mobility, not the dominant force an earlier,
mis-calibrated version of this experiment suggested (that version let
the offshore account compound without bound and implied almost the
entire base fled, which is not credible). Two defensible claims follow:
most of the wealth tax\textquotesingle s compression is genuine
redistribution rather than measurement artefact (the domestic-true wedge
is at most a point of Gini), and the compression is front-loaded, so a
low single-digit rate captures almost all of the achievable equality
while a higher rate mostly just pushes capital offshore. A careful
evaluation should still report the true figure rather than the domestic
one.

\hypertarget{c-concentration-from-heterogeneous-returns}{%
\subsection{C \textperiodcentered{} Concentration from heterogeneous
returns}\label{c-concentration-from-heterogeneous-returns}}

\textbf{Question.} Does the microfounded engine produce a realistic,
stationary wealth distribution without the imposed kernel?

\begin{figure}[htbp]
\centering
\includegraphics[width=0.9\linewidth]{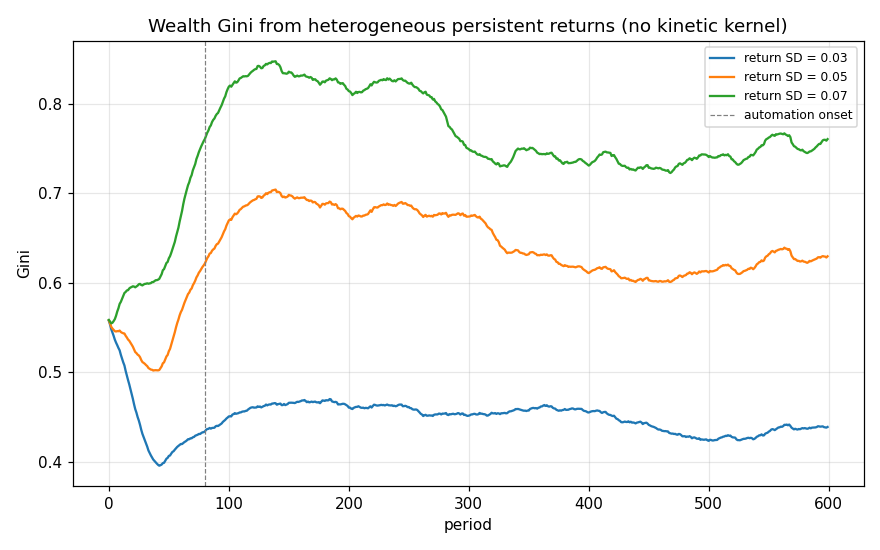}
\caption*{\textbf{Figure 3} Wealth Gini over time for three levels of
return dispersion. The tail emerges from heterogeneous persistent
returns plus demographic turnover, with no kinetic kernel.}
\end{figure}

\textbf{Result.} The engine produces a heavy but stationary tail. Return
dispersion of 0.03, 0.05 and 0.07 maps to terminal Gini of about 0.44,
0.63 and 0.75, with top-1 per cent shares of about 12, 24 and 39 per
cent (and top-10 per cent shares of about 35, 53 and 68 per cent), all
empirically plausible for wealth. The distribution settles rather than
running away to full condensation.

\textbf{Interpretation.} Concentration is now an economic output driven
by a calibrated primitive (the dispersion of returns to wealth), not an
artefact of an imposed mean-reverting process. This removes the deepest
methodological objection to the first-generation model.

\hypertarget{d-formal-stability}{%
\subsection{D \textperiodcentered{} Formal stability}\label{d-formal-stability}}

\textbf{Question.} Is the steady state stable in a provable sense,
rather than by trajectory inspection?

\begin{figure}[htbp]
\centering
\includegraphics[width=0.9\linewidth]{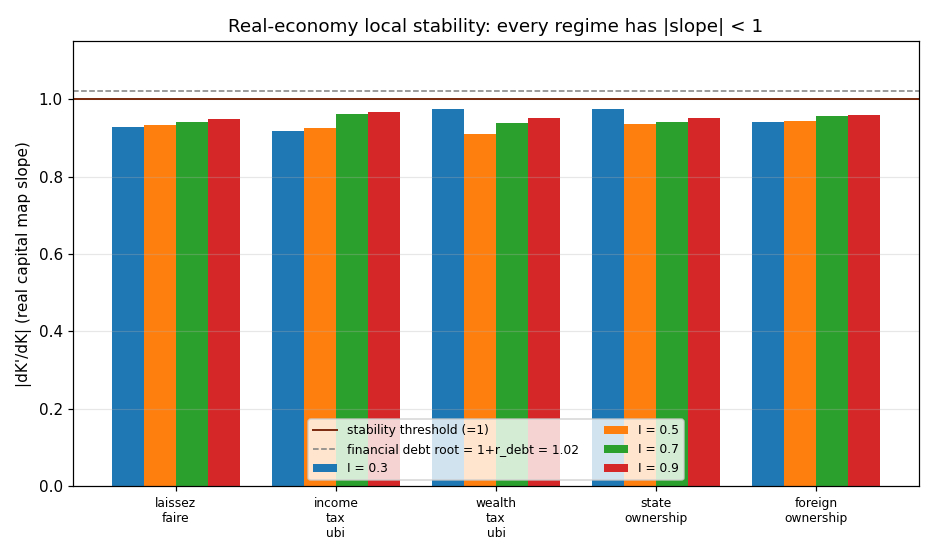}
\caption*{\textbf{Figure 4} Local slope of the aggregate capital map,
\textbar dK$'$/dK\textbar, at the deterministic-skeleton fixed point,
across five regimes and four automation levels. Every bar is below the
stability threshold of one; the dashed line marks the structural
financial root 1 + r\_debt = 1.02.}
\end{figure}

\textbf{Result.} The real economy is locally stable everywhere: the
slope of the one-dimensional capital map at its fixed point lies between
about 0.89 and 0.98 in every regime and at every automation level,
comfortably inside the unit circle. This is the formal counterpart of
the global convergence shown numerically (the capital-output ratio
settles to the same value from any initial stock). Separately, the
deposit block carries a structural root at exactly 1 + r\_debt = 1.02:
any bond balance compounds at the interest rate, so it collapses to one
when the rate is set to zero.

\textbf{Interpretation.} Real stability and fiscal solvency are distinct
questions. The capital stock has a stable rest point under every policy,
so the real economy always converges; what differs across regimes is the
financial block, where the r-greater-than-g root means that a regime
running a persistent primary deficit puts public debt on an explosive
path. This is the formal, root-level statement of H1: the real economy
is never the source of instability, and a regime is fiscally sustainable
only if it generates the surplus or the asset income to neutralise the
debt root, which the adequately-taxed regimes do and the under-taxed
welfare states do not.

\hypertarget{e-global-sensitivity-which-parameters-drive-the-results}{%
\subsection{E \textperiodcentered{} Global sensitivity: which parameters drive the
results}\label{e-global-sensitivity-which-parameters-drive-the-results}}

\textbf{Question.} Are the headline outcomes robust, or do they hinge on
the least-certain parameters (the behavioural elasticities we had to
assume)?

\begin{figure}[htbp]
\centering
\includegraphics[width=0.9\linewidth]{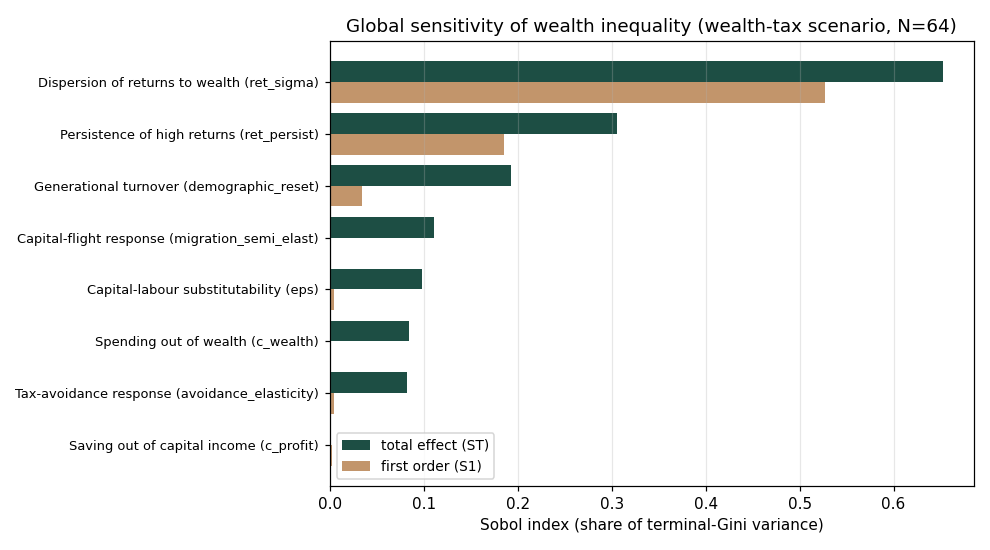}
\caption*{\textbf{Figure 5} First-order (S1) and total-order (ST) Sobol
indices decomposing the variance of the terminal wealth Gini across
eight parameters (wealth-tax scenario). Higher bars are parameters that
matter more.}
\end{figure}

\textbf{Result.} At a base sample of N=64 (about 1,150 model runs) the
wealth Gini is driven mainly by the dispersion of returns to wealth
(total Sobol index about 0.61) and the persistence of the return type
(about 0.24), with demographic turnover and the
consumption-out-of-wealth propensity secondary (about 0.11 and 0.07) and
the behavioural-tax elasticities small (avoidance and migration around
0.05 to 0.06 each in this sample). These behavioural indices are the
least precisely estimated: at N=64 the Sobol Monte Carlo error is
non-trivial (some first-order estimates come back slightly negative, the
usual small-index artefact), and an independent run at the same sample
size returned values nearer 0.13, so the honest statement is that the
behavioural elasticities are an order of magnitude smaller than the
return-dispersion driver but pinned only to within a factor of two or so
at this sample size; a publication-grade figure would raise N.
Government net worth, by contrast, is driven almost entirely by the CES
elasticity of substitution (total index about 1.0, every other parameter
near zero), the cleanest result in the analysis.

\textbf{Interpretation.} This is reassuring for the inequality results
and disciplining for the fiscal ones. The distributional findings are
governed by the microfounded concentration primitives (the empirical
dispersion and persistence of returns to wealth) rather than by the
behavioural elasticities that are hardest to pin down, so they do not
rest on the least-certain numbers, even allowing for the Monte Carlo
error on those elasticities. The fiscal results depend overwhelmingly on
the CES elasticity, which is exactly why holding it in the empirically
defensible gross-complements range (section 4) matters.

\hypertarget{f-source-taxation-of-foreign-owned-automation-income}{%
\subsection{F \textperiodcentered{} Source taxation of foreign-owned automation
income}\label{f-source-taxation-of-foreign-owned-automation-income}}

\textbf{Question.} A recurring concern for governments is that the
income from highly-automated production is earned in their jurisdiction
but owned, and ultimately repatriated, abroad, so a residence-based
wealth or income tax never reaches it. If instead the income is taxed at
SOURCE, where the output is produced, before it leaves, what happens to
the ownership of the capital stock, to revenue, and to the domestic
distribution?

The model levies a source tax at rate tax\_repat on the foreign
owners\textquotesingle{} full attributed share of after-corporate-tax
profit. Crucially it falls on more than the repatriated dividend: in a
growth phase profit is largely reinvested, so a dividend-only
withholding tax would miss most of the income. The tax therefore has a
cash leg on the part distributed abroad and an in-kind equity leg on the
part reinvested in the foreign owner\textquotesingle s name, the latter
diverting that slice of new equity to the state. The proceeds are
rebated to residents per capita. It is run on the foreign-ownership
scenario (60\% foreign at the outset).

\begin{figure}[htbp]
\centering
\includegraphics[width=0.9\linewidth]{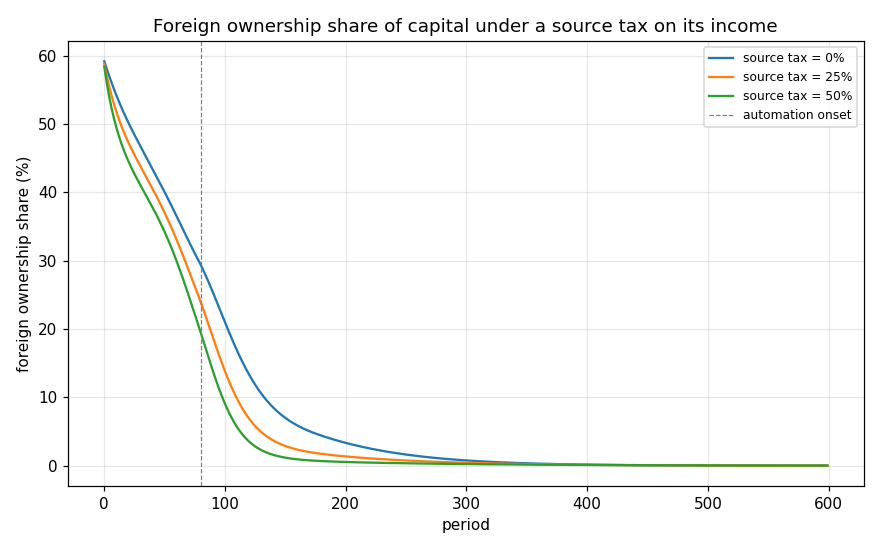}
\caption*{\textbf{Figure 6} Foreign ownership share of the capital stock
over time, under a source tax of 0\%, 25\% and 50\% on the foreign
owners\textquotesingle{} attributed automation income (proceeds rebated
to residents). A higher rate brings the share down faster: the automated
capital stock is domesticated sooner.}
\end{figure}

\textbf{Result.} The source tax raises little long-run revenue (well
under one per cent of output even over the automation-transition window)
and leaves the terminal wealth Gini (about 0.56) and government solvency
(net worth about six times output) essentially unchanged. What it does
change is the SPEED of domestication: the foreign ownership share
crosses five per cent markedly sooner the higher the rate (around period
170 untaxed, 130 at a 25\% rate, 113 at a 50\% rate). The recycling mode
(rebate to citizens versus retain as revenue) barely matters here,
because the distributable foreign income is small: most of the foreign
owners\textquotesingle{} return is reinvested, not paid out.

\textbf{Interpretation.} In this model a source tax on foreign-owned
automation income is a TRANSITION instrument, not a permanent
distributional lever. Domestic savers and the surplus state buy out the
foreign stake regardless, so the capital stock ends up domestically
owned with or without the tax; the tax mainly accelerates that, bringing
the automated capital home faster and converting part of the foreign
owners\textquotesingle{} reinvested earnings into a public stake along
the way. Two qualifications matter for governments. First, the result is
conditional on the model\textquotesingle s dilution dynamics: it does
not sustain a sticky foreign ownership share. Under structurally
persistent foreign ownership (sticky FDI, which would require modelling
continuous capital inflows) the source tax would be the ONLY instrument
able to reach foreign owners, since the residence-based wealth and
income taxes cannot, and its revenue and equality effects would then
persist rather than fade. Second, the in-kind equity leg is what gives
the tax its bite: a conventional dividend withholding tax, taxing only
repatriated profit, would capture almost nothing while earnings are
being reinvested, which is precisely the automation-investment phase
governments would most want to tax. The policy reading is that taxing
automation income at source is valuable mainly for the speed and the
politics of domestication, and only becomes a first-order revenue and
equality instrument if foreign ownership is genuinely persistent and the
levy reaches reinvested, not just repatriated, earnings.

\hypertarget{a-domestic-owner-first-the-simpler-benchmark-the-us-case}{%
\subsection{A domestic owner first: the simpler benchmark (the US
case)}\label{a-domestic-owner-first-the-simpler-benchmark-the-us-case}}

Before the foreign owner that occupies the rest of this section, take
the simpler case the United States is in: the AI is owned at home. The
same two-channel economy applies, but the AI IP rent no longer leaves as
a licence fee, it accrues to resident owners as capital income (the
model\textquotesingle s foreign-owned-IP share set to zero). Nothing
leaves the country, so the worry is not lost ownership but how the rent
is distributed at home.

\begin{figure}[htbp]
\centering
\includegraphics[width=0.9\linewidth]{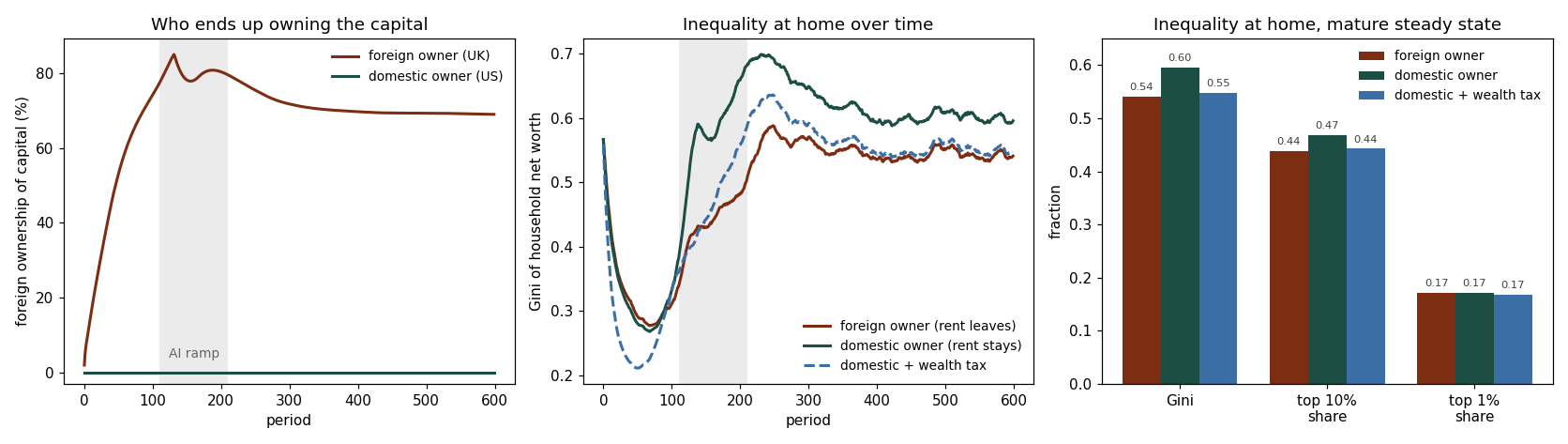}
\caption*{\textbf{Figure 25} The two owner regimes, the same economy,
with only the owner\textquotesingle s domicile changed. Left: foreign
ownership of the domestic capital stock. With a foreign owner (the UK
case) the reinvested rent carries ownership toward seventy per cent;
with a domestic owner (the US case) it stays flat, because the rent
never leaves to be recycled into foreign-held equity. Centre: inequality
of household net worth over time. The domestic owner keeps the rent at
home but concentrates it among capital owners, so home inequality runs
higher than under the foreign owner, where the surplus simply leaves the
country; an illustrative five per cent wealth tax, which can reach the
home-held rent because the owner is resident, compresses it back to
about the foreign-owner level. Right: three measures at the mature
steady state. The owner-domicile effect shows up in the Gini (about 0.54
versus 0.60) and the top-decile share (about 44 versus 47 per cent), but
not in the top-percentile share (about 17 per cent in every case),
because the rent accrues pro-rata to equity and so concentrates among
the broad owning class rather than the very apex.}
\end{figure}

\textbf{Interpretation.} Two things follow, and they are the mirror
image of the foreign-owner results below. First, ownership does not
drift abroad, so the national-ownership problem that dominates the rest
of this section does not arise. Second, the rent concentrates among the
domestic households that own the AI, so the problem becomes domestic
inequality, and the instruments that bite are the ordinary domestic
ones: a wealth tax, capital-gains and dividend taxation, and competition
policy. The concentration is broad rather than apex: it raises the Gini
and the top-decile share but leaves the top-percentile share essentially
unchanged (Figure 25, right), because the rent accrues in proportion to
equity holdings, so it enriches the owning class as a whole rather than
only the very richest. Crucially, a wealth tax \emph{does} reach this
rent, because the owner is resident and the IP is onshore, which is
exactly the lever that fails against a foreign owner (subsection N and
section 8). The cost is the one specific to the
provider\textquotesingle s home: this is the regime in which the
territoriality objection of section 1 has force, since taxing a domestic
champion can push its activity to a lighter-touch jurisdiction. The
policy menu here is therefore the familiar one of domestic progressive
capital taxation, traded off against competitiveness.

This benchmark is deliberately compact, fixing the simpler case so the
contrast is clear. The harder and more novel case, and the one a country
like the United Kingdom actually faces, is a \emph{foreign} owner, where
the rent crosses a border and the domestic levers above no longer reach
it. Experiments G to P take that case, and it is the
paper\textquotesingle s main contribution.

\hypertarget{g-ai-versus-robotic-automation-taxed-with-different-instruments}{%
\subsection{G \textperiodcentered{} AI versus robotic automation, taxed with different
instruments}\label{g-ai-versus-robotic-automation-taxed-with-different-instruments}}

\textbf{Question.} The single-channel model treats all automation alike.
In reality the two dominant forms sit in very different places for tax.
Robotic automation is physical capital located where it operates, so its
income is taxable at source. AI automation is split: the compute
(servers, data centres) is locatable physical capital, but the value
sits in mobile intellectual property that earns a monopoly rent and is
recognised wherever the parent chooses. If a foreign multinational runs
the AI layer, charges a deductible licence fee, and books the rent
abroad, which instruments can a host state actually use, and what do
they reach?

The production block is replaced by a nested three-cluster CES
(production\_v4): a routine cluster pairing routine labour with robotic
capital, a cognitive cluster pairing cognitive labour with AI compute,
and a top CES between them with a higher elasticity, so robots and AI
substitute more with each other than with their own labour. Robots
displace routine workers, AI displaces the higher-skilled cognitive
workers. The two capital stocks accumulate from the same investment
pool, allocated by a damped move toward equalising their net returns,
with AI compute depreciating faster. A markup peels a share of cognitive
value added as an IP rent that leaves to the foreign owner as a
deductible licence fee, so it never enters the corporate base. Four
instruments are compared on this structure: a source tax on robotic
income, a withholding on the rent, a digital-services levy on AI
revenue, and the home-located share of compute.

{\setlength{\tabcolsep}{4pt}%
\begin{longtable}[]{@{}>{\raggedright\arraybackslash}p{0.26\linewidth}>{\centering\arraybackslash}p{0.12\linewidth}>{\centering\arraybackslash}p{0.14\linewidth}>{\centering\arraybackslash}p{0.12\linewidth}>{\centering\arraybackslash}p{0.10\linewidth}>{\centering\arraybackslash}p{0.13\linewidth}@{}}
\toprule\noalign{}
scenario & AI rent leaving (\% output) & capital-sector revenue (\% output) & govt net worth ($\times$ output) & wealth Gini & foreign ownership of capital \\
\midrule\noalign{}
\endhead
\bottomrule\noalign{}
\endlastfoot
foreign AI rent, untaxed & 13.1 & 0.0 & 0.9 & 0.544 & 0.69 \\
+ robot tax (15\%) & 13.1 & 1.5 & 1.5 & 0.540 & 0.64 \\
+ AI digital-services levy (10\%) & 13.1 & 3.7 & 2.8 & 0.535 & 0.51 \\
+ rent withholding (30\%, rebated) & 13.2 & 4.0 & 0.9 & 0.506 & 0.40 \\
compute offshore (s\_home 0.2) & 13.1 & 0.0 & 0.9 & 0.544 & 0.69 \\
full domestic toolkit & 13.1 & 5.2 & 3.2 & 0.531 & 0.48 \\
\end{longtable}}

Steady-state averages (periods 250 to 600, five seeds). Capital-sector
revenue aggregates corporate tax, the robot tax, the digital levy, and
the rent withholding. Foreign ownership is the rest-of-world equity
share of the capital stock at the horizon.

\begin{figure}[htbp]
\centering
\includegraphics[width=0.9\linewidth]{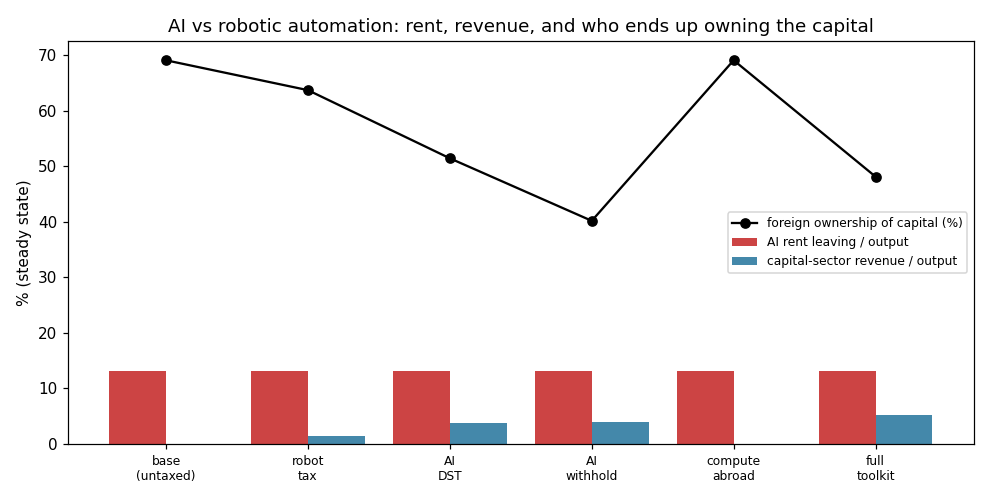}
\caption*{\textbf{Figure 7} Under a foreign-owned AI rent, what each
instrument reaches in the mature economy: the rent leaving the country
(red), the capital-sector revenue collected (blue), and the foreign
share of the capital stock (black line). The digital levy is the
effective revenue instrument; the rebated withholding is the most
equalising and most curbs foreign ownership; corporate tax and the
compute-location share reach almost nothing in steady state.}
\end{figure}

\textbf{Result.} The AI layer generates a permanent rent of about
thirteen per cent of output that, untaxed, leaves the country every
period. Corporate tax reaches almost none of it, because the rent is a
deductible licence fee and the competitive capital profit is competed to
zero in the mature economy. The robot tax cleanly collects on the
physical robotic income (about 1.5 per cent of output at a fifteen per
cent rate) but cannot touch the rent. The digital-services levy is the
effective instrument on the rent, collecting 3.7 per cent of output and
lifting government net worth from 0.9 to 2.8 times output. The rebated
withholding collects a similar amount but, because it is handed straight
back to citizens, raises no net public saving while delivering the
lowest inequality. The compute-location share moves the transition but
not the steady state, because the durable surplus there is the rent,
which the levy reaches wherever the servers sit. The deepest result is
the foreign-ownership column: left untaxed, the AI owner recycles its
rent into buying domestic equity and ends up holding about seventy per
cent of the capital stock; taxing the rent slows that buy-up to roughly
half (levy) or forty per cent (rebated withholding), because it leaves
less rent to reinvest.

\begin{figure}[htbp]
\centering
\includegraphics[width=0.9\linewidth]{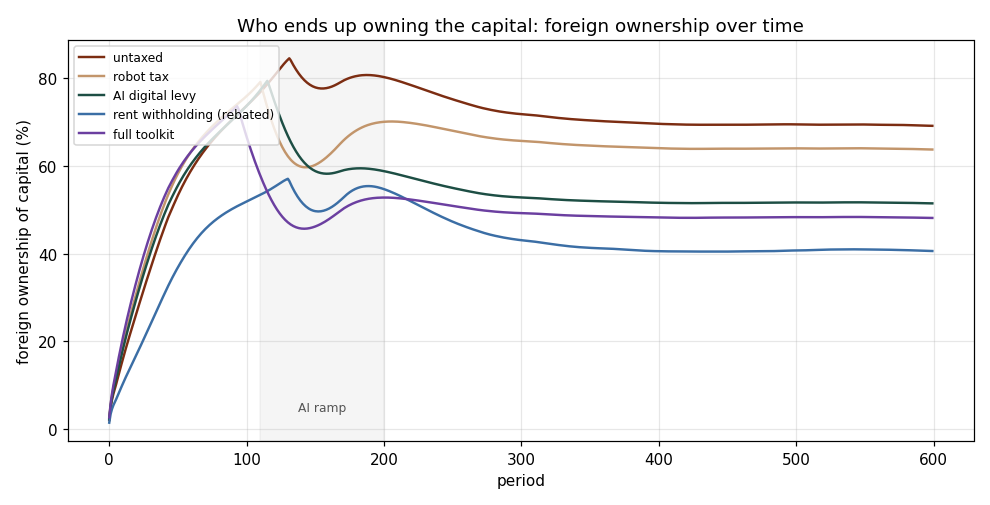}
\caption*{\textbf{Figure 8} Who ends up owning the capital. Foreign
ownership of the domestic capital stock over time under five policies,
with the AI ramp shaded. Left untaxed, the foreign owner recycles its
rent into domestic equity and climbs toward eighty per cent of the stock
during the transition, settling near seventy per cent. Each instrument
that reaches the rent leaves less to reinvest and so holds ownership
lower in the long run: the digital levy to about half, the full toolkit
to just under half, and the rebated withholding furthest, to about forty
per cent. The robot tax, which misses the rent, barely shifts the path.}
\end{figure}

\begin{figure}[htbp]
\centering
\includegraphics[width=0.9\linewidth]{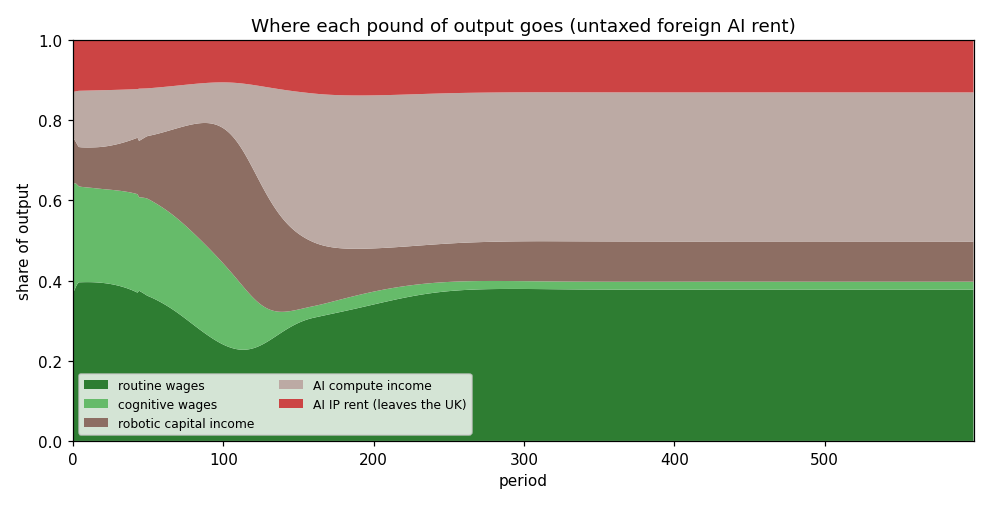}
\caption*{\textbf{Figure 9} Where each pound of output goes, over time,
under the untaxed baseline. As automation proceeds the labour share
falls (the cognitive wage share nearly vanishes as AI takes the
cognitive cluster, while routine wages hold up better), and the gains
accrue to AI compute income and to the AI IP rent. The rent (red, about
thirteen per cent of output in the mature economy) is the band that
leaves the country every period; it is the surplus the digital levy and
the withholding are designed to reach, and the part that ordinary
capital and corporate taxes miss.}
\end{figure}

\textbf{Interpretation.} Splitting the two channels changes the policy
conclusion. A robot tax is administrable and lands squarely, but it
taxes the wrong thing: the physical, competitive, low-rent part of
automation, while the durable surplus is the mobile AI rent it cannot
reach. Only a levy on AI revenue, or a withholding that reaches the
licence fee itself, touches that rent, and the corporate tax and any
compute-onshoring incentive are second-order in the mature economy (they
matter during the transition profit window, not after). The most
consequential finding is about ownership rather than revenue: a
persistently extracted, reinvested AI rent does not merely drain a flow,
it lets the foreign owner accumulate the domestic capital stock, so the
instruments\textquotesingle{} deeper value is curbing that drift. Two
caveats bound this, and subsection I takes both up directly. The
ownership result is conditional on the closure: with no goods-trade
channel, the rest of the world can only spend its rent receipts on host
assets, so the model shows the fully-reinvested pole; adding a trade
balance (Figure 15) lets the rent be repatriated as real resources,
which turns the single ownership number into a range and drains the flow
without necessarily transferring ownership. And the rent itself is a
calibrated markup rather than an estimated one, so the magnitudes
illustrate the mechanism rather than forecast a number, though the
ranking of instruments is robust to the markup (Figure 13).

\hypertarget{h-worked-examples-what-each-policy-means-for-britain-its-people-and-the-foreign-owner}{%
\subsection{H \textperiodcentered{} Worked examples: what each policy means for Britain, its
people, and the foreign
owner}\label{h-worked-examples-what-each-policy-means-for-britain-its-people-and-the-foreign-owner}}

To make the table concrete, read the host economy as the United Kingdom.
The robots in the routine cluster are supplied and largely owned by a
foreign humanoid-robotics maker, a Tesla (Optimus) or a 1X (NEO); the AI
in the cognitive cluster is supplied by a foreign frontier-model
company, an OpenAI or an Anthropic, whose model weights and brand are
the intellectual property. The foreign owner is resident in the United
States or China. UK workers operate alongside both, UK consumers and
firms buy the automated output, and the UK Treasury tries to tax the
value. Each scenario below is a specific policy (a parameter setting)
and what it does to four parties: the UK economy, the UK population, the
UK state, and the foreign owner.

\textbf{Baseline: a foreign-owned AI rent, left untaxed} (mu\_frac 0.25,
no robot tax, no DST, compute at home). The AI company prices its
service well above the cost of the compute and the cognitive labour it
uses, and books the difference, about thirteen per cent of national
output every period, as an IP licence fee paid from the UK operation to
the parent abroad. \emph{UK economy:} output still rises through
automation, but a permanent slice leaves as a deductible fee, so
measured domestic income is lower than the production would suggest.
\emph{UK population:} cognitive workers (the higher-skilled half) see AI
take a growing share of their cluster; measured domestic inequality
actually falls, because the person capturing the rent sits in San
Francisco or Shenzhen, not in the UK top tail. \emph{UK state:} the
corporate tax base is nearly empty, because the rent is a deductible
cost and the competitive return is competed away; the
government\textquotesingle s sovereign fund slips to under one times
output. \emph{Foreign owner:} collects the rent and recycles it into
buying UK equity, ending up owning roughly seventy per cent of the
domestic capital stock. The quiet result is loss of ownership, not just
a tax gap.

\textbf{Policy 1: a robot tax} (robot\_tax 0.15). A Gates-style levy on
the income of the physical robots operating in Britain. \emph{UK economy
and state:} it lands cleanly and raises about 1.5 per cent of output,
because robots cannot be moved out of the jurisdiction they work in; the
fund recovers toward 1.5 times output. \emph{UK population:} modest
extra revenue to fund transfers. \emph{Foreign owner:} the robot maker
pays, but the AI rent is untouched, so the owner still accumulates most
of the capital stock. The lesson: a robot tax is administrable and fair
on the physical channel, but it taxes the part of automation that
carries little rent and misses the part that carries most.

\textbf{Policy 2: a digital-services levy on AI} (dst\_ai 0.10). A tax
on AI revenue collected in Britain, in the spirit of the
UK\textquotesingle s existing 2\% digital-services tax, which was
designed to tax revenue precisely because revenue is far harder to move
offshore than profit. \emph{UK state:} this is the effective instrument,
collecting about 3.7 per cent of output and lifting the fund back toward
2.8 times output, because it reaches the rent the corporate tax cannot.
\emph{UK economy and population:} the rent leaking abroad shrinks and
inequality edges down. \emph{Foreign owner:} with less rent to recycle,
its accumulation of UK capital slows from seventy to about fifty per
cent. This is the lever that actually bites on the AI side, and it is
the one that has angered the United States in the real DST disputes.

\textbf{Policy 3: a withholding on the rent, rebated to citizens}
(tax\_repat 0.30, rebated). Britain intercepts a share of the licence
fee as it leaves and hands it straight back to residents per head.
\emph{UK population:} the most equalising option, because it is a direct
transfer to citizens; the wealth Gini falls furthest. \emph{UK state:}
raises no net public saving, since the proceeds are rebated rather than
banked, so the fund does not recover. \emph{Foreign owner:} the buy-up
of UK capital is curbed the most, to about forty per cent, because the
rebate strips the most rent out of the reinvestment loop. A real version
would run into treaty caps on withholding rates, so the 30\% here is
illustrative of the reach rather than a currently-available rate.

\textbf{Policy 4: where the compute sits} (s\_home, 1.0 versus 0.2).
Whether the AI is served from data centres in Britain or Europe, or from
the United States. \emph{Steady state:} in the mature economy this
barely moves revenue, because the durable surplus is the mobile IP rent
(reached by the DST), not the competitive return on the servers.
\emph{Transition:} it matters more while capital returns are briefly
high, and it is the lever a sovereign-compute policy pulls. The reading
for Britain is that onshoring compute is worth doing for resilience and
for the transition tax base, but it is not a substitute for a levy that
reaches the rent itself.

\textbf{Policy 5: the full domestic toolkit} (onshored compute, robot
tax and DST together). \emph{UK state:} raises the most, about 5.2 per
cent of output, and restores the fund to over three times output.
\emph{Foreign owner:} ownership drift is held to about half the capital
stock. \emph{Reading:} the instruments are complements, not substitutes:
the robot tax catches the physical channel, the DST catches the AI rent,
and onshored compute widens the transition base, so a state that wants
both revenue and to keep ownership at home uses all three rather than
betting on one.

\hypertarget{when-the-rent-rises-with-automation-pricing-power-and-output-capture}{%
\subsection{When the rent rises with automation: pricing power and
output
capture}\label{when-the-rent-rises-with-automation-pricing-power-and-output-capture}}

\textbf{Question.} The baseline holds the rent at a fixed share of
cognitive value added, which keeps it at a roughly constant eighth of
output throughout. That is deliberately conservative, and it embeds an
assumption worth testing: that automation changes the composition inside
the cognitive cluster (AI compute income replacing cognitive wages)
without changing the owner\textquotesingle s cut or the
cluster\textquotesingle s share of the whole. There are two reasons the
rent share could instead climb as automation deepens. First, pricing
power: as the model becomes more essential and the market more
concentrated, the owner can charge a higher markup, not a fixed one.
Second, output capture: AI need not stay a priced input inside an
otherwise-human firm. AI-native producers can do the whole job and
displace the firms that used human workers, so the AI cluster takes a
growing share of total output rather than a fixed slice of one cluster.
The early evidence in software, where AI is capturing a rising share of
development work rather than merely assisting it, is of the second kind.
Two off-default levers represent these: a markup that rises with the
degree of AI automation, and a top-level weight on the cognitive cluster
that rises with it.

\begin{figure}[htbp]
\centering
\includegraphics[width=0.9\linewidth]{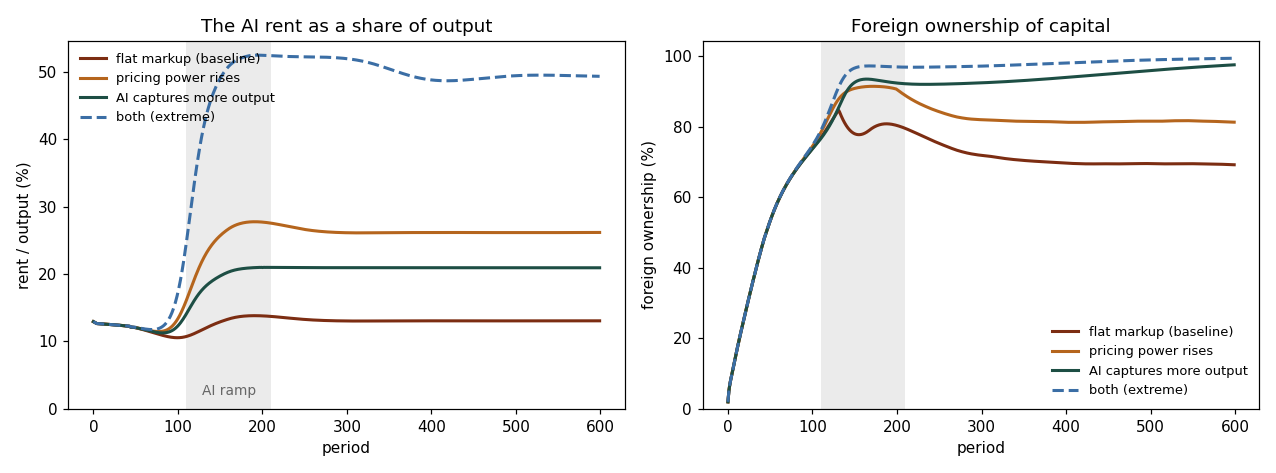}
\caption*{\textbf{Figure 26} The rent need not stay a fixed share. Left:
the AI rent as a share of output. Held flat (the baseline) it settles
near an eighth; with pricing power rising over the automation ramp it
roughly doubles to about a quarter; with the AI cluster capturing a
growing share of total output it reaches about a fifth; with both it
approaches half of output. Right: the consequence for ownership. The
flat-rent case plateaus near seventy per cent foreign ownership, but a
rising rent accelerates the drift toward the whole capital stock,
eighty-two per cent under pricing power and ninety-five to ninety-nine
per cent once the AI cluster captures output or both channels operate.}
\end{figure}

\textbf{Interpretation.} The flat-markup baseline used everywhere else
in the paper is therefore the conservative floor, not the central case.
If the rent rises with automation, which is the more likely direction
and is arguably already visible, every result in the paper sharpens
rather than softens: the surplus at stake is larger, the foreign owner
accumulates the domestic capital stock faster and more completely, and
the cost of delay and the case for reaching the rent early both grow. A
rent-importer that finds the flat-rent ownership drift to seventy per
cent uncomfortable should note that the realistic risk is closer to the
whole stock.

One honest structural limit bounds how literally to read the
output-capture channel. The model still has a single representative firm
with a markup, so the capture lever raises the cognitive
cluster\textquotesingle s share of output in reduced form rather than
mechanising the displacement. It approximates the economics of AI-native
production eating into the human cluster, but it does not model
competition between distinct AI-native and human-incumbent firms, in
which an AI producer would capture the entire output of the firms it
displaces rather than a markup on a cluster. That heterogeneous-firm
treatment is the proper way to model the distinction between AI used as
an input and AI used as the business, and it is the natural next
structural extension; this subsection bounds the consequence of moving
in that direction without yet building the mechanism. The
owner\textquotesingle s domicile is orthogonal to all of this: the
rising-rent levers change how large the surplus is, while the
foreign-owned-IP share (the domestic-owner benchmark) changes where it
goes, and a rising rent makes the domicile question more consequential,
not less.

\hypertarget{robots-embody-ip-too-the-rent-is-on-the-model-not-the-machine}{%
\subsection{Robots embody IP too: the rent is on the model, not the
machine}\label{robots-embody-ip-too-the-rent-is-on-the-model-not-the-machine}}

\textbf{Question.} The baseline puts the entire rent on the cognitive
cluster and treats the robotic channel as purely competitive:
reproducible hardware earning a normal return that ordinary instruments,
and a hardware robot tax, can reach. That is the cleanest case for the
robot-tax debate, but it is not the only one. A robot is hardware plus
the IP that runs it, the control policy or foundation model and the
proprietary design, and as robots become more capable that IP is the
scarce, foreign-held part in the same way the AI model is, while the
metal is a commodity anyone can buy. An off-default lever (robot\_ip)
puts a foreign-held markup on the robotic cluster, peeled from its value
added exactly as the AI markup is peeled from the cognitive cluster and
routed through the same licence-fee, withholding and digital-levy
machinery. The hardware robot tax continues to fall only on the
competitive remainder.

\begin{figure}[htbp]
\centering
\includegraphics[width=0.9\linewidth]{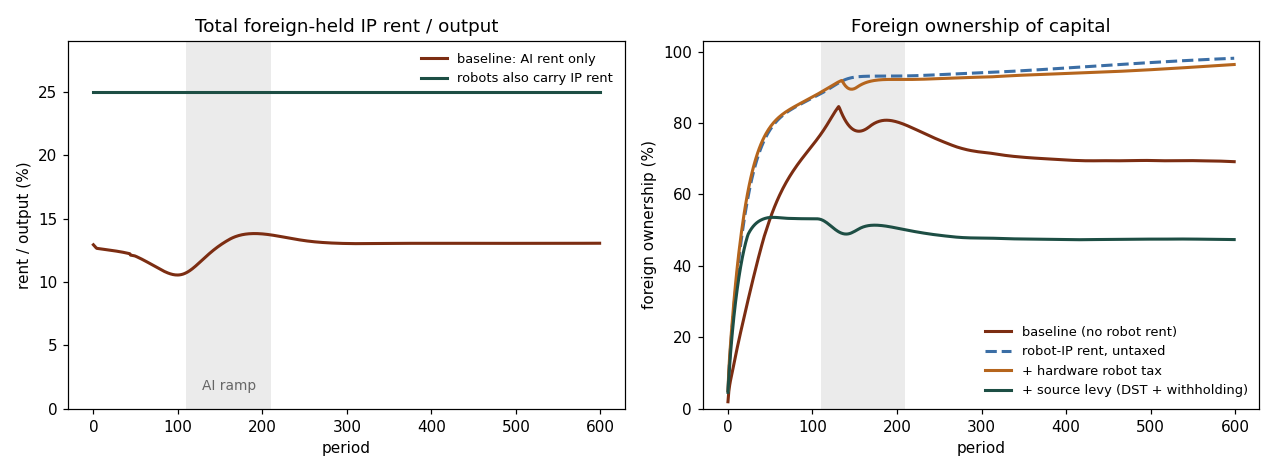}
\caption*{\textbf{Figure 27} Robots carry IP too. Left: the total
foreign-held IP rent as a share of output. With the rent confined to the
AI cluster (the baseline) it settles near an eighth; giving robots an IP
markup of the same size as the AI one roughly doubles it, to about a
quarter. Right: foreign ownership of the capital stock under that
robot-IP rent. Untaxed it drifts toward the whole stock, and a hardware
robot tax barely changes that path (the two lines almost coincide),
because it falls on the competitive hardware return and the rent has
already been peeled away from it. The same source levy that reaches the
AI rent (a digital levy plus a withholding) reaches the embodied robot
rent and holds ownership well below even the no-robot-rent baseline.}
\end{figure}

\textbf{Interpretation.} With a robot markup equal to the AI one, the
total foreign-held rent rises from about an eighth of output to about a
quarter, and the ownership drift accelerates from roughly seventy per
cent of the capital stock toward almost the whole of it. The instrument
result is the point. The hardware robot tax barely touches the drift,
because it lands on the competitive hardware return while the rent sits
on the IP, but the source levy reaches the robot rent exactly as it
reaches the AI rent, because both are recognised as the same
cross-border licence fee. The ranking is therefore unchanged, and the
robot tax\textquotesingle s role is reinforced as a thin complement on
the physical channel, not the instrument that reaches the surplus. The
deeper reading is that the economically meaningful distinction is not
robots versus AI but competitive hardware versus rent-bearing IP. In the
limit the two channels converge into a single thing, a model that does a
growing share of total labour, physical and cognitive alike, with the
durable rent sitting on the model wherever it is embodied. Confining the
rent to the cognitive cluster, as the rest of the paper does, therefore
understates the surplus rather than overstating it.

The same structural limit as the rising-rent levers applies. The robot
rent is a markup on a single representative firm\textquotesingle s
robotic cluster, foreign-held in the same proportion as the AI rent,
rather than the outcome of competition between rent-bearing and
commodity robot producers; and the incentive cost of taxing it is kept
on the AI side, so the figure is conservative about the efficiency cost
of reaching the larger rent. The lever is off by default, so every other
result in the paper is the special case in which robots earn no rent.

\hypertarget{i-robustness-and-two-open-economy-extensions}{%
\subsection{I \textperiodcentered{} Robustness and two open-economy
extensions}\label{i-robustness-and-two-open-economy-extensions}}

The two-channel interpretation (subsection G) carried two explicit
caveats: the conclusions might depend on the size assumed for the rent,
and the ownership result was the fully-reinvested pole of a closed
economy. This subsection addresses both, first by checking robustness to
the rent size and to simulation noise, then by relaxing the closure and
adding the missing incentive cost. Each extension is a single parameter
that is off by default, so every result above is the special case in
which it is zero.

\textbf{Robustness.} The rent markup is calibrated, not estimated, so
the first question is whether the ranking of instruments survives a
different rent. Sweeping the markup from a tenth to two fifths of
cognitive value added moves the levels (a larger rent means more foreign
ownership and more revenue to collect) but leaves the ordering intact at
every point: the digital levy and the full toolkit always reach the
rent, the robot tax always misses most of it, and the untaxed baseline
always lets ownership drift furthest. The conclusion is about which
lever reaches the rent, and that does not depend on how large the rent
is.

\begin{figure}[htbp]
\centering
\includegraphics[width=0.9\linewidth]{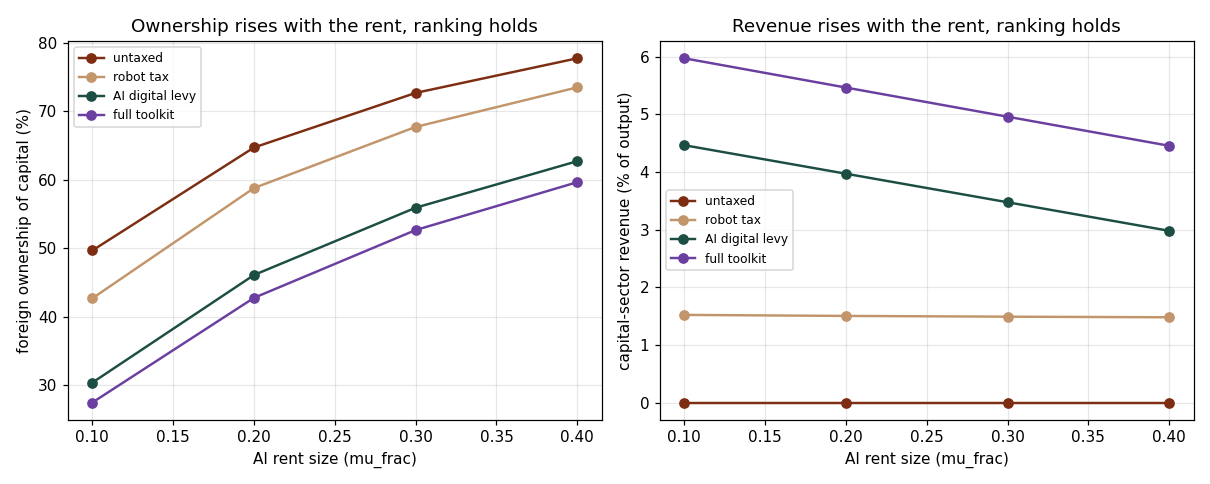}
\caption*{\textbf{Figure 13} Robustness to the rent size. Foreign
ownership of the capital stock (left) and capital-sector revenue (right)
as the AI markup mu\_frac varies from 0.10 to 0.40, for four policies.
The levels rise with the rent but the ordering of the instruments is
preserved throughout, so the qualitative conclusions do not hinge on the
calibrated markup.}
\end{figure}

Simulation noise is the second robustness question. Re-running the five
headline policies over sixteen seeds, the capital-sector revenue is
effectively deterministic (confidence intervals under a tenth of a
percentage point) and the foreign-ownership share varies by at most
about two tenths of a percentage point across seeds. The headline
numbers are properties of the structure, not artefacts of a particular
draw.

\begin{figure}[htbp]
\centering
\includegraphics[width=0.9\linewidth]{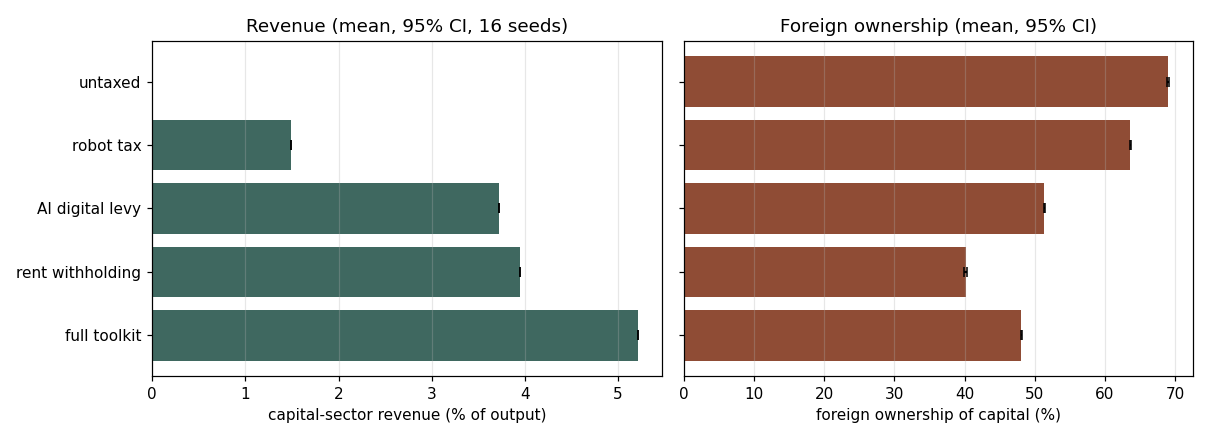}
\caption*{\textbf{Figure 14} Seed dispersion. Capital-sector revenue
(left) and foreign ownership (right) as means with 95 per cent intervals
over sixteen seeds. The intervals are within the marker width,
confirming the headline comparisons are not seed-sensitive.}
\end{figure}

\textbf{Open economy.} The ownership result was the closed pole: with no
goods-trade channel the foreign owner can only spend its rent on host
assets, so it accumulates the capital stock. Adding a current-account
leak, the share of the after-tax rent the owner repatriates as real
goods rather than reinvesting, turns the single number into a range. At
one extreme (no repatriation) the owner reaches about seventy per cent
of the stock, as before; at the other (full repatriation) it reaches
none, because the rent leaves as imports and buys no equity. The
realistic case lies between, and the same channel carries a real cost:
repatriating the rent as goods means less is invested at home, so the
steady-state capital stock and output fall to about three-quarters of
the closed economy. The drain and the ownership transfer are therefore
distinct harms, and trade policy moves the economy along this line
rather than removing the problem.

\begin{figure}[htbp]
\centering
\includegraphics[width=0.9\linewidth]{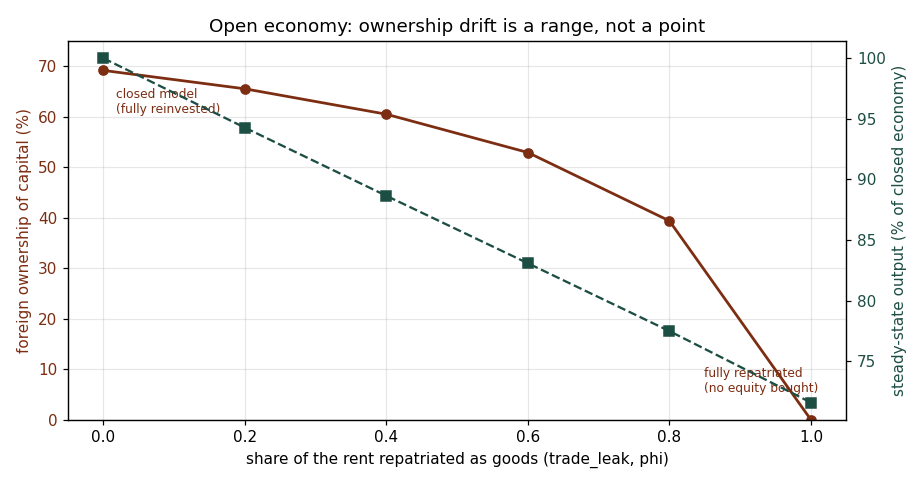}
\caption*{\textbf{Figure 15} The ownership result as a range. As the
repatriated share of the rent rises from zero (the closed,
fully-reinvested model) to one (the rent taken entirely as goods),
foreign ownership of the capital stock falls from about seventy per cent
to zero, while steady-state output falls to roughly seventy per cent of
the closed economy. Repatriation drains the flow and shrinks domestic
capital instead of transferring ownership: the two are separate costs,
and the truth lies between the poles.}
\end{figure}

\textbf{Incentive cost.} The instruments above reached the rent without
the owner reacting, which made taxing it look free. Allowing the AI
owner to deploy less capability as its net-of-tax rent falls gives the
levy an efficiency cost. Under the digital levy and a repatriation
withholding, a modest supply response lowers steady-state output to
about four-fifths of the untaxed level, while the revenue collected as a
share of output barely moves. The reading is not that the levy fails to
raise money, it still does, but that it shrinks the pie it taxes, so the
case for reaching the rent must be set against an output cost rather
than treated as a free transfer. The size of that cost is illustrative:
the model\textquotesingle s accumulation channel amplifies any
persistent output loss, so the per-period deadweight is capped to keep
the long-run magnitude credible, and pinning it down empirically is left
to the next phase.

\begin{figure}[htbp]
\centering
\includegraphics[width=0.9\linewidth]{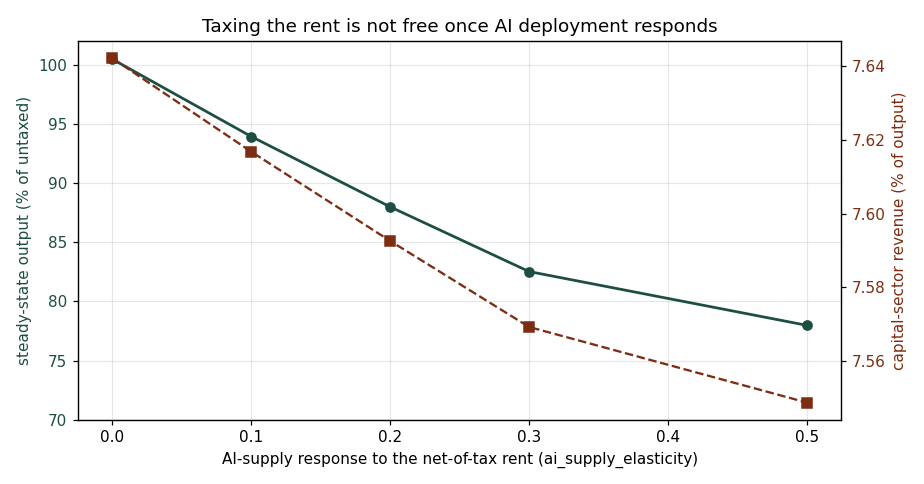}
\caption*{\textbf{Figure 16} The cost of taxing the rent once AI
deployment responds. As the AI-supply elasticity to the net-of-tax rent
rises, steady-state output falls to about seventy-eight per cent of the
untaxed level (teal), while capital-sector revenue as a share of output
is almost unchanged (sienna). Taxing the rent is no longer free: it
leaves the take broadly intact but lowers the level of output, a
trade-off the policy choice must weigh.}
\end{figure}

\textbf{Calibration.} The two extensions above are now anchored to data
rather than set arbitrarily, with the full sweeps (Figures 15 and 16)
bracketing the range. The repatriation share is set to a half:
balance-of-payments evidence puts reinvested earnings at roughly forty
to sixty per cent of foreign direct-investment income for mature host
economies (higher for young investments, by the FDI life-cycle), so the
complementary repatriated share sits near a half; and because the AI
rent is a licence fee, which is by nature a distributed flow rather than
reinvested earnings, a half is if anything conservative. The AI-supply
elasticity is set to 0.5, the low end of the empirical user-cost
elasticity of investment, which the literature places at about $-$0.5 to
$-$1.0 (Hassett and Hubbard 2002; firm-level estimates cluster $-$0.25 to
$-$0.75). These are central values for illustration, not estimates for
this market, which has no tax-variation history to estimate from; the
point is that the mechanism\textquotesingle s direction is now
disciplined by plausible magnitudes.

\hypertarget{j-reinstatement-and-the-labour-share}{%
\subsection{J \textperiodcentered{} Reinstatement and the labour
share}\label{j-reinstatement-and-the-labour-share}}

\textbf{Labour market.} The model so far lets automation push the labour
share down with no countervailing force, so that fall is partly
mechanical. Adding a reinstatement margin (new labour-intensive tasks
created in the wake of automation, in the sense of Acemoglu and
Restrepo) lets the share recover. New tasks are taken to be as
productive as the automated ones, so reinstatement is output-neutral at
the technology level and moves income from capital to labour rather than
destroying it; the only general-equilibrium effect on output is
indirect, through a lower aggregate saving rate as income shifts to
higher-spending households. The result is that the labour-share path is
conditional, not mechanical: it dips through the AI ramp and then
settles at a level set by how strongly new tasks offset automation, from
about forty per cent with no reinstatement to about sixty per cent in
the balanced case. The trade-off is that a more labour-heavy economy
saves and accumulates less, so steady-state output eases as
reinstatement strengthens.

\begin{figure}[htbp]
\centering
\includegraphics[width=0.9\linewidth]{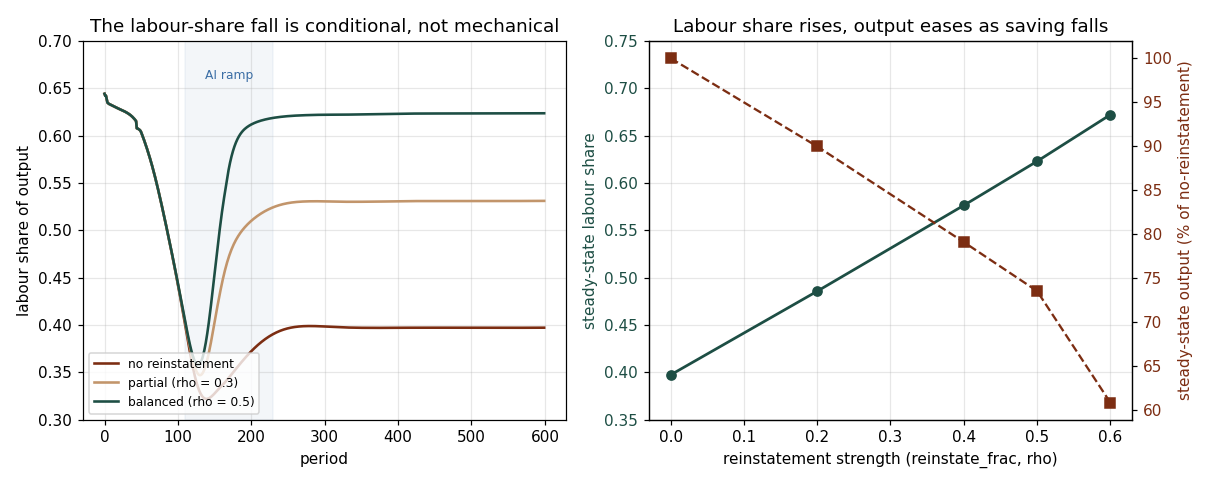}
\caption*{\textbf{Figure 17} The reinstatement margin. Left: the labour
share over time for three reinstatement strengths, with the AI ramp
shaded; the share dips during automation and recovers to a level set by
how many new tasks offset it, so the fall is not mechanical. Right: in
the mature economy the labour share rises with reinstatement (teal, from
about 0.40 to 0.67) while output eases (sienna) because the aggregate
saving rate falls as income moves to labour. This is the first half of
the labour-market extension; turning displacement into measured
unemployment, rather than a lower wage share, is the next step.}
\end{figure}

\hypertarget{k-testing-two-claims-from-the-ai-tax-debate}{%
\subsection{K \textperiodcentered{} Testing two claims from the AI-tax
debate}\label{k-testing-two-claims-from-the-ai-tax-debate}}

\textbf{Erosion.} Two specific objections from the debate (section 1)
can be put to the model directly. The first is
Friedman\textquotesingle s "the base shrinks the moment you tax it",
echoed by the IMF\textquotesingle s warning that taxing AI "could stifle
productivity". Sweeping the digital levy from zero to forty per cent,
with and without the AI-supply response, the response does erode the
base and does cost output: with it switched on, steady-state output
falls to about seventy-eight per cent of the untaxed level, and revenue
comes in roughly five to twenty-five per cent below the no-response
case. But revenue still rises with the rate throughout, it does not peak
and turn down. So the model supports the IMF\textquotesingle s "there is
a cost" while rejecting the strong form of Friedman\textquotesingle s
claim: in the credible range the rent is durable enough that reaching it
still pays, at an output cost that has to be weighed rather than a
self-defeating tax. The qualifier is that the per-period deadweight is
capped (section 6I), so this rules out a revenue-maximising peak only
within that range, not in general.

\begin{figure}[htbp]
\centering
\includegraphics[width=0.9\linewidth]{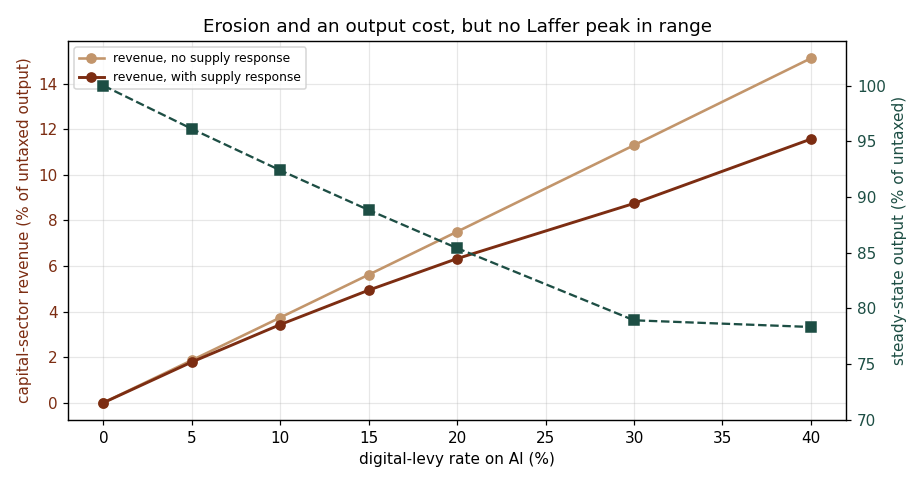}
\caption*{\textbf{Figure 18} Testing "the base shrinks the moment you tax
it". Capital-sector revenue against the digital-levy rate, without the
AI-supply response (sand) and with it (sienna); the widening gap is the
erosion. Output with the response (teal, right axis) falls to about
seventy-eight per cent of untaxed. Revenue still rises monotonically
with the rate, so within this range the levy is costly but not
self-defeating.}
\end{figure}

\textbf{Territoriality.} The second is the territorial objection,
Friedman\textquotesingle s "a provider-level tax is a subsidy for
foreign inference" and Luckey\textquotesingle s "it just makes foreign
models more attractive". Holding the levy fixed and moving the AI
compute offshore (the share located at home, s\_home, from 1.0 down to
0.2), the host\textquotesingle s take is flat: the digital levy stays at
about 3.7 per cent of output and the withholding at about 3.9 per cent
at every server location, while the corporate base is empty throughout.
The reason is that a levy on AI value recognised, and rent repatriated,
within the host does not depend on where the servers physically sit,
whereas a levy on the provider or the compute does. The territorial
objection is therefore really an argument about the choice of base: it
bites for a United States provider-level token tax (the home of the
providers) and not for a rent-importing host taxing recognised value.
The flatness reflects how a recognised-value base is defined, which is
exactly the point rather than an artefact.

\emph{For Britain concretely:} a UK digital levy on the AI revenue an
OpenAI or an Anthropic earns in Britain still pays even as the firm
trims UK deployment in response (the erosion is real but the take keeps
rising), and it collects the same whether those models are served from
data centres in Oregon or in London (the territorial neutrality). The
contrast is with a United States per-token tax on the same firms, which
would miss the value consumed in Britain entirely. This is why the
territorial objection, decisive for Washington, does not transfer to a
rent-importing host like the UK.

\begin{figure}[htbp]
\centering
\includegraphics[width=0.9\linewidth]{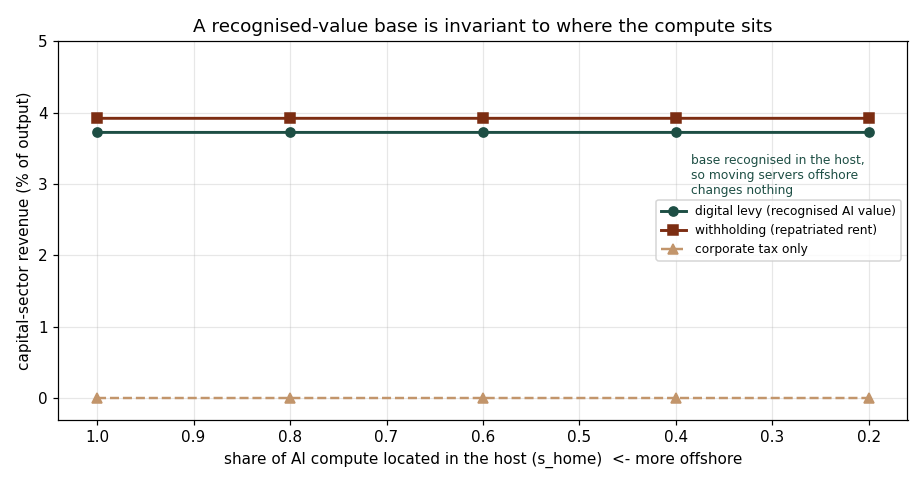}
\caption*{\textbf{Figure 19} Testing "a domestic tax just sends the base
offshore". Capital-sector revenue as the AI compute moves from fully
home-located to mostly offshore (right to left). A digital levy on
recognised AI value (teal) and a withholding on the repatriated rent
(sienna) are invariant to compute location; only a base tied to where
the servers sit would fall. For a rent-importing host the territorial
objection does not bite.}
\end{figure}

\hypertarget{l-unemployment-from-automation-and-funding-the-safety-net}{%
\subsection{L \textperiodcentered{} Unemployment from automation, and funding the safety
net}\label{l-unemployment-from-automation-and-funding-the-safety-net}}

\textbf{Employment.} The reinstatement subsection moved the labour-share
fall off its mechanical path; this one turns part of that fall into
measured unemployment rather than a uniformly lower wage. The device is
an extensive-margin overlay: production and accumulation are untouched
(the displaced tasks are done by capital, which is why output holds),
but a fraction of the labour displaced by automation, net of
reinstatement, is recorded as out of work rather than as everyone
earning less, with the displacement falling on the lowest-skill workers
in each cluster first (the entry-level pattern the debate describes).
The unchanged wage bill is then concentrated on those still employed,
and the out-of-work receive a state-funded benefit. Because it is a
relabelling of the existing labour share plus a transfer, the deposit
and net-worth invariants hold to machine precision.

Two results follow. First, whether automation produces mass unemployment
or a manageable rate is not a property of the technology but of the
balance between displacement and the creation of new tasks. With no
reinstatement and a high pass-through the rate runs toward half the
workforce; with balanced reinstatement and a moderate pass-through it
settles in the ten-to-twenty per cent band that Amodei warned of, after
a larger transition spike. The model cannot say which case is realistic
(the pass-through is a parameter, not an estimate), but it shows the
optimists and the pessimists are arguing about that balance, not about
whether the technology is powerful. Second, the safety net the proposals
call for (McMorrow\textquotesingle s reinvestment fund,
Weinberg\textquotesingle s lockbox) is affordable only if its funding
reaches the rent: leaning on the eroding income-tax base alone, public
net worth settles low, while adding the digital levy and the withholding
funds the transfers and rebuilds the sovereign fund to over twice
output. The instrument debate and the worker-protection debate are
therefore the same debate.

\emph{For Britain concretely:} if AI displaces UK cognitive roles (the
entry-level analyst, the junior drafter) faster than new roles appear,
the displaced need a benefit, and the UK income-tax base is itself
shrinking as those wages disappear. The only funding that keeps pace is
a levy reaching the AI rent that OpenAI or Anthropic books out of
Britain: the same digital levy and withholding that reach the rent for
revenue also pay for the safety net and rebuild the
Treasury\textquotesingle s balance sheet. A British retraining or
reinvestment fund, in other words, is affordable precisely to the extent
that Britain taxes the rent the automation creates.

\begin{figure}[htbp]
\centering
\includegraphics[width=0.9\linewidth]{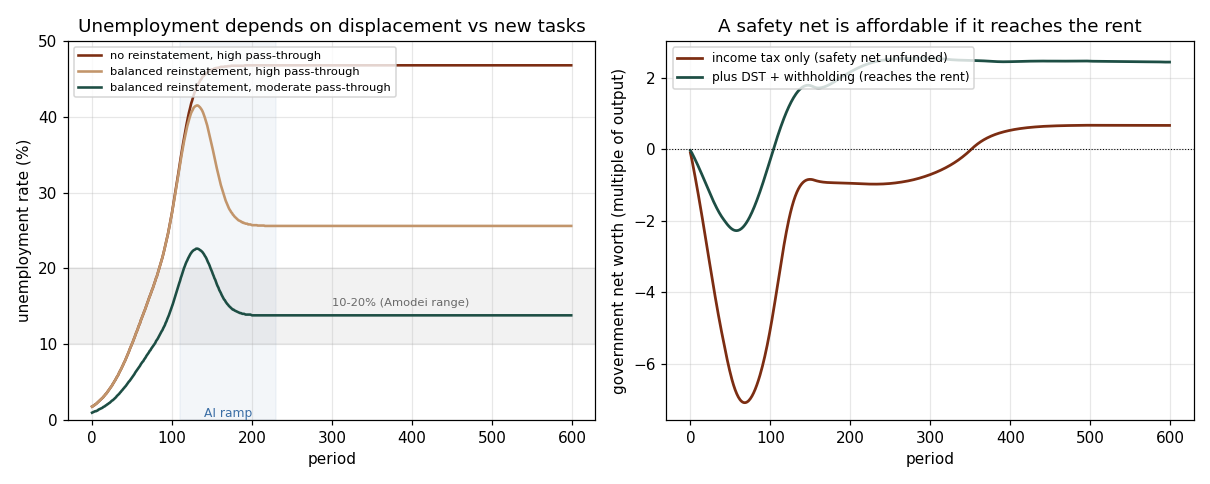}
\caption*{\textbf{Figure 20} Left: the unemployment rate over time under
three displacement-versus-reinstatement balances; balanced reinstatement
with a moderate pass-through settles in the 10 to 20 per cent band
(shaded), while weak reinstatement runs far higher. Right: government
net worth over time when the safety net is funded by income tax alone
(sienna) versus when a digital levy and a withholding also reach the
rent (teal); only the rent-reaching case rebuilds the public balance
sheet. Production and output are identical across these cases, the
overlay is on the extensive margin only.}
\end{figure}

\hypertarget{m-calibration-and-sensitivity}{%
\subsection{M \textperiodcentered{} Calibration and
sensitivity}\label{m-calibration-and-sensitivity}}

\textbf{Calibration.} The mechanisms added across the open-economy and
labour-market work introduce six parameters. None is estimated for this
market, which has no history to estimate from, but each can be given a
literature-anchored central value and a plausible range, so the results
below are read as a disciplined range rather than a point. The table
gathers the anchors in one place.

{\renewcommand{\_}{\textunderscore\allowbreak}%
\begin{longtable}[]{@{}>{\raggedright\arraybackslash}p{0.21\linewidth}>{\centering\arraybackslash}p{0.10\linewidth}>{\centering\arraybackslash}p{0.13\linewidth}>{\raggedright\arraybackslash}p{0.43\linewidth}@{}}
\toprule\noalign{}
Parameter & Central & Range & Anchor \\
\midrule\noalign{}
\endhead
\bottomrule\noalign{}
\endlastfoot
AI rent size (mu\_frac) & 0.25 & 0.15--0.40 & Upper-tail markup
estimates (De Loecker, Eeckhout \& Unger 2020); the ranking of
instruments is robust to it (Figure 13). \\
Repatriation share (phi) & 0.50 & 0.30--0.70 & Reinvested earnings are
about 40--60\% of FDI income for mature hosts, so the repatriated
complement is near a half; the AI rent is a licence fee (a distributed
flow), so a half is conservative. \\
AI-supply elasticity (eta) & 0.50 & 0.25--1.0 & User-cost elasticity of
investment of about $-$0.5 to $-$1.0 (Hassett \& Hubbard 2002); firm-level
estimates cluster $-$0.25 to $-$0.75. \\
Reinstatement (rho) & 0.50 & 0.30--0.70 & Acemoglu \& Restrepo (2019):
displacement and reinstatement were roughly balanced before 1987 (stable
labour share) and reinstatement was weaker after; a half is the recent
regime. \\
Unemployment pass-through & 0.30 & 0.10--0.50 & Acemoglu \& Restrepo
(2020) find automation lowers both employment and wages, so the
extensive-margin share is intermediate. The least-anchored parameter. \\
Benefit replacement rate & 0.50 & 0.37--0.58 & OECD net replacement rate
is 58\% in the initial spell and 37\% long-term for a single worker at
the average wage (Society at a Glance 2024). \\
\end{longtable}}

A seventh parameter, the UBI labour-supply response, is set small: the
OpenResearch guaranteed-income trial (Vivalt et al. 2024) found a
one-thousand-dollar monthly transfer cut labour-force participation by
two to four percentage points, mostly on the extensive margin,
consistent with the earlier negative-income-tax experiments, so the
model uses a modest value rather than the large response that would
otherwise dominate.

\textbf{Sensitivity.} Running the fully-instrumented economy (the
toolkit, reinstatement and unemployment all on) and moving each
parameter across its range, two things stand out. First, a basic sanity
check passes: each outcome\textquotesingle s dominant driver is the
parameter it should depend on, ownership on the repatriation share,
revenue on the rent size, the labour share on reinstatement, and
unemployment on the pass-through. Second, the results sort cleanly into
robust and parameter-sensitive. Revenue and the labour share move
modestly relative to their levels, and the qualitative conclusions
(which lever reaches the rent, the labour-share fall being conditional)
hold throughout. Foreign ownership and especially unemployment are the
sensitive ones: unemployment swings by more than twenty points across
the pass-through range, which is both the least-anchored and the most
consequential parameter. The honest reading is that the
paper\textquotesingle s qualitative claims are robust, while the precise
ownership and unemployment magnitudes should be taken as ranges set by
parameters that are plausibly bounded but not estimated.

\begin{figure}[htbp]
\centering
\includegraphics[width=0.9\linewidth]{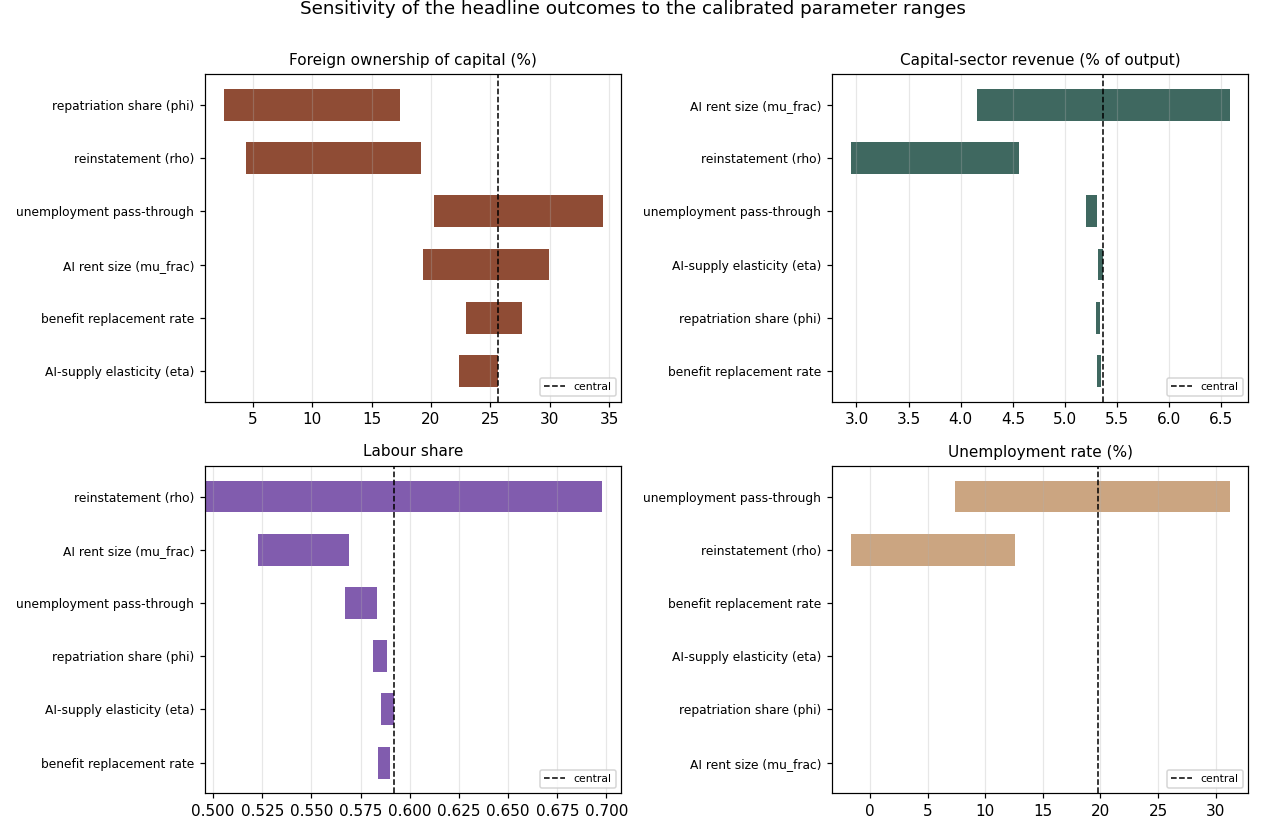}
\caption*{\textbf{Figure 21} Sensitivity of the headline outcomes to the
calibrated parameter ranges. Each bar is the swing in an outcome as one
parameter spans its range (others held at the centre, dashed line); bars
are sorted by size. Revenue and the labour share are comparatively
robust; foreign ownership and unemployment lean on the repatriation
share and the pass-through respectively, which is why those magnitudes
are reported as ranges.}
\end{figure}

\hypertarget{n-the-optimising-foreign-owner}{%
\subsection{N \textperiodcentered{} The optimising foreign
owner}\label{n-the-optimising-foreign-owner}}

\textbf{Strategic owner.} With the open-economy and labour-market
mechanisms calibrated, a further extension makes the foreign owner
strategic. The territoriality test (subsection K, Figure 19) held the
owner passive and found the host\textquotesingle s take invariant to
where the compute sits. That is the right answer to the wrong version of
the objection. The sharper question is not where the servers are but
where the rent is \emph{booked}: an owner facing a levy on the
recognised rent can move the recognition itself to a low-tax entity, the
standard profit-shifting move. The model now lets a share of the rent
recognition shift offshore, rising with the host\textquotesingle s wedge
and capped because some of the rent is tied to recognised domestic use,
so the shifted portion escapes both the digital levy and the
withholding. With the elasticity set to zero this is the non-strategic
owner of the earlier sections exactly.

Two things follow (Figure 22). The host\textquotesingle s take now bends
below the non-strategic line, with the gap, the rent booked offshore,
widening as the rate rises: the reach is real but bounded by how cheaply
the owner can shift, calibrated here to a profit-shifting
semi-elasticity of around 0.8 (Heckemeyer and Overesch 2017). And a more
mobile owner both pays less and owns more: at the full-toolkit wedge,
raising the owner\textquotesingle s mobility cuts the
host\textquotesingle s revenue from about 7.6 to 5.3 per cent of output
and lifts foreign ownership from about 34 to 48 per cent, because the
rent that escapes the levy stays with the owner to reinvest. The honest
refinement of the territoriality result is therefore this: a
rent-importing host has more reach than the United-States-centred debate
assumes, because moving the servers does not erode a recognised-value
base, but that reach is not unconditional, because moving where the
value is booked does. The policy implication is less "the base cannot
move" than "the base moves through transfer pricing, not server
location", which points at recognition and anti-avoidance rules rather
than at where the compute is hosted.

\emph{For Britain concretely:} the AI owner (an OpenAI or an Anthropic,
resident in the United States or China) cannot dodge a UK levy on
recognised AI value by serving its model from Oregon instead of London,
which is the reassuring half of the result. But it can book the UK
licence fee through a low-tax intermediary, the familiar
transfer-pricing route, and that does shrink what the UK Treasury
collects. The reading for Britain is that a digital levy has to be
written against recognised UK value and paired with transfer-pricing and
recognition rules, otherwise the base leaks through the accounting even
though it cannot leak through the data-centre map.

\begin{figure}[htbp]
\centering
\includegraphics[width=0.9\linewidth]{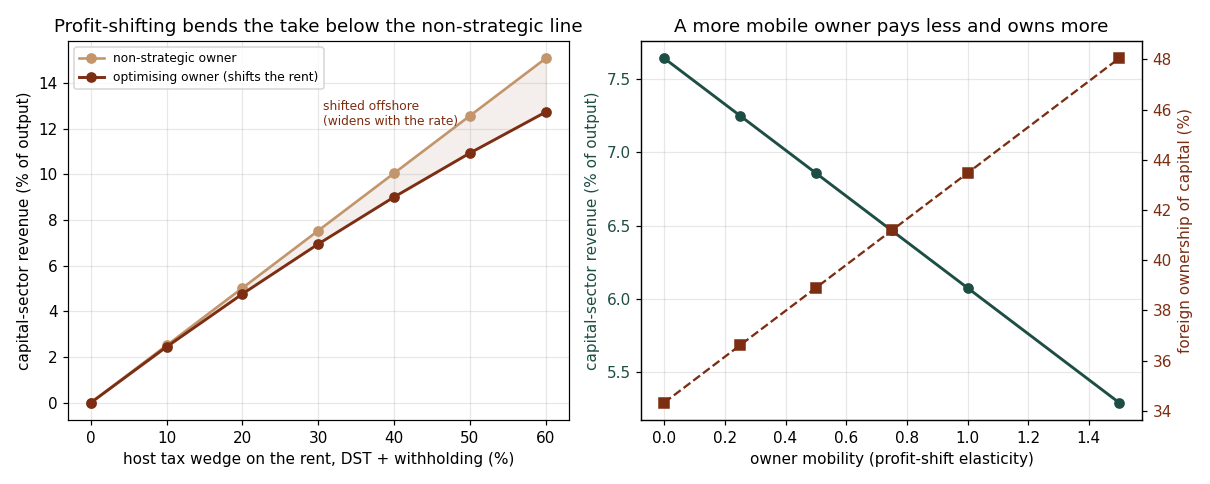}
\caption*{\textbf{Figure 22} The optimising foreign owner. Left: the
host\textquotesingle s take against its tax wedge on the rent, for a
non-strategic owner (sand) and one that shifts the rent recognition
offshore (sienna); the shaded gap is the rent booked elsewhere, widening
with the rate. Right: at the full-toolkit wedge, as the
owner\textquotesingle s mobility rises, the host\textquotesingle s
revenue falls (teal) and foreign ownership rises (sienna), since the
shifted rent stays with the owner to reinvest. Reach survives server
relocation but is bounded by profit-shifting of the recognition.}
\end{figure}

\hypertarget{o-the-contestable-frontier-competition-as-an-alternative-to-tax}{%
\subsection{O \textperiodcentered{} The contestable frontier: competition as an alternative
to
tax}\label{o-the-contestable-frontier-competition-as-an-alternative-to-tax}}

\textbf{Competition.} The rent has so far been a fixed share of
cognitive value added. That is right only if the frontier stays
monopolistic. A further extension makes the markup respond to
contestability: open-weight catch-up and more providers price part of
the rent away, so the effective rent is the markup net of competition.
With contestability at zero this is the monopolistic frontier of the
earlier sections; at one the rent is fully commoditised. This matters
because it turns competition policy into a lever that sits alongside
taxation, and the two behave very differently.

As contestability rises the rent shrinks and, with it, foreign
ownership: the rent was the durable surplus the foreign owner
accumulated, so competing it away takes ownership from about 69 per cent
toward zero (Figure 23, left). The labour share barely moves, because
the rent was peeled from both the cognitive wage and the domestic AI
capital income, so returning it benefits both factors rather than labour
alone, a useful corrective to the intuition that killing the rent
straightforwardly reflates wages. The instrument contrast is the point
(Figure 23, right): competition and taxation both cut foreign ownership,
but they are opposites for the public finances. Taxation captures the
rent, so revenue climbs as ownership falls; competition destroys the
rent, so there is nothing left to tax and revenue stays at its low
floor. They are substitutes for the ownership and distributional goal
and opposites for the fiscal one. This is the model\textquotesingle s
way of stating a choice the debate gestures at: the IMF and Brookings
preference for reviewing IP and innovation incentives, and
Friedman\textquotesingle s expectation that prices will be competed down
anyway, point toward the competition lever, which is first-best if the
concern is that the rent should not exist; a levy is the instrument if
the concern is that, while it exists, the host should capture rather
than merely dissipate it. The honest limit is that contestability here
is an exogenous market or policy state, not yet entry responding
endogenously to the size of the rent, which would be the fuller
treatment.

\emph{For Britain concretely:} if open-weight models (a Llama, a
DeepSeek) commoditise the frontier, the UK-derived rent of an OpenAI or
an Anthropic shrinks on its own, and foreign ownership of British
capital falls without the UK levying anything. But that same erosion
leaves the Treasury with nothing to tax. So Britain faces a real choice
of route, not just of rate: backing open models and competition serves
the ownership and consumer-price goal, while a digital levy serves the
fiscal goal, and only the levy puts money in the public purse.

This is not a hypothetical lever, and it may already be the live one.
China\textquotesingle s leading labs (DeepSeek,
Alibaba\textquotesingle s Qwen, Moonshot, Baidu) release their strongest
models as free open weights, a strategy a US government commission
report of March 2026 reads as deliberately undercutting the American
providers\textquotesingle{} keep-it-behind-an-API-and-charge model and
diffusing Chinese models worldwide. By early 2026 Chinese open-weight
models led the most-downloaded charts, their token usage on at least one
routing platform had overtaken the US models, and one venture-capital
partner estimated that, among the start-ups it sees which run on open
weights (perhaps a fifth to a third of its deal flow), roughly four in
five build on a Chinese model; OpenAI\textquotesingle s own move to
release open weights was, on its chief executive\textquotesingle s
account, a response to the prospect that the world would otherwise be
built on Chinese open source. In the model\textquotesingle s terms this
is the contestability lever pushed from the outside: if open weights
commoditise the frontier, the rent erodes whether or not any host taxes
it, which removes the ownership-drift problem and the tax base at the
same time. This is something the model already represents rather than
something it would need a new mechanism to show, and it implies a
rent-importer has a second, non-fiscal route to the ownership goal
(backing open models) that a country hosting the providers would be far
more reluctant to take, since it would be commoditising its own
champions.

\begin{figure}[htbp]
\centering
\includegraphics[width=0.9\linewidth]{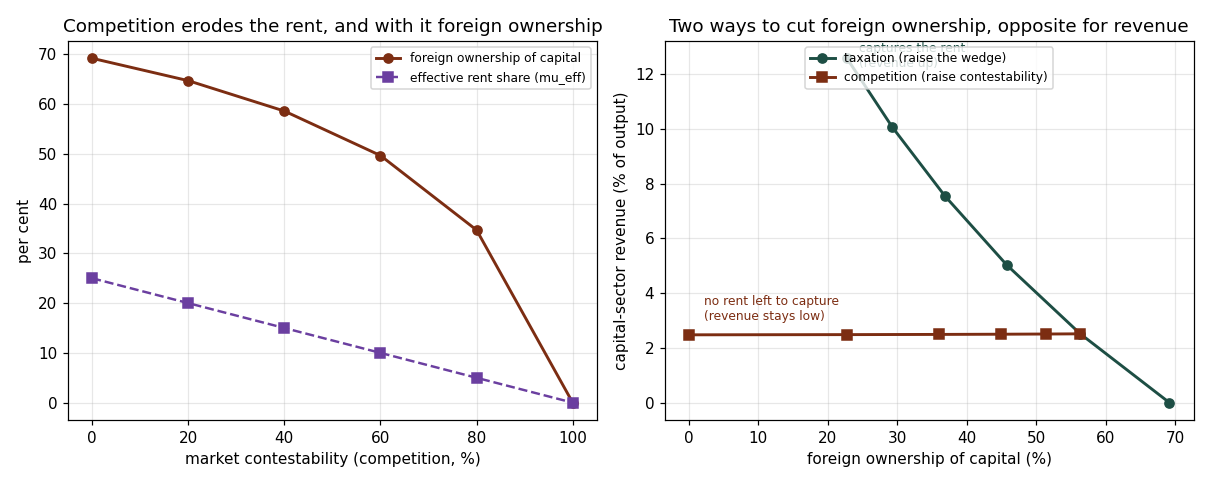}
\caption*{\textbf{Figure 23} The contestable frontier. Left: as market
contestability rises, the effective rent share (purple) falls and
foreign ownership (sienna) collapses with it. Right: both taxation and
competition cut foreign ownership, but taxation raises revenue as it
does so (it captures the rent) while competition leaves no rent to tax
(revenue stays at its floor). The levers are substitutes for ownership
and opposites for revenue.}
\end{figure}

\hypertarget{p-valuing-the-rent-the-capitalised-ip-and-why-a-market-price-adds-nothing}{%
\subsection{P \textperiodcentered{} Valuing the rent: the capitalised IP, and why a market
price adds
nothing}\label{p-valuing-the-rent-the-capitalised-ip-and-why-a-market-price-adds-nothing}}

\textbf{Valuation.} The paper has worked in flows: the rent per period,
the tax per period, the ownership share. A natural question for the
ownership story is what the rent is worth as a \emph{stock},
capitalised, because that is the owner\textquotesingle s true wealth and
it is where any wealth-tax or asset-price argument would have to bite.
This subsection capitalises the after-host-tax rent the model produces,
at a required return, as a valuation overlay. It was developed and
validated in a separate sandbox and is kept off the shipped model, so
none of the earlier results move; and it is a reduced form, a discount
rate and a premium rather than estimated ones, so the magnitudes are
illustrative rather than calibrated.

The capitalised IP is worth about 2.2 years of output at the baseline
rent and scales with it (Figure 24, left, purple). Because it is the
foreign owner\textquotesingle s asset and sits above the book capital
stock, it lifts foreign ownership measured at market value above book,
from about 69 to 75 per cent at the baseline and more at a larger rent
(the shaded gap). A source levy on the rent erodes this capitalised
wealth, from about 2.2 to 1.3 years of output across the wedge (Figure
24, right): a digital-services levy or a withholding does not merely
take a slice of this year\textquotesingle s rent, it lowers the present
value of the owner\textquotesingle s claim. A residence-based wealth
tax, by contrast, cannot reach the IP at all, because it is foreign-held
and offshore, which is the same structural reason the
paper\textquotesingle s instrument is a source levy rather than a tax on
domestic wealth.

It is worth saying why a fuller, market-clearing price was not pursued.
The intuition behind it, the owner bidding up the price of ownership as
it accumulates, turns out to be degenerate in this economy. The domestic
equity yield is slightly negative in the mature steady state, even with
no rent, because the competitive return on capital is competed away by
automation and the only surplus, the rent, is siphoned to the foreign
owner. A Tobin\textquotesingle s q on domestic equity therefore sits
below replacement cost and \emph{falls} as the rent rises, the opposite
of a bidding-up story, while the asset that does hold the value, the
rent-bearing IP, is singly held and so has no market to clear; pricing
it simply returns the fundamental already shown here. The reproduction
check on this attempt passed cleanly (a switch recovers the book model
exactly, with the accounting invariants holding to machine precision),
so this is a clean negative result rather than a failure. It is the
third independent route to the same fact: in an automated economy of
this kind the durable surplus is the foreign-held IP rent, not a return
on the domestic capital stock, and that is precisely why a market price
adds nothing and a source levy on the rent is the instrument that
reaches the surplus.

\emph{For Britain concretely:} the AI owner\textquotesingle s UK-derived
IP is worth a multiple of a year\textquotesingle s rent, and it sits on
a balance sheet in San Francisco or Shenzhen, out of reach of any UK
wealth tax. A UK source levy, by contrast, lowers the present value of
that claim directly, because it taxes the rent the claim capitalises.
This is the asset-side version of the whole paper\textquotesingle s
point: the thing worth owning is offshore, so the instrument that
reaches it is a tax at the British border on the value recognised there,
not a tax on wealth held in Britain.

\begin{figure}[htbp]
\centering
\includegraphics[width=0.9\linewidth]{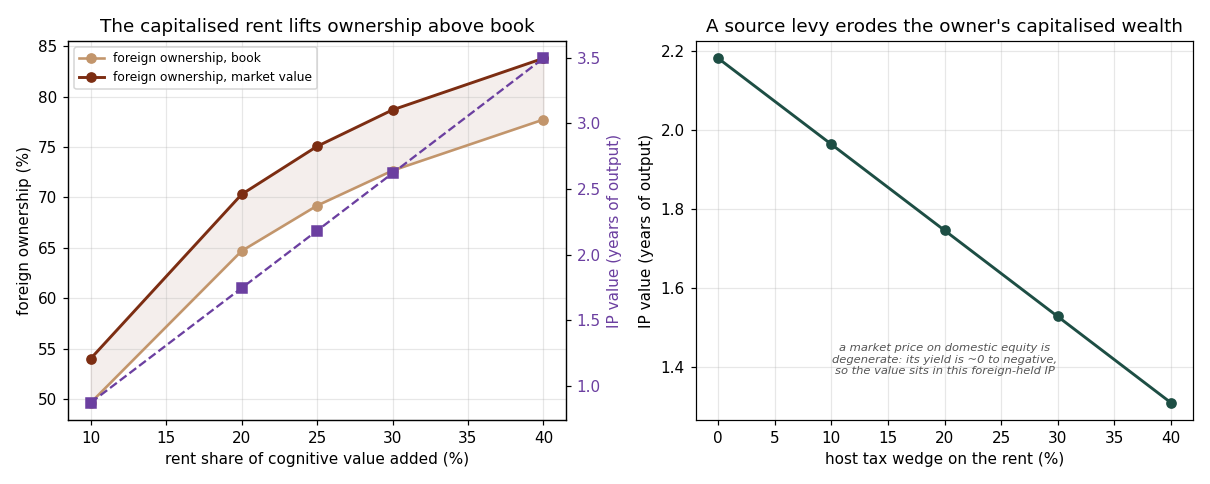}
\caption*{\textbf{Figure 24} Valuing the rent as a stock. Left: across
the rent share, the capitalised IP value in years of output (purple,
right axis) and foreign ownership measured at book (sand) versus at
market value (sienna); the capitalised rent lifts the
owner\textquotesingle s market-value share above book, and the gap
widens with the rent. Right: a source levy on the rent erodes the
present value of the owner\textquotesingle s claim. A market price on
domestic equity is degenerate here because its yield is around zero to
negative, so the value sits entirely in the foreign-held IP.}
\end{figure}

\hypertarget{carried}{}
\hypertarget{further-results-condensation-speed-leakage-and-the-cost-of-waiting}{%
\section{Further results: condensation speed, leakage, and the cost of
waiting}\label{further-results-condensation-speed-leakage-and-the-cost-of-waiting}}

Three further dynamics complete the picture. Each is reported in the
same way: what the experiment does, what the result is, and why it
matters. They concern the dynamics of the transition (how fast
inequality concentrates, how foreign ownership drains the fiscal base,
and how the timing of intervention matters), and they are robust across
the model\textquotesingle s generations; two of the figures are retained
from earlier runs of the programme and illustrate the mechanism rather
than re-stating the section 6 quantities.

\hypertarget{speed-of-condensation-rises-with-automation}{%
\subsection{Speed of condensation rises with
automation}\label{speed-of-condensation-rises-with-automation}}

\emph{What we do:} vary the depth of automation and track how fast the
wealth distribution concentrates when concentration is coupled to the
capital share. \emph{Result:} deeper automation produces both higher
terminal inequality and faster condensation toward it. \emph{Why it
matters:} the speed, not just the level, is policy-relevant, because a
faster condensation leaves a shorter window in which intervention is
cheap, which is the theme the timing result below makes precise.

\begin{figure}[htbp]
\centering
\includegraphics[width=0.9\linewidth]{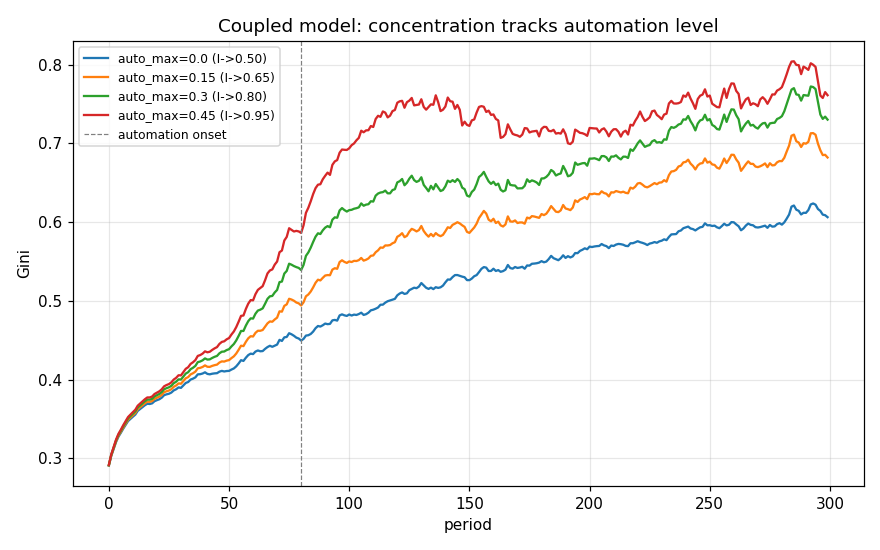}
\caption*{\textbf{Figure 10} Deeper automation produces both higher and
faster wealth condensation when concentration is coupled to the capital
share.}
\end{figure}

\hypertarget{foreign-ownership-leaks-wealth-and-erodes-the-fiscal-base}{%
\subsection{Foreign ownership leaks wealth and erodes the fiscal
base}\label{foreign-ownership-leaks-wealth-and-erodes-the-fiscal-base}}

\begin{figure}[htbp]
\centering
\includegraphics[width=0.9\linewidth]{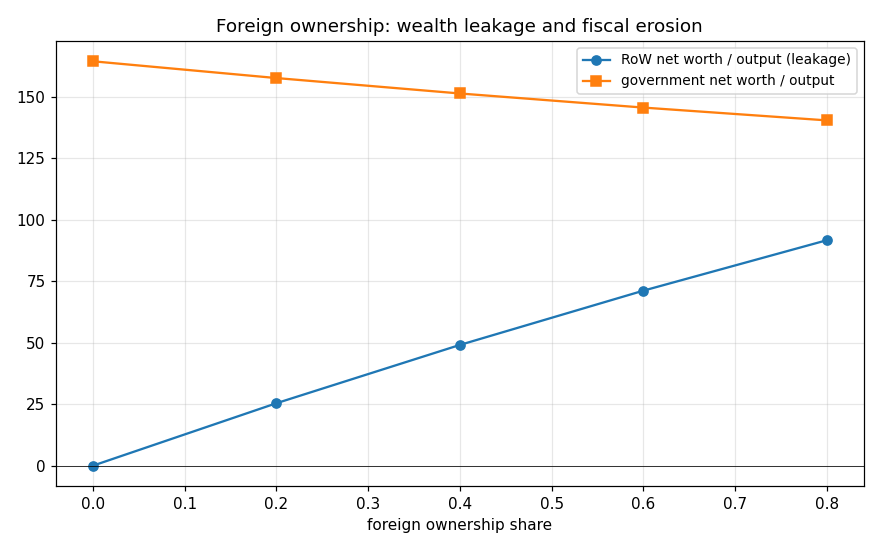}
\caption*{\textbf{Figure 11} Repatriated wealth rises and government net
worth falls with the foreign ownership share. The model makes this share
endogenous to the tax (experiment B), and the two-channel extension
makes it endogenous to the AI rent (Figure 8).}
\end{figure}

\emph{What we do:} start the rest of the world with a foreign equity
stake and let the ownership share evolve under the saving and tax rules,
tracking repatriated income and government net worth. \emph{Result:} the
dedicated foreign-ownership scenario starts the rest of the world at a
60\% equity stake but does not pin it. Because new issuance is
subscribed by whoever is saving, and the big domestic saver under a 30\%
income tax is the state, the foreign stake dilutes over the horizon and
is, in effect, gradually transferred to the domestic public sector: by
the terminal period the scenario looks like the income-tax-plus-UBI
economy with a sovereign equity fund. \emph{Why it matters:} the leakage
of repatriated profit while the stake lasts still erodes the fiscal base
during the transition, which is exactly the drain the two-channel AI
rent makes permanent unless an instrument reaches it (section 6G).
Sustaining a constant foreign share would require modelling persistent
FDI inflows, whose domestic counterpart is an ever-growing gross
creditor position under r\textgreater g; that is flagged as future work
rather than imposed, since imposing it introduces a spurious
balance-sheet explosion.

\hypertarget{intervention-timing-exhibits-hysteresis}{%
\subsection{Intervention timing exhibits
hysteresis}\label{intervention-timing-exhibits-hysteresis}}

\begin{figure}[htbp]
\centering
\includegraphics[width=0.9\linewidth]{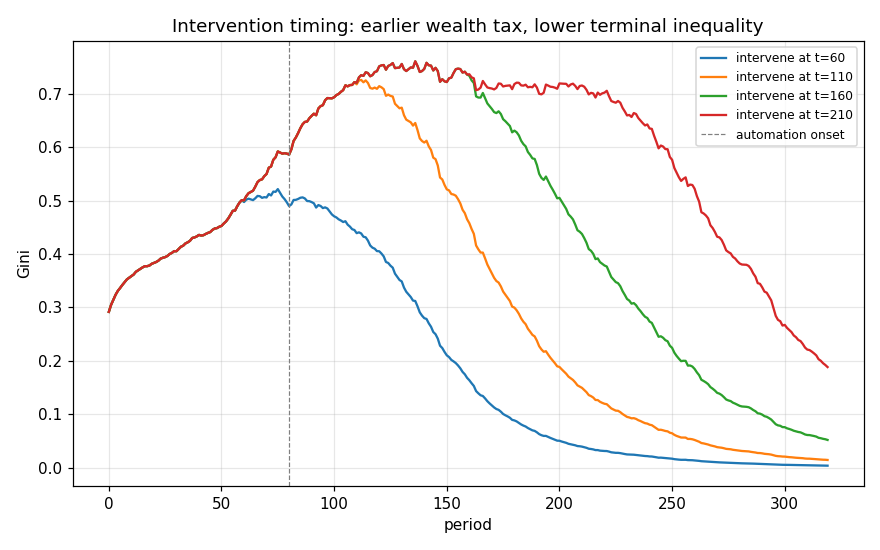}
\caption*{\textbf{Figure 12} The later a wealth tax is introduced, the
higher the terminal inequality; the cost of waiting is convex.}
\end{figure}

\emph{What we do:} introduce the same redistributive tax at
progressively later dates and compare terminal inequality.
\emph{Result:} the later the intervention, the higher the terminal
inequality, and the penalty grows more than linearly with delay.
\emph{Why it matters:} once concentration has compounded it is expensive
to reverse, so the timing of a policy is itself a lever; combined with
the first result (automation speeds condensation up), it implies the
window for cheap intervention narrows precisely as automation deepens.

\hypertarget{policy}{}
\hypertarget{policy-implications-for-the-uk}{%
\section{Policy implications for the
UK}\label{policy-implications-for-the-uk}}

This section reads the results as conditional policy implications. The
model is positive, not normative, so it cannot say what the United
Kingdom \emph{should} want. If the objectives are taken to be a long-run
stable economy and contained inequality, which is the assumption made
here, then the experiments point fairly consistently in one direction.
The magnitudes are illustrative (the AI rent is a calibrated markup, not
an estimate) and the ownership results are conditional on the
no-goods-trade closure, so what follows is about the direction and
ranking of instruments rather than precise numbers.

\hypertarget{two-regimes-two-policy-menus}{%
\subsection{Two regimes, two policy
menus}\label{two-regimes-two-policy-menus}}

The single most important determinant of the right policy is one the
debate usually leaves implicit: whether the AI is owned at home or
abroad. The two regimes call for almost opposite instruments, which is
why the same proposal can look right to an American commentator and
wrong to a British one.

When the owner is domestic (the United States), the rent stays in the
country and the problem is its concentration among domestic owners. The
instruments that reach it are the ordinary domestic ones: progressive
capital, dividend and capital-gains taxes, a wealth tax (which works
here precisely because the owner is resident and the IP onshore), and
competition policy. The binding constraint is competitiveness, since
taxing a domestic champion risks pushing its activity offshore, which is
where the territoriality objection genuinely bites. The benchmark in
section 6 shows the shape of it, with a domestic wealth tax compressing
the home inequality the rent creates.

When the owner is foreign (the United Kingdom, with a United States or
Chinese owner), the rent leaves as a licence fee and the problem is lost
ownership and an eroding base. Now the domestic levers fail: a wealth
tax cannot reach offshore IP, and ordinary corporate tax misses a
deductible fee. What works instead is a source levy, a digital-services
tax or a withholding written against value recognised at home, and the
binding constraint is not server location (the take is invariant to it)
but where the rent is booked, so the levy must be paired with
anti-avoidance rules. This is the case the rest of this section is
about.

The asymmetry is the paper\textquotesingle s organising point: the same
AI rent calls for domestic progressive taxation in the country that owns
it and a source levy in the country that imports it, and the
rent-importer has more reach against relocation, not less, than the
provider\textquotesingle s home. The remainder of this section develops
the foreign-owner menu, the harder and less familiar one.

\hypertarget{reading-the-model-against-the-ai-tax-proposals}{%
\subsection{Reading the model against the AI-tax
proposals}\label{reading-the-model-against-the-ai-tax-proposals}}

The live proposals (section 1) are, at bottom, arguments about the tax
base, and the two-channel model is a way to ask which base reaches the
surplus. Four mappings stand out, and they cut for and against different
parts of the debate.

\textbf{The base matters more than the label, and the durable surplus is
the rent.} The model\textquotesingle s clearest result is that the
lasting surplus from automation is the mobile AI IP rent, a markup on
the cognitive cluster, not the competitive return on the physical
machines. A levy that tracks AI revenue (the digital levy here) or the
repatriated rent (the withholding) reaches it; ordinary corporate tax
reaches an almost empty base, and a robot tax lands on the low-rent
physical channel. If robots themselves carry an embodied-IP rent (the
model running them), an off-default lever confirms it is reached by the
same source levy and missed by the hardware robot tax, so this ranking
is unchanged and the durable surplus is simply larger (Appendix A.2).
Read against the debate, this is why a per-token charge is a weak
instrument: a token tracks an internal accounting unit, not the rent,
which is Friedman\textquotesingle s neutrality objection in the
model\textquotesingle s own terms. The base that bites is a share of AI
value, or of the rent as it is recognised, which is closer to
Friedman\textquotesingle s own suggested alternatives (a gross-receipts
or inference-revenue base) and to Amodei\textquotesingle s mooted
revenue levy than to a token count.

\textbf{The base does shrink when taxed, but the alternatives reach
less.} Friedman\textquotesingle s strongest point is that the base
erodes the moment it is taxed. The model now contains exactly that
channel (the AI-supply response, section 6I): taxing the rent is not
free, and a high enough rate lowers output. But the same model shows
that the instruments which avoid this objection, ordinary corporate tax
and the robot tax, reach almost none of the durable surplus. The honest
conclusion is not that taxing the rent is costless, it is that the
choice is among imperfect instruments, and a revenue-or-rent base
dominates a token or physical-capital base on reach, at a cost that has
to be weighed rather than wished away.

\textbf{Territoriality bites for the provider\textquotesingle s home,
far less for the rent\textquotesingle s host.} The territorial objection
(Luckey, and Friedman\textquotesingle s fourth point), that a domestic
levy just sends activity to untaxed foreign providers, is decisive when
the taxing country is the home of the providers: a United States
per-token tax on United States firms captures none of the value that
routes through non-US inference. The model\textquotesingle s
open-economy and compute-location results say something more specific
for a host that imports the rent. What such a host can reach does not
depend on where the servers sit (the s\_home result: onshoring compute
barely moves the steady state) but on whether it taxes the value
recognised, and the rent repatriated, within its borders. A revenue or
withholding levy on AI value consumed in the jurisdiction is therefore
much harder to route around than a provider-level token tax, because the
consumption and the licence fee are in the host whatever the
provider\textquotesingle s domicile. Luckey is thus right for the United
States and wrong as a general claim, an asymmetry a UK or EU policymaker
should hold onto. The reach is bounded rather than unconditional,
though: an optimising owner can still move where the rent is booked,
even if not where the servers sit (subsection N), so the binding margin
is transfer pricing and recognition rules, not compute location.

\textbf{The worker-fund question is the rebate-versus-bank choice.}
McMorrow earmarks the proceeds for an AI Workforce Reinvestment Fund and
retraining; Weinberg\textquotesingle s lockbox does the same for
displaced workers; Warren would bank an energy-excise take for public
goods. The model frames this as the one distributional choice it cannot
make for the policymaker (see the package below): a rebate or worker
fund delivers the lowest measured inequality but builds no public
saving, while banking the levy in a sovereign fund rebuilds public net
worth and re-domesticates ownership but does less for near-term
inequality. The reinstatement result (section 6J) adds a wrinkle: the
displacement these funds are meant to cushion is not mechanical, since
how far the labour share falls depends on whether new tasks are created,
so a fund that finances genuinely new roles (McMorrow\textquotesingle s
apprenticeship framing) acts on the same margin the model identifies,
not only on the transfer.

None of this settles whether to tax AI. It does suggest the productive
disagreement is not "tokens, yes or no" but the choice of base and the
question of ownership, and that a rent-importing economy has more reach
than the United-States-centred debate assumes.

\hypertarget{the-proposals-one-by-one}{%
\subsection{The proposals, one by one}\label{the-proposals-one-by-one}}

The mappings above are organised by theme. It is worth also walking the
named proposals one at a time, because the model has something specific
to say about most of them, and is honestly not equipped to judge a few,
and the distinction is itself a result.

Take the token-base proposals first. Mark Cuban\textquotesingle s
per-token charge, and the "Token Tax" that Mallory McMorrow built into a
worker-protection plan, both make the unit of tax the token a model
emits. The model\textquotesingle s verdict on that base is unambiguous
and lines up with Dave Friedman\textquotesingle s critique: a token is
an internal accounting unit, not a unit of value, and the durable
surplus here is a rent earned on the model and the brand behind it, not
a quantity of tokens. A levy on recognised AI value, or a withholding on
the rent as it crosses the border, reaches that surplus; a token count
does not, because it tracks the wrong thing and a provider can redefine
it. Where the model has to stop short is exactly where
Friedman\textquotesingle s case is most mechanical. It contains no
tokens and no tokenizer, so it cannot reproduce his argument that the
base is endogenous to the taxpayer, nor his observation that per-token
prices have fallen roughly two-hundredfold a year, which would make any
fixed per-token rate quickly either confiscatory or trivial. The model
reaches his conclusion, do not tax tokens, by a different route, and
should be read as agreeing with it rather than as having tested those
two particular failure modes.

Dario Amodei enters the debate twice, and the two should be kept apart.
His widely-reported warning is about displacement, that AI could remove
a large share of entry-level white-collar work and push unemployment
into double digits. His own tax suggestion, by contrast, is a modest
levy of around three per cent on model revenue, not on tokens. The model
is favourable to that revenue framing: a revenue base sits close to the
base that actually bites, because it scales with the value the rent is
extracted from in a way a token count never could. On the displacement
warning, the labour-market block reproduces the ten-to-twenty per cent
unemployment band, but as a conditional result, with the outcome turning
on whether new tasks appear as fast as old ones vanish rather than on
how capable the technology is. The limitation here is the sharpest in
the paper. The speed at which displacement passes through to joblessness
is the least anchored parameter in the model and the one the outcome is
most sensitive to, which is why that band is a range and not a forecast,
and why the revenue-maximising rate for a levy like
Amodei\textquotesingle s is not something the model can pin down.

The robot tax, associated with Bill Gates and defended by economists
such as Edmund Phelps, fares better on administrability than on aim.
Because robots are physical and cannot be spirited abroad, a tax on them
is clean to levy and hard to avoid, and the model confirms it lands
without leakage. But it lands on the wrong channel. The robotic side
earns an ordinary competitive return that automation drives towards
zero, so it carries very little of the lasting surplus, while the rent
sits on the AI side and leaves as a licence fee. A robot tax is
therefore a sensible complement on the physical channel, not the
centrepiece. The caveat is that this ranking depends on the split the
model draws between the AI and robotic clusters, which is a calibrated
assumption, and on robots themselves not relocating.

Palmer Luckey\textquotesingle s territorial objection, that a domestic
AI tax simply makes untaxed foreign models more attractive and invites a
surveillance apparatus to enforce it, is the claim the model engages
most directly, and the answer is strikingly asymmetric. For a country
that hosts the providers, such as the United States, the objection has
force. For a country that imports the rent, such as the United Kingdom,
it does not: the model\textquotesingle s revenue is invariant to where
the compute is physically located, because a levy on value recognised at
home does not care whether the servers sit in London or Oregon. The
qualification, which an experiment added rather than assumed, is that
the reach is bounded not by server location but by where the rent is
booked. A strategic owner can still route the licence fee through a
low-tax entity, so the levy has to be written against recognised
domestic value and paired with transfer-pricing and anti-avoidance
rules. On Luckey\textquotesingle s enforcement-and-surveillance concern
the model is silent, and the profit-shifting it does represent rests on
a calibrated mobility rather than an estimated one.

The proposals that earmark the proceeds, McMorrow\textquotesingle s
Workforce Reinvestment Fund and the ten per cent lockbox floated by
Gabriel Weinberg, are really about the spending side, and the model has
two things to say about them. The first is a funding constraint: the
transfers these schemes promise are affordable over the transition only
if the revenue that fills them reaches the rent, because the income-tax
base that would otherwise pay for them is itself shrinking as the wages
disappear. The instrument debate and the worker-protection debate are
the same debate. The second is that the choice between banking the
proceeds and rebating them to citizens is the one genuinely political
fork in the whole analysis: banking builds public net worth and brings
ownership of the automated economy back home, while rebating delivers
lower inequality sooner but no public stake. A fund that finances
genuinely new kinds of work, rather than only cushioning lost ones, also
acts on the reinstatement margin that governs unemployment, though
whether retraining creates new tasks is an input to the model rather
than something it derives.

That leaves the cautious institutional positions, and the proposals the
model is not equipped to score. The International Monetary
Fund\textquotesingle s warning that an AI-specific tax could blunt
productivity is supported, in the sense that the model does show a real
efficiency cost once deployment responds to the levy; but the broader
bases the Fund prefers, ordinary corporate tax in particular, reach
almost none of the durable surplus in this economy, so the caution cuts
both ways. Two proposals fall outside what the model can represent at
all, and it is better to say so than to stretch. Elizabeth
Warren\textquotesingle s excise on data-centre electricity addresses an
energy base, and the model has no explicit energy or compute-cost
channel, so it cannot evaluate it. The Brookings preference for folding
an AI charge into a consumption-side value-added tax, with
business-to-business use exempted, is related in spirit to the source
levy the model does study, since both tax value where it is recognised
and both resist relocation, but the model does not represent a
value-added tax with a business exemption, so it cannot compare that
design directly. These are the clean edges of what the exercise can
claim.

\hypertarget{tax-the-rent-not-just-the-machines}{%
\subsection{Tax the rent, not just the
machines}\label{tax-the-rent-not-just-the-machines}}

The single clearest implication is that the instrument has to reach the
AI IP rent, because that is where the durable surplus sits once the
economy has automated. A robot tax is administrable and lands cleanly on
the physical robotic income, but it taxes the competitive, low-rent
channel and misses the rent entirely; the ordinary corporate tax reaches
almost nothing, because the rent is a deductible licence fee and the
competitive capital return is competed to zero in the mature economy.
Only a digital-services-style levy on AI revenue, or a withholding on
the rent as it is repatriated, touches the surplus. If the UK wants the
automation transition to fund anything, the levy on AI value (in the
spirit of the existing digital-services tax, which already taxes revenue
precisely because revenue resists shifting) is the load-bearing
instrument, and the robot tax is a useful complement rather than the
centrepiece. There is one alternative to taxing the rent rather than
complementing it, namely competing it away (subsection O): promoting a
contestable frontier shrinks the rent and so curbs foreign ownership
without a levy, but it leaves nothing to tax, so it serves the ownership
and consumer-welfare goal rather than the fiscal one.

\hypertarget{watch-ownership-not-only-revenue}{%
\subsection{Watch ownership, not only
revenue}\label{watch-ownership-not-only-revenue}}

The deepest result is about who ends up owning the capital stock, not
about the tax take. An untaxed, reinvested foreign rent does not merely
drain a flow; it lets the foreign owner accumulate domestic equity,
climbing toward seventy per cent of the capital stock in the mature
economy (Figure 8). For a goal of long-run stability and domestic
resilience, that ownership drift is arguably more important than the
revenue line, because it determines whether the returns to the automated
economy accrue at home or abroad. The instruments that reach the rent
also curb the drift, the rebated withholding most of all. A policy that
banks the proceeds in a sovereign equity fund (rather than rebating
them) does double duty here: it both raises net public saving and
re-domesticates ownership, which is the combination the full-toolkit
scenario achieves.

\hypertarget{if-the-rent-rises-every-recommendation-matters-more}{%
\subsection{If the rent rises, every recommendation matters
more}\label{if-the-rent-rises-every-recommendation-matters-more}}

The recommendations here are read off the conservative case in which the
rent stays a fixed share of output. The rising-rent results (section 6,
Figure 26) say the more likely direction is up, through greater pricing
power and through AI-native production capturing a growing share of
output, and that this changes the policy calculus in degree rather than
in kind. Re-running the instruments against a risen rent confirms the
ranking is unchanged: a source levy still reaches it and a robot or
corporate tax still misses it, because a larger rent is still a rent.
What changes is how much is at stake. Under a rent driven to a quarter
of output by pricing power, leaving it untaxed lets foreign ownership
run to about eighty per cent of the capital stock, while the digital
levy paired with a withholding pulls that back to the low forties and
lifts public net worth several-fold, a wider gap than the same
instruments open in the flat-rent case. In the extreme, where the rent
approaches half of output, the untaxed economy does not merely transfer
ownership but strains its public finances to the point of insolvency,
whereas the full toolkit keeps it solvent and holds foreign ownership
below sixty per cent. The instruments do not change; their value, and
the cost of forgoing them, scales with the rent.

Three of the recommendations are promoted by this. Watching ownership
rather than only revenue moves from prudent to essential, because a
rising rent is precisely the case in which the revenue captured
understates the ownership transferred. Intervening early matters more,
because the drift the early levy is meant to curb is faster and runs
further. And competition policy gains weight as a first-best, because
preventing the concentration that lets the markup rise is the one route
that stops the rent growing in the first place rather than taxing it
after the fact; the open-weight dynamic in subsection O is, in this
light, not only an erosion of the rent but a brake on its rise. The
banking-versus-rebate choice tilts marginally toward banking, since a
near-total ownership drift is the case in which rebuilding a domestic
stake is most valuable. The retaliation trade-off discussed below
sharpens in the same direction: a larger prize raises both the revenue
forgone by folding and the value of holding the line.

\hypertarget{intervene-early-and-prefer-stock-to-flow-taxes}{%
\subsection{Intervene early, and prefer stock to flow
taxes}\label{intervene-early-and-prefer-stock-to-flow-taxes}}

Two further results bear on design. First, the timing result (section 7)
shows the cost of waiting is convex and the window for cheap
intervention narrows as automation deepens, so a credible instrument
introduced early dominates a larger one introduced late. Second, across
the policy frontier (section 6A) the stock taxes (a wealth tax) occupy
the favourable region on both inequality and solvency, while pure flow
taxes leave inequality near the laissez-faire level; a behavioural
wealth tax still redistributes at realistic rates even after its base
erosion and the modest capital flight it induces. For contained
inequality, a wealth-type instrument is therefore worth more than its
headline rate suggests, provided it is paired with the rent-reaching
levies above so that the automated surplus is itself in the tax base.

\hypertarget{sovereign-compute-helps-but-is-not-the-lever-that-reaches-the-rent}{%
\subsection{Sovereign compute helps, but is not the lever that reaches
the
rent}\label{sovereign-compute-helps-but-is-not-the-lever-that-reaches-the-rent}}

Onshoring the AI servers (raising s\_home) is worth doing for resilience
and for the transition tax base, when capital returns are briefly high,
but it barely moves the steady state on its own, because the durable
surplus is the mobile IP rent rather than the competitive return on the
servers. The reading is that a sovereign-compute push and a levy on the
rent are complements, not alternatives: compute policy widens the base
that the levy then taxes, and keeps the physical infrastructure (and its
resilience) at home, but it does not substitute for an instrument that
reaches the rent wherever the servers sit.

\hypertarget{the-binding-constraint-in-practice-digital-taxes-and-retaliation}{%
\subsection{The binding constraint in practice: digital taxes and
retaliation}\label{the-binding-constraint-in-practice-digital-taxes-and-retaliation}}

One thing the model deliberately leaves out has to be put back for the
policy to be realistic. The source levy it recommends is, in the real
world, a digital-services tax or a withholding, and the United States
treats these as discriminatory taxes on its firms and retaliates against
them. The instruments are live and contested: the US ran Section 301
tariff investigations into these digital taxes from 2019 to 2021
(covering France, Austria, Italy, Spain, Turkey and the United Kingdom,
with twenty-five per cent tariffs determined and then suspended pending
the OECD process), and in early 2025 a presidential memorandum directed
their renewal while Congress drafted a retaliatory tax (the proposed
Section 899) aimed at countries with such measures. The pressure has
largely worked: Canada repealed its digital tax, India dropped its
equalisation levy, Italy agreed to reform, and the United
Kingdom\textquotesingle s measure has been reported as a likely
concession, while France, pushing to raise its rate, has been warned to
expect proportionate retaliation. So the model\textquotesingle s result,
that a rent-importing host technically can reach the surplus and that
server relocation does not defeat it, comes with a political price tag
the model does not carry: using the instrument means accepting the risk
of trade retaliation.

This sharpens the recommendation rather than overturning it, and it
makes the timing argument concrete. The prize at stake grows as
automation deepens, because the rent compounds into ownership and
capital\textquotesingle s share of income rises, so the value a country
forgoes by folding to avoid retaliation increases over time. A country
that wants to keep the rent and slow the loss of ownership by taxing it
will, on current evidence, have to be willing to absorb retaliation to
do so, and the case for absorbing it strengthens the further automation
proceeds. The competition result above is the important escape from the
bind: backing open-weight competition erodes the rent without a levy,
and so without provoking retaliation, at the cost of collecting no
revenue. The blunt choice for a rent-importer is therefore three-way,
not two: tax the rent and accept the retaliation, foster competition and
forgo the revenue, or fold and accept the loss of ownership.

\hypertarget{a-coherent-package-with-one-explicit-choice-left-to-politics}{%
\subsection{A coherent package, with one explicit choice left to
politics}\label{a-coherent-package-with-one-explicit-choice-left-to-politics}}

Putting these together, the model\textquotesingle s conditional
recommendation for the stated goals is a package rather than a single
tax: a digital levy on AI value as the centrepiece (reaching the rent),
a robot tax as a clean complement on the physical channel, an
onshored-compute policy to widen the transition base and keep
infrastructure at home, and a behavioural wealth tax to hold inequality
down, with the proceeds banked in a sovereign fund rather than fully
rebated. The instruments are complements: in the experiments the full
toolkit raises the most revenue and holds foreign ownership to about
half the stock. The one genuine choice the model cannot make is
distributional: a rebated withholding delivers the lowest inequality but
builds no public saving, while a banked levy builds the fund and
re-domesticates ownership but does less for measured inequality in the
near term. That trade between paying citizens now and accumulating a
public stake is a political judgement about whether the state or the
household should hold the claim on the automated economy, and the model
can only lay out the consequences of each, not choose between them.

\hypertarget{did-the-experiments-change-the-recommendation}{%
\subsection{Did the experiments change the
recommendation?}\label{did-the-experiments-change-the-recommendation}}

It is worth being explicit about how much of this recommendation was
already in place after the core two-channel experiments (subsections F
to I) and how much the later extensions moved it. The short answer is
that the centrepiece did not change, the case for it was hardened
against its strongest objections, and one genuine rider was added, with
the menu of complements widened around it.

\textbf{What did not change.} The load-bearing recommendation, a source
levy that reaches the AI rent, a focus on ownership and not only
revenue, and the bank-versus-rebate distributional choice, was set by
the two-channel and open-economy experiments and survived every later
test intact. None of the extensions displaced the digital levy as the
centrepiece or overturned the ownership story.

\textbf{What hardened it.} The debate tests (subsection K) put the two
strongest objections directly to the model. Friedman\textquotesingle s
claim that the base self-destructs the moment it is taxed came out as
costly but not self-defeating in the credible range; the territorial
objection came out as decisive for the United States but not for a
rent-importing host. The capitalisation result (subsection P) reinforced
the same conclusion from the asset side: the surplus is foreign-held
intellectual property that a domestic wealth tax cannot reach, while a
source levy lowers its present value directly. These did not change the
recommendation, they removed the main reasons to doubt it.

\textbf{The one genuine change.} The optimising-owner extension
(subsection N) qualified the territorial result. The earlier finding
that reach does not depend on where the compute sits is correct, but the
owner can still move where the rent is \emph{booked}. This adds a rider
that was not present at the two-channel stage: the levy has to be
written against recognised domestic value and paired with
transfer-pricing and anti-avoidance rules, or the base leaks through the
accounts even though it cannot leak through the data-centre map. This is
the clearest case of an experiment changing the policy, from "the base
cannot move" to "the base moves through recognition, so that channel has
to be closed too".

\textbf{What widened the menu.} The contestable-frontier result
(subsection O) added competition policy as a genuine alternative route
to the ownership goal, first-best if the aim is that the rent should not
exist, but raising no revenue, so it complements rather than replaces
the levy. The unemployment and safety-net result (subsection L) tied the
worker-protection debate to the instrument debate by showing the
transfers the proposals call for are affordable only from a
rent-reaching levy. Neither displaced the centrepiece; both attached new
commitments to it. The net effect is a recommendation that is more
robust and slightly more demanding than at the two-channel stage: the
same digital levy, now explicitly paired with anti-avoidance rules, with
competition policy offered as an ownership-only alternative, and with
the safety net\textquotesingle s funding shown to rest on the very same
instrument.

\leavevmode\vadjust pre{\hypertarget{summary}{}}%
\textbf{What raised the stakes.} The rising-rent results (Figure 26) did
not change the instrument ranking, which holds because a larger rent is
still a rent reached by the same source levy. What they changed is the
magnitude of the problem and the urgency of the response: a rent that
grows with automation drives the ownership drift toward the whole
capital stock and, untaxed in the extreme, toward insolvency, so the
ownership focus and the early-action argument are not merely supported
but made central. This is the sense in which the flat-rent results
elsewhere are a floor; the recommendation is the same, but the
consequence of ignoring it is larger, and competition policy gains a
second rationale as the route that keeps the rent from rising at all.

\hypertarget{summary-of-findings}{%
\section{Summary of findings}\label{summary-of-findings}}

\begin{longtable}[]{@{}>{\raggedright\arraybackslash}p{0.22\linewidth}>{\raggedright\arraybackslash}p{0.16\linewidth}>{\raggedright\arraybackslash}p{0.52\linewidth}@{}}
\toprule\noalign{}
Finding & Status in v3 & Detail \\
\midrule\noalign{}
\endhead
\bottomrule\noalign{}
\endlastfoot
Stock taxes vs flow taxes & Holds, refined & Stock taxes still occupy
the favourable region; flow taxes leave inequality at the laissez-faire
level. \\
Wealth-tax "dominance" & Reframed & Still redistributes at realistic
rates; a modest, rate-increasing share of apparent equality is capital
flight. \\
Composition effect & Corrected & Measured vs true Gini wedge stays under
a point of Gini at all rates; compression is front-loaded; offshore
share scales \textasciitilde2\% per pp. \\
Concentration engine & Microfounded & Tail emerges from heterogeneous
persistent returns, not an imposed kernel. \\
Robustness (Sobol) & New & Inequality driven by the dispersion and
persistence of returns, not the behavioural elasticities; fiscal results
driven by the CES elasticity. \\
Fiscal solvency (H1) & Supported under this closure & Real economy is
locally stable in every regime (capital-map slope \textless{} 1);
solvency turns on the r\textgreater g debt root, which under-taxed
welfare states (15\% income tax + UBI) hit and adequately-taxed regimes
neutralise. This is a numerical result under the residual-investment,
full-employment closure, not a general theorem. \\
Transition realism & Fixed & Differential-saving closure gives a
bounded, convergent path (no explosion); K/Y rises from
\textasciitilde2.9 to \textasciitilde8--10 as the economy automates and
settles. \\
Two-channel robustness & New & The ranking of instruments holds across
the rent size (Figure 13); headline revenue and ownership are tight
across seeds (Figure 14). \\
Rising rent & New & If pricing power grows or the AI cluster captures a
rising output share, the rent climbs from about an eighth of output
toward a quarter or more and foreign ownership accelerates toward the
whole capital stock (Figure 26). The instrument ranking is unchanged,
but the stakes and the urgency of reaching the rent rise sharply; the
flat-rent results elsewhere are a conservative floor. \\
Robot-IP rent & New & If robots carry their own foreign-held IP rent
(the model running them) rather than being purely competitive hardware,
the total rent roughly doubles and ownership drifts toward the whole
stock (Figure 27). A hardware robot tax misses it; the same source levy
reaches it; the ranking is unchanged. The meaningful distinction is
competitive hardware versus rent-bearing IP, not robots versus AI, so
confining the rent to AI is conservative. \\
Open-economy ownership & New & With a current-account leak the ownership
drift is a range, from \textasciitilde70\% (closed) to 0\% (full
repatriation), and repatriation cuts domestic output to
\textasciitilde70--75\% (Figure 15). \\
Cost of taxing the rent & New & With an AI-supply response the levy is
no longer free: output falls to \textasciitilde78\% of untaxed while the
revenue share is broadly unchanged (Figure 16). \\
Reinstatement and the labour share & New & New tasks make the
labour-share fall conditional, not mechanical: it recovers from
\textasciitilde40\% to \textasciitilde60\% as reinstatement strengthens,
at some cost to output via lower saving (Figure 17). \\
Unemployment and the safety net & New & Extensive-margin displacement
gives a 10--20\% unemployment rate under balanced reinstatement; the
safety net is affordable only if funded from a rent-reaching levy
(Figure 20). \\
Optimising foreign owner & New & An owner that shifts where the rent is
booked bounds the host\textquotesingle s take (it bends below the
non-strategic line) and drifts ownership up; reach survives server
relocation but not profit-shifting of the recognition (Figure 22). \\
Contestable frontier & New & Competition erodes the markup: it cuts
foreign ownership like a tax does, but destroys the rent rather than
capturing it, so it is a substitute for the ownership goal and the
opposite for revenue (Figure 23). \\
Capitalised rent (IP value) & New & The rent capitalised as a stock is
worth about 2.2 years of output and lifts foreign ownership at market
value above book; a source levy erodes it, while a market price on
domestic equity is degenerate because the surplus is the foreign-held
IP, not the domestic capital (Figure 24). \\
\end{longtable}

\hypertarget{caveats}{}
\hypertarget{caveats-and-remaining-work}{%
\section{Caveats and remaining work}\label{caveats-and-remaining-work}}

The model\textquotesingle s claims are bounded by its assumptions. The
limitations below are ordered by how much each one constrains what the
results can claim, with the first group the priorities for the next
iteration. They concern the specific results; the foundational
assumptions of the modelling approach itself, and which conclusions are
robust to them, are catalogued separately in Appendix A.

\hypertarget{what-most-bounds-the-headline-claims}{%
\subsection{What most bounds the headline
claims}\label{what-most-bounds-the-headline-claims}}

\begin{itemize}
\tightlist
\item
  \textbf{Open-economy closure.} \emph{Partially addressed.} A
  parametric current-account leak now lets the foreign owner repatriate
  a share of its after-tax rent as real goods rather than reinvesting
  it, so the ownership drift in Figure 8 is reported as a range between
  the closed pole and full repatriation (Figure 15) rather than as a
  single upper bound. What remains is that the repatriation share is a
  parameter, not an estimated behavioural choice: there is still no
  exchange rate, no relative-price adjustment, and no optimising foreign
  owner deciding how much to repatriate, so the range is bounded
  correctly but its interior point is not yet pinned down. This stays
  the largest structural caveat on the most striking result.
\item
  \textbf{The AI rent: its size and its incentive effects.}
  \emph{Partially addressed.} The ranking of instruments is now shown to
  be robust to the rent size across a markup sweep (Figure 13), so the
  qualitative conclusions no longer hinge on the calibrated mu\_frac,
  and an AI-supply elasticity to the net-of-tax rent now gives the levy
  an efficiency cost, so taxing the rent is no longer free (Figure 16).
  Two things still bound this: the markup and the supply elasticity are
  now anchored to plausible central values (mature-FDI reinvestment
  shares for the repatriation rate, the $-$0.5 to $-$1.0 user-cost
  elasticity for the supply response, see the calibration note in
  section 6I) but are not estimated for this specific market, which has
  no tax-variation history to estimate from; and the deadweight loss is
  capped per period because the model\textquotesingle s accumulation
  channel amplifies any persistent output loss, so the magnitude of the
  cost is indicative rather than calibrated. The model can now claim a
  net rather than gross benefit in direction, and a disciplined range,
  but not yet a point estimate of the cost.
\item
  \textbf{Task creation and unemployment.} \emph{Addressed, with
  calibration caveats.} Both halves of the labour-market extension are
  now in: a reinstatement margin makes the labour-share fall conditional
  rather than mechanical (Figure 17), and an extensive-margin overlay
  turns part of that fall into measured unemployment with a funded
  safety net (Figure 20), so the model can speak to the displacement
  debate directly. The unemployment is structural (workers displaced by
  capital, output held) rather than cyclical, which suits an automation
  model but means it does not capture demand-driven joblessness. The
  remaining limits are calibration, not structure: the displacement
  pass-through, the benefit replacement rate and the UBI labour-supply
  response are anchored to literature ranges (the calibration table in
  section 6M) but not estimated for this market, and the sensitivity
  analysis (Figure 21) shows the unemployment rate is the most
  parameter-sensitive outcome in the model, swinging on the
  pass-through, so it is reported as a range set by the
  displacement-versus-reinstatement balance rather than a point.
\end{itemize}

\hypertarget{what-would-sharpen-the-results}{%
\subsection{What would sharpen the
results}\label{what-would-sharpen-the-results}}

\begin{itemize}
\tightlist
\item
  \textbf{AI as a producer, not only a priced input.} The model treats
  AI as a factor inside one representative firm that pays a markup, and
  the rising-rent subsection lets that markup and the cognitive
  cluster\textquotesingle s output share grow in reduced form. It does
  not yet model competition between distinct AI-native and
  human-incumbent firms, in which an AI-native producer captures the
  entire output of the firms it displaces rather than a markup on a
  cluster. A heterogeneous-firm version, with the owner of the AI-native
  firm either domestic or foreign, is the natural way to model the
  difference between AI used inside a company and AI used as the
  company, and is the most consequential remaining structural extension.
\item
  \textbf{Endogenous markup.} Addressed (subsection O): the markup now
  falls with market contestability, so competition policy can be set
  against taxation as a route to the rent. What would further sharpen it
  is making entry respond endogenously to the size of the rent, rather
  than treating contestability as an exogenous state.
\item
  \textbf{A market-clearing equity price.} Attempted (subsection P),
  with a negative result: a market price on domestic equity is
  degenerate, because the competitive return is competed away and the
  durable surplus is the foreign-held IP, so the value sits in the
  capitalised rent the reduced-form valuation already reports. A genuine
  asset-price and bidding-up dynamic would need the rent-bearing firm
  restructured as a traded asset with competing buyers, which remains
  future work.
\item
  \textbf{An explicit welfare criterion.} Policy comparisons remain
  positive (inequality and solvency), not welfare-optimal; a social
  welfare function would move the policy section from directional
  rankings to normative recommendations and let any "preferable regime"
  language be stated properly.
\item
  \textbf{An optimising foreign owner.} Addressed (subsection N): the
  owner now shifts where the rent is recognised in response to the
  host\textquotesingle s wedge, which bounds the territorial reach and
  makes transfer pricing the binding margin. What would further sharpen
  it is endogenising the compute-location and repatriation choices as
  well, and anchoring the profit-shifting elasticity to estimates rather
  than a calibrated value.
\end{itemize}

\hypertarget{realism-and-robustness}{%
\subsection{Realism and robustness}\label{realism-and-robustness}}

\begin{itemize}
\tightlist
\item
  \textbf{International-tax architecture.} The withholding sits above
  typical treaty caps, and the model ignores treaties, Pillar Two, and
  the retaliation a unilateral digital levy invites (the actual US/China
  flashpoint); constraining instruments to treaty-feasible ranges and
  adding a strategic dimension would make the policy claims
  implementable.
\item
  \textbf{Empirical validation.} The two-channel results are internally
  consistent (the stock-flow invariants and the production Euler
  condition hold to machine precision) but are not yet matched to
  external moments (observed factor shares, FDI stocks, AI-sector
  margins); calibrating to these would turn the magnitudes from
  illustrative into estimated.
\item
  \textbf{Credit, leverage, housing and asset prices.} There is still no
  credit or leverage and no housing or asset-price sector; the former is
  where the closest prior art locates instability, the latter where much
  of the empirical capital-share movement actually lives.
\item
  \textbf{Behavioural realism elsewhere.} The UBI is a transfer with no
  labour-supply response, which the experimental evidence suggests is
  first-order for welfare; the behavioural elasticities are calibrated
  to the literature but are not structural (the Sobol analysis in
  section 6E reports how much the results lean on them); and the
  two-channel headline numbers are now reported with seed dispersion
  (Figure 14), which is small.
\end{itemize}

\hypertarget{refs}{}

\section*{Declaration of generative AI use}
During the preparation of this work the author used large language models, Claude Opus 4.8 (Anthropic) and GPT-5.5 (OpenAI), as coding and writing assistants: to help implement, debug, and refactor the simulation and analysis code and to produce figures, and to draft and edit the manuscript. The author reviewed and edited all such output and takes full responsibility for the model, the experiments, the results, and the conclusions presented here.

\section*{References}
\small
\begingroup\setlength{\parindent}{0pt}
\hangindent=1.5em\hangafter=1 Acemoglu, D. and Restrepo, P. (2019). Automation and New Tasks: How Technology Displaces and Reinstates Labor. \textit{Journal of Economic Perspectives}, 33(2), 3--30.\par\addvspace{0.45em}
\hangindent=1.5em\hangafter=1 Acemoglu, D. and Restrepo, P. (2020). Robots and Jobs: Evidence from US Labor Markets. \textit{Journal of Political Economy}, 128(6), 2188--2244.\par\addvspace{0.45em}
\hangindent=1.5em\hangafter=1 Acemoglu, D. and Restrepo, P. (2022). Tasks, Automation, and the Rise in U.S. Wage Inequality. \textit{Econometrica}, 90(5), 1973--2016.\par\addvspace{0.45em}
\hangindent=1.5em\hangafter=1 Benhabib, J., Bisin, A. and Zhu, S. (2011). The Distribution of Wealth and Fiscal Policy in Economies with Finitely Lived Agents. \textit{Econometrica}, 79(1), 123--157.\par\addvspace{0.45em}
\hangindent=1.5em\hangafter=1 Brülhart, M., Gruber, J., Krapf, M. and Schmidheiny, K. (2022). Behavioral Responses to Wealth Taxes: Evidence from Switzerland. \textit{American Economic Journal: Economic Policy}, 14(4), 111--150.\par\addvspace{0.45em}
\hangindent=1.5em\hangafter=1 Caiani, A., Godin, A., Caverzasi, E., Gallegati, M., Kinsella, S. and Stiglitz, J. E. (2016). Agent based-stock flow consistent macroeconomics: Towards a benchmark model. \textit{Journal of Economic Dynamics and Control}, 69, 375--408.\par\addvspace{0.45em}
\hangindent=1.5em\hangafter=1 Carvalho, L. and Di Guilmi, C. (2020). Technological unemployment and income inequality: a stock-flow consistent agent-based approach. \textit{Journal of Evolutionary Economics}, 30(1), 39--73.\par\addvspace{0.45em}
\hangindent=1.5em\hangafter=1 De Loecker, J., Eeckhout, J. and Unger, G. (2020). The Rise of Market Power and the Macroeconomic Implications. \textit{Quarterly Journal of Economics}, 135(2), 561--644.\par\addvspace{0.45em}
\hangindent=1.5em\hangafter=1 Fagereng, A., Guiso, L., Malacrino, D. and Pistaferri, L. (2020). Heterogeneity and Persistence in Returns to Wealth. \textit{Econometrica}, 88(1), 115--170.\par\addvspace{0.45em}
\hangindent=1.5em\hangafter=1 Foster, L., Haltiwanger, J. and Tuttle, C. (2022). Rising Markups or Changing Technology? \textit{NBER Working Paper} 30491.\par\addvspace{0.45em}
\hangindent=1.5em\hangafter=1 Gates, B. (2017). Quoted in K. J. Delaney, ``The Robot That Takes Your Job Should Pay Taxes, Says Bill Gates.'' \textit{Quartz}, 17 February.\par\addvspace{0.45em}
\hangindent=1.5em\hangafter=1 Godley, W. and Lavoie, M. (2007). \textit{Monetary Economics.} Palgrave Macmillan.\par\addvspace{0.45em}
\hangindent=1.5em\hangafter=1 Guerreiro, J., Rebelo, S. and Teles, P. (2022). Should Robots Be Taxed? \textit{The Review of Economic Studies}, 89(1), 279--311.\par\addvspace{0.45em}
\hangindent=1.5em\hangafter=1 Hassett, K. A. and Hubbard, R. G. (2002). Tax Policy and Business Investment. In \textit{Handbook of Public Economics}, vol. 3, 1293--1343. Elsevier.\par\addvspace{0.45em}
\hangindent=1.5em\hangafter=1 Heckemeyer, J. H. and Overesch, M. (2017). Multinationals' Profit Response to Tax Differentials: Effect Size and Shifting Channels. \textit{Canadian Journal of Economics}, 50(4), 965--994.\par\addvspace{0.45em}
\hangindent=1.5em\hangafter=1 Information Technology and Innovation Foundation (ITIF) (2025). \textit{Digital Tax Policy} (country knowledge-base series, 11 February) and \textit{The Tortured Logic of Digital Services Taxes} (7 July). ITIF, Washington DC.\par\addvspace{0.45em}
\hangindent=1.5em\hangafter=1 International Monetary Fund (2024). Broadening the Gains from Generative AI: The Role of Fiscal Policies. \textit{IMF Staff Discussion Note} SDN/2024/002.\par\addvspace{0.45em}
\hangindent=1.5em\hangafter=1 Jakobsen, K., Kleven, H., Kolsrud, J., Landais, C. and Muñoz, M. (2024). Taxing Top Wealth: Migration Responses and their Aggregate Economic Implications. \textit{NBER Working Paper} 32153 (forthcoming, \textit{American Economic Review}).\par\addvspace{0.45em}
\hangindent=1.5em\hangafter=1 Kaldor, N. (1957). A Model of Economic Growth. \textit{The Economic Journal}, 67(268), 591--624.\par\addvspace{0.45em}
\hangindent=1.5em\hangafter=1 Moll, B., Rachel, L. and Restrepo, P. (2022). Uneven Growth: Automation's Impact on Income and Wealth Inequality. \textit{Econometrica}, 90(6), 2645--2683.\par\addvspace{0.45em}
\hangindent=1.5em\hangafter=1 Oberfield, E. and Raval, D. (2021). Micro Data and Macro Technology. \textit{Econometrica}, 89(2), 703--732.\par\addvspace{0.45em}
\hangindent=1.5em\hangafter=1 OECD (2024). \textit{Society at a Glance 2024: OECD Social Indicators}. OECD Publishing, Paris.\par\addvspace{0.45em}
\hangindent=1.5em\hangafter=1 Parnreiter, C., Steinwärder, L. and Kolhoff, K. (2024). Uneven Development through Profit Repatriation. \textit{Antipode}, 56(6), 2343--2367.\par\addvspace{0.45em}
\hangindent=1.5em\hangafter=1 Pasinetti, L. L. (1962). Rate of Profit and Income Distribution in Relation to the Rate of Economic Growth. \textit{The Review of Economic Studies}, 29(4), 267--279.\par\addvspace{0.45em}
\hangindent=1.5em\hangafter=1 Rognlie, M. (2015). Deciphering the Fall and Rise in the Net Capital Share. \textit{Brookings Papers on Economic Activity}, 2015(1), 1--69.\par\addvspace{0.45em}
\hangindent=1.5em\hangafter=1 Vivalt, E., Rhodes, E., Bartik, A., Broockman, D., Krause, P. and Miller, S. (2024). The Employment Effects of a Guaranteed Income: Experimental Evidence from Two U.S. States. \textit{NBER Working Paper} 32719.\par\addvspace{0.45em}
\endgroup\normalsize

\appendix
\renewcommand{\thesection}{A}
\hypertarget{appendix-a-assumptions-and-limitations}{%
\section{Appendix A: Assumptions and
limitations}\label{appendix-a-assumptions-and-limitations}}

This appendix catalogues what the model assumes and where it is most
likely to mislead. Section 10 lists caveats specific to individual
results and the work that would sharpen them; this appendix is the
foundational list, the assumptions built into the modelling approach
itself. The short version: treat the structural orderings as reliable
and the magnitudes as illustrative, and pair the model with the things
it cannot see.

\hypertarget{a.1-what-kind-of-model-this-is}{%
\subsection{A.1 What kind of model this
is}\label{a.1-what-kind-of-model-this-is}}

Stock-flow-consistent and agent-based models of this kind are
consistency engines, not forecasting machines (the Godley and Lavoie
tradition). Their value is that money cannot appear or vanish, every
assumption is explicit and can be swept, and the channel driving each
result can be switched on and off. They answer conditional questions
("if the rent is mobile and foreign-held, which tax reaches it?") well,
and unconditional questions ("what will ownership be in year X?") badly.
Box\textquotesingle s aphorism is literal here: the model is wrong, and
it is useful for orderings and mechanisms rather than point estimates.

\hypertarget{a.2-the-central-assumption-that-the-rent-exists}{%
\subsection{A.2 The central assumption: that the rent
exists}\label{a.2-the-central-assumption-that-the-rent-exists}}

The entire analysis rests on a large, durable, separable, foreign-held
AI rent. This is calibrated, not estimated. The frontier labs lose money
today, so the rent is a bet on a future durable margin, and our own
competition result shows it can erode if the frontier commoditises. It
is also not cleanly observable: separating an IP rent from a competitive
return and from ordinary profit is a contested measurement problem, so
the surplus the model treats as a clean markup is, in practice, reached
only through proxies (a revenue or consumption base, a withholding on
the licence fee, or a residual-profit allocation such as Pillar
One\textquotesingle s Amount A) and adversarial transfer-pricing rules.
If the rent is smaller, more competed, or harder to characterise than
assumed, the problem the paper addresses shrinks with it.

A related assumption is that the rent sits only on the AI side. The
model treats the robotic channel as purely competitive (reproducible
hardware earning a normal return), so all the rent is on the cognitive
cluster. This is conservative, because a robot is hardware plus IP, and
the IP that runs it (the control or foundation model, plus the
proprietary design) is scarce and foreign-held in the same way the AI
model is. An off-default lever (robot\_ip) puts an embodied-IP rent on
the robotic cluster and confirms it behaves exactly like the AI rent: it
raises the total foreign-held rent (from about an eighth of output to
about a quarter at a robot markup equal to the AI one), accelerates the
ownership drift, and is reached by the same source levy while a hardware
robot tax misses it, leaving the instrument ranking unchanged (section
6, Figure 27). The economically meaningful distinction is therefore
competitive hardware versus rent-bearing IP, not robots versus AI;
confining the rent to the AI cluster understates rather than overstates
it.

\hypertarget{a.3-production-and-technology}{%
\subsection{A.3 Production and
technology}\label{a.3-production-and-technology}}

Output is an aggregate nested CES function with a small number of
substitution elasticities. Whether automation raises or lowers the
labour share turns on whether the relevant elasticity exceeds one, which
is empirically contested and may not be a stable structural object at
all (the Cambridge-capital objection that aggregate capital is not
always coherent). The task framework (Acemoglu and Restrepo) is the
modern repair and supplies the reinstatement margin, but
Rognlie\textquotesingle s caution that much of the measured
capital-share rise is housing and markup rather than productive capital
is a live objection. Crucially, the automation path is imposed as an
exogenous logistic ramp: the model gives the consequences if AI
automates cognitive work on that schedule, not whether or how far it
will. The largest real-world uncertainty is an input, not a result.

\hypertarget{a.4-closure-and-finance}{%
\subsection{A.4 Closure and finance}\label{a.4-closure-and-finance}}

Several closure choices bite directly on the headline numbers. Output
sits at potential, so there is no demand-side recession from
displacement, arguably the outcome the displacement literature fears
most. Saving equals investment by construction, equity is valued at
book, and there is no inflation, monetary policy, banking, credit cycle
or asset market, so financial fragility and bubbles are absent. Most
consequentially, the baseline is a closed economy, so the foreign owner
can only recycle its rent into domestic assets, which is what
mechanically produces the dramatic "buys the whole capital stock"
result; adding a trade balance turns that single number into a range.
The ownership headline should therefore be read as the fully-reinvested
pole, not a central estimate.

\hypertarget{a.5-behaviour-and-expectations}{%
\subsection{A.5 Behaviour and
expectations}\label{a.5-behaviour-and-expectations}}

Agents follow fixed behavioural rules rather than re-optimising when
policy changes, which invites the Lucas critique: in reality the
powerful party, the foreign rent-owner, would respond hardest, and most
of that response is unmodelled (the one channel we did open, the owner
choosing where to book the rent, mattered and forced the anti-avoidance
rider). There is also no genuine uncertainty or expectations formation,
so there is no precautionary behaviour, no investment boom on
anticipated AI returns, and no front-running of a pre-announced tax.
Beliefs about where AI is heading, a dominant force in the real story,
are absent.

\hypertarget{a.6-aggregation-who-loses-is-invisible}{%
\subsection{A.6 Aggregation: who loses is
invisible}\label{a.6-aggregation-who-loses-is-invisible}}

The economy has one good, no input-output structure across sectors, a
coarse cognitive-versus-routine labour split, no spatial or regional
dimension, and (in the baseline) a single representative firm rather
than competing AI-native and incumbent firms. AI\textquotesingle s
impact is highly uneven across industries, regions and skills, and none
of that texture survives. The model speaks to the labour share and the
wealth distribution in aggregate, not to which workers or places are
hit, and the rising-rent output-capture channel is a reduced-form proxy
for firm-level displacement rather than a mechanism (see section 6 and
section 10).

\hypertarget{a.7-the-distribution-engine}{%
\subsection{A.7 The distribution
engine}\label{a.7-the-distribution-engine}}

Wealth concentration is generated by a heterogeneous-returns
random-growth process (Benhabib, Bisin and Zhu; Fagereng and
co-authors). The inequality results lean on the assumed level and
dispersion of returns to wealth, which is among the least-settled inputs
empirically and one of the main drivers of the headline distributional
numbers. The finding that stock taxes compress inequality while flow
taxes do not is partly a property of how this engine is built, and
should be read as internally consistent rather than independently
established.

\hypertarget{a.8-what-the-model-is-silent-on}{%
\subsection{A.8 What the model is silent
on}\label{a.8-what-the-model-is-silent-on}}

The silences are where a policymaker is most exposed, because they never
appear as caveats inside the model. It assumes the levy can be imposed
and collected, but the binding real-world constraint is political
economy: the owner lobbies, threatens to withdraw, and its home
government retaliates, which the paper treats as narrative rather than
as a modelled strategic game. It focuses on the distribution of the
surplus and largely ignores the consumer-surplus gains if AI makes goods
cheaper, so it may understate the size-of-pie benefits. And it has no
welfare function, so any "preferable regime" language is the
authors\textquotesingle{} gloss, not the model\textquotesingle s output.

\hypertarget{a.9-the-deepest-limitation-a-moving-structure}{%
\subsection{A.9 The deepest limitation: a moving
structure}\label{a.9-the-deepest-limitation-a-moving-structure}}

Structural modelling assumes deep parameters (elasticities,
propensities) that survive the policy change being studied. The sharper
worry here is not the Lucas critique narrowly but that the entire
subject is a regime change, in what a firm is, how value is produced,
and what labour does, modelled with parameters calibrated to the world
before that change. If AI genuinely transforms production, the structure
itself is moving, and a model fitted to the recent past has uncertain
ground beneath it. This is the strongest single reason to read the work
as disciplined conditional reasoning rather than prediction.

\hypertarget{a.10-what-to-trust-and-what-to-discount}{%
\subsection{A.10 What to trust, and what to
discount}\label{a.10-what-to-trust-and-what-to-discount}}

Putting the above together, the conclusions sort into two groups.

\begin{itemize}
\tightlist
\item
  \textbf{Robust (structural, close to accounting facts).} Which base
  reaches the rent and why (a source levy over a robot, corporate or
  token tax); the territoriality asymmetry (decisive for the
  provider\textquotesingle s home, weak for a rent-importing host); and
  the direction that an untaxed, reinvested, foreign-held rent transfers
  ownership over time. These depend on the sign and presence of a
  mechanism, not on magnitudes, and survive sweeping the uncertain
  parameters.
\item
  \textbf{Fragile (calibrated, assumption-sensitive).} Every magnitude:
  the rent at an eighth of output, the seventy per cent ownership share,
  the rising-rent quarter or half, and above all the unemployment range,
  which the sensitivity analysis flags as the least anchored output.
  These should be read as illustrative ranges set by parameters that are
  bounded but not estimated.
\item
  \textbf{Outside the model.} Enforceability and political economy,
  demand-side macro, the pace and reach of AI capability, and welfare. A
  recommendation can be correct inside the model and still fail in the
  world for these reasons, so the model is a complement to judgement on
  them, not a substitute.
\end{itemize}

The intended use follows from this split: base the recommendations on
the orderings and the framework, report the numbers as ranges, and weigh
them against the constraints the model cannot see.

\end{document}